\newcommand*{\ATLASLATEXPATH}{latex/}
\pdfoutput=1
\documentclass[cernpreprint, texlive=2016, UKenglish]{\ATLASLATEXPATH atlasdoc}
\PreprintIdNumber{CERN-EP-2017-274}
\usepackage[subfigure]{\ATLASLATEXPATH atlaspackage}
\usepackage{\ATLASLATEXPATH atlasbiblatex}
\usepackage{\ATLASLATEXPATH atlascontribute}
\usepackage[process=true]{\ATLASLATEXPATH atlasphysics}
\usepackage{tocloft}
\usepackage{multirow}
\usepackage{tabularx}
\usepackage{threeparttable}
\graphicspath{{logos/}{figures/}}
\usepackage{METPaper2015-defs}
\usepackage{atlasmetdefs}
\usepackage{mathtools}
\AtlasTitle{Performance of missing transverse momentum reconstruction with the ATLAS detector using proton--proton collisions at $\sqrt{s}=\SI{13}{\TeV}$}

\author{The ATLAS Collaboration}

 \usepackage{authblk}
 \author{The ATLAS Collaboration}

\AtlasRefCode{PERF-2016-07}

\AtlasJournalRef{Eur. Phys. J. C 78 (2018) 903}
\AtlasDOI{10.1140/epjc/s10052-018-6288-9}

\AtlasAbstract{%
The performance of the missing transverse momentum (\met) reconstruction with the ATLAS detector is evaluated using data collected in 
proton--proton collisions at the LHC at a \cms energy of \unit{13}{\TeV} in 2015.
To reconstruct \met, fully calibrated electrons, muons, photons, hadronically decaying \tauleps, and jets reconstructed from 
calorimeter energy deposits and charged-particle tracks are used. 
These are combined with the soft hadronic activity measured by reconstructed charged-particle tracks not associated with the hard objects. 
Possible double counting of contributions from reconstructed charged-particle tracks from the inner detector, 
energy deposits in the calorimeter, and reconstructed muons from the muon spectrometer is avoided by applying a signal ambiguity 
resolution procedure which rejects already used signals when combining the various \met contributions. 
The individual terms as well as the overall reconstructed \met are evaluated with various performance metrics for scale 
(linearity), resolution, and sensitivity to the data-taking conditions. 
The method developed to determine the systematic uncertainties of the \met scale and resolution is discussed. 
Results are shown based on the full 2015 data sample corresponding to an integrated luminosity of \ilum.

}

\hypersetup{pdftitle={ATLAS draft},pdfauthor={The ATLAS Collaboration}}

\makeatletter
\renewcommand{\numberline}[1]{\@cftbsnum #1\@cftasnum~\@cftasnumb}%
\makeatother

\begin{document}

\maketitle

\tableofcontents

\clearpage
\section{Introduction}
\label{sec:intro}
The missing transverse momentum (\met) is an important observable serving as an experimental proxy for the transverse momentum carried by 
undetected particles produced in proton--proton (\pp) collisions measured with the \ATLAS detector \cite{PERF-2007-01} at the Large Hadron Collider (\LHC).
It is reconstructed from the signals of detected particles in the final state. 
A value incompatible with zero may indicate not only the production of Standard Model (SM) neutrinos but also the production of new particles suggested in models for physics beyond the SM that escape the \ATLAS detector without being detected.
The reconstruction of \met is challenging because it involves all detector subsystems and requires the most complete and unambiguous representation 
of the hard interaction of interest by calorimeter and tracking signals. 
This representation is obscured by limitations introduced by the detector acceptance and by signals and signal remnants from additional \pp interactions occurring in the same, previous and subsequent \LHC bunch crossings (\pu) relative to the triggered hard-scattering.
\ATLAS has developed successful strategies for a high-quality \met reconstruction focussing on the minimisation of effects introduced by \pu for the data recorded between 2010 and 2012 (\LHC \runOne) \cite{Aad:2012re,Aad:2016nrq}. 
These approaches are the basis for the \met reconstruction developed for the data collected in 2015 (\LHC \runTwo) that is described in this paper, together with results from performance evaluations and the determination of systematic uncertainties.

This paper is organised as follows. The subsystems forming the \ATLAS detector are described in \secRef{sec:detector}. 
The \met reconstruction is discussed in \secRef{sec:etmiss-reco}. 
The extraction of the data samples and the generation of the Monte Carlo (\MC) simulation samples are presented in \secRef{sec:samples}.
The event selection is outlined in \secRef{sec:event-selection}, followed by results for \met performance in \secRef{sec:perf}.
\SecRef{sec:uncertainties} comprises a discussion of methods used to determine systematic uncertainties associated with the \met measurement, and the presentation of the corresponding results. \SecRef{sec:etmiss-variants} describes variations of the \met reconstruction using calorimeter signals for the soft hadronic event activity, or reconstructed charged-particle tracks only.
The paper concludes with a summary and outlook in \secRef{sec:conclusion}. 
The nomenclature and conventions used by \ATLAS for \met-related variables and descriptors can be found in \appRef{app:nomenclature}, while the composition of \met reconstruction variants is presented in \appRef{app:composition}. 
An evaluation of the effect of alternative jet selections on the \met reconstruction performance is given in \appRef{app:jet}.

\section{ATLAS detector}
\label{sec:detector}
\newcommand{\AtlasCoordFootnote}{%
ATLAS uses a right-handed coordinate system with its origin at the nominal interaction point (IP)
in the centre of the detector and the $z$-axis along the beam pipe.
The $x$-axis points from the IP to the centre of the LHC ring,
and the $y$-axis points upwards.
Cylindrical coordinates $(r,\phi)$ are used in the transverse plane,
$\phi$ being the azimuthal angle around the $z$-axis.
The pseudorapidity is defined in terms of the polar angle $\theta$ as $\eta = -\ln \tan(\theta/2)$.
Angular distance is measured in units of $\Delta R \equiv \sqrt{(\Delta\eta)^{2} + (\Delta\phi)^{2}}$.\label{foot:geom}}

The ATLAS experiment at the LHC features a multi-purpose particle
detector with a forward--backward symmetric cylindrical geometry and a nearly full ($4\pi$) 
coverage in solid angle.\footnote{\AtlasCoordFootnote}
It consists of an inner detector (\ID) tracking system in a \unit{2}{\text{T}} axial magnetic field provided by a superconducting solenoid.
The solenoid is surrounded by electromagnetic and hadronic calorimeters, and a muon spectrometer (\MS).
The \ID covers the pseudorapidity range $\abseta < 2.5$, and consists of a silicon
pixel detector, a silicon microstrip detector and
a transition radiation tracker for $\abseta < 2.0$.
During the \LHC shutdown between \runOne and \runTwo, a new tracking layer, known as the
insertable B-layer~\cite{ATLAS-TDR-19}, was added between the previous innermost pixel layer and a new, narrower beam pipe.

The high-granularity lead/liquid-argon (\LAr) sampling electromagnetic
calorimeter covers the region $\abseta < 3.2$. 
The regions $\abseta < 1.37$ and $1.5 < \abseta < 1.8$ are instrumented with \presamplers in front of the \LAr calorimeter in the same cryostat.
A steel/scintillator-tile calorimeter (\Tile) provides hadronic coverage in the central pseudorapidity range $\abseta < 1.7$.
\LAr technology is also used for the hadronic calorimeters in the \ec region $1.5 < \abseta < 3.2$ and for electromagnetic and hadronic energy
measurements in the forward calorimeters covering  $3.2 < \abseta < 4.9$.

The \MS surrounds the calorimeters.
It consists of three large superconducting air-core toroidal magnets, precision
tracking chambers providing precise muon tracking out to $\abseta = 2.7$,
and fast detectors for triggering in the region $\abseta < 2.4$.

A two-level trigger system is used to select events \cite{TRIG-2016-01}.
A low-level hardware trigger reduces the data rate, and a high-level software trigger selects events with interesting final states. 
More details of the \ATLAS detector can be found in \citRef{PERF-2007-01}.

\section{\texorpdfstring{\met}{ETmiss} reconstruction}
\label{sec:etmiss-reco}
The reconstructed \met in \ATLAS is characterised by two contributions. 
The first one is from the \emph{hard-event} signals comprising fully reconstructed and calibrated particles and jets (hard objects). 
The reconstructed particles are electrons, photons, \tauleps, and muons. 
While muons are reconstructed from \ID and \MS tracks, electrons and \tauleps are identified combining calorimeter signals with tracking information. 
Photons and jets are principally reconstructed from calorimeter signals, with possible signal refinements from reconstructed tracks.
The second contribution to \met is from the \emph{soft-event} signals consisting of reconstructed charged-particle tracks (soft signals) associated with the \hs vertex defined in \appRef{app:nomenclature} but not with the hard objects.

\ATLAS carries out a dedicated reconstruction procedure for each kind of particle
as well as for jets, casting a particle or jet hypothesis on the origin of (a group of) detector signals. 
These procedures are independent of one another. 
This means that e.g.~the same calorimeter signal used to reconstruct an electron is likely also used to reconstruct a jet, thus potentially introducing 
double counting of the same signal when reconstructing \met. 
This issue is addressed by the explicit \emph{signal ambiguity resolution} in the object-based \met reconstruction originally introduced in 
\citMultiRef{Aad:2012re}{and}{Aad:2016nrq}, and by its 2015 implementation described in 
\secMultiRef{subsec:etmiss-basics}{and}{subsec:etmiss-definition}.

Additional options for the 
set of signals used to reconstruct \met are available and discussed in detail in \secRef{sec:etmiss-variants}. 
One of these alternative options is the calorimeter-based \met reconstruction discussed in \secRef{subsec:etmiss-calo},
which uses a soft event built from clusters of topologically connected calorimeter cells (\topos) \cite{Aad:2016upy}.
Another option is the track-based missing transverse momentum, which differs from \met only in the use of tracks in place of jets. 
It is described in more detail in \secRef{subsec:etmiss-track}.

\subsection{\texorpdfstring{\met}{Etmiss} basics}
\label{subsec:etmiss-basics}
The missing transverse momentum reconstruction provides a set of observables constructed from the components \pxy of the transverse momentum vectors (\pTvec) of the various contributions. 
The missing transverse momentum components \metxy serve as the basic input for most of these observables. They are given by
\begin{align}
	\metxy &=  - \sum_{i\in\{\text{hard objects}\}} \pxyi{i} - \sum_{j\in\{\text{soft signals}\}} \pxyi{j} \label{eq:metbasics:comp}\,.	
\end{align}   
The set of observables constructed from \metxy is 
\begin{align}
	\metvec & = (\metx,\mety)\,,                                            \label{eq:metbasics:vector} \\
	\met    &= |\metvec| = \sqrt{(\metx)^{2}+(\mety)^{2}}\,, \label{eq:metbasics:met}   \\
	\metphi & = \tan^{-1}(\mety/\metx)\,.                                  \label{eq:metbasics:phi}
\end{align}
The vector \metvec provides the amount of the missing transverse momentum via its magnitude \met, and its direction in the transverse plane in terms of the azimuthal angle \metphi. 
Consequently, \met is non-negative by definition. 
However, in an experimental environment where not all relevant \pT from the hard-scatter interaction can be reconstructed and used in \eqRef{eq:metbasics:comp}, and the reconstructed \pT from each contribution is affected by the limited resolution of the detector, an \emph{observation bias} towards non-vanishing values for \met is introduced
even for final states without genuine missing transverse momentum generated by undetectable particles.  

The scalar sum of all transverse momenta ($\pT = |\pTvec|$) from the objects contributing to \met reconstruction is given by
\begin{align}
	\sumet &=  \sum_{i\in\{\text{hard objects}\}} \pTi{i} + \sum_{j\in\{\text{soft signals}\}} \pTi{j}\,. \label{eq:metbasics:sumet}
\end{align}
In the context of \met reconstruction, \sumet is calculated in addition to the sum given in \eqRef{eq:metbasics:comp}, and the derived quantities defining \met given in \eqMultiRef{eq:metbasics:vector}{to}{eq:metbasics:phi}. 
It provides a useful overall scale for evaluating the hardness of the \hs event in the transverse plane, and thus provides a measure for the event activity in physics analyses and \met reconstruction performance studies.

In the calculation of \metxy and \sumet the contributing objects need to be reconstructed from mutually exclusive detector signals. 
This rule avoids multiple inclusions of the same signal in all constructed observables. 
The implementation of this rule in terms of the signal ambiguity resolution 
requires the definition of a sequence for selected contributions, 
in addition to a rejection mechanism based on common signal usage between different objects. 
Similarly to the analysis presented in \citRef{Aad:2016nrq}, the most commonly used order for the \met reconstruction sequence for the hard-object 
contribution starts with electrons (\electron), followed by photons (\photon), then hadronically decaying \tauleps (\taupm), and finally jets.
Muons (\muon) are principally reconstructed from \ID and \MS tracks alone, with corrections based on their energy loss in the calorimeter, leading to 
little or no signal overlap with the other reconstructed particles in the calorimeter. 
 
In the sequence discussed here, all electrons passing the selection enter the \met reconstruction first. 
The lower-priority reconstructed particles (\photon, \taupm) are fully rejected if they share their calorimeter signal with a higher-priority object that has already entered the \met reconstruction.
Muons experience energy loss in the calorimeters, but only non-isolated muons overlap with other hard objects, most likely jets or \tauleps. 
In this case the muon's energy deposit in the calorimeter cannot be separated from the overlapping jet-like objects with the required precision, 
and the calorimeter-signal-overlap resolution based on the shared use of \topos cannot be applied. 
A discussion of the treatment of isolated and non-isolated muons is given in \secRef{subsec:muon-sel}.

Generally, jets are rejected if they overlap with accepted higher-priority particles.
To avoid signal losses for \met reconstruction in the case of partial or marginal overlap, and to suppress the accidental
inclusion of jets reconstructed from calorimeter signals from large muon energy losses or \pu, the more refined overlap resolution strategies 
described in \secMultiRef{subsec:jet-sel}{and}{subsec:muon-jet} are applied. 

Excluding \ID tracks associated with any of the accepted hard objects contributing to \met, \ID tracks from the hard-scatter collision vertex are used to construct the soft-event signal for the results presented in this paper.

\newlength{\bwidth}
\newlength{\cwidth}
\newlength{\nwidth}
\newlength{\rwidth}
\newlength{\dwidth}
\newlength{\twidth}

\newlength{\xheight}
\setlength{\xheight}{\heightof{x}}
\settowidth{\bwidth}{\LAND{{\abseta < 1.37}}{{1.52 < \abseta < 2.47}}}
\settowidth{\cwidth}{$\lor$}
\addtolength{\bwidth}{-\cwidth}
\settowidth{\nwidth}{\text{\topo}}
\settowidth{\rwidth}{$\mathbb{P}$}
\settowidth{\dwidth}{Variables}
\setlength{\twidth}{\dimexpr 0.98\textwidth}
\addtolength{\twidth}{\dimexpr -\rwidth-2\tabcolsep} 
\addtolength{\twidth}{\dimexpr -\nwidth-2\tabcolsep}
\addtolength{\twidth}{\dimexpr -\bwidth-2\tabcolsep} 
\addtolength{\twidth}{\dimexpr -\dwidth-2\tabcolsep} 	
\addtolength{\twidth}{\dimexpr -2\tabcolsep}

\setlength{\fboxsep}{1pt}
\setlength{\fboxrule}{0.5pt}

\begin{table}[hp!] \centering
	\caption[]{Overview of the contributions to \met{} and \sumet{} from hard objects such as electrons (\electron), photons (\photon), 
		     hadronically decaying \tauleps (\tauhad), muons (\muon), and jets, together with the signals for the soft term. The configuration shown 
		     is the one used as reference for the performance evaluations presented in this paper. The table is ordered descending in priority $\mathbb{P}$
		     of consideration for \met{} reconstruction, with (\ref{ref:epm}) being the first and (\ref{ref:jet}) being the last calculated hard-object contribution.
		     The soft-event contribution is constructed at the lowest priority (\ref{ref:track}), after all hard objects are considered.
		     The transverse (longitudinal) impact parameter \dzero (\zzsth) used to select the \ID tracks contributing to \metsft and \sumetsft in 
                     $\mathbb{P} = (\ref{ref:track})$ is measured relative to the \hs vertex. All variables are explained in 
                     \secRef{subsec:etmiss-definition}. The angular distance \deltaR between objects is defined as        
                     $\deltaR = \sqrt{(\Delta\eta)^{2}+(\Delta\phi)^{2}}$.\vspace*{1mm}
		     \label{tab:contrib:default}}

\renewcommand{\arraystretch}{1.45}

\setcounter{myrefctr}{0}

\begin{threeparttable}[b]
\begin{tabular}{p{\rwidth}>{$}p{\nwidth}<{$}>{$}p{\bwidth}<{$}>{\centering}p{\dwidth}p{\twidth}}

       \hline \hline
       \multirow{2}{*}{$\mathbb{P}$}    & \MCC{4}{Objects contributing to \met and \sumet}                         \tabularnewline \cline{2-5} 
                                        & \LCC{1}{Type} & \LCC{1}{Selections}               & Variables & Comments \tabularnewline \hline
	\LCC{1}{\myrefstep{ref:epm}} 	& \electron                                                                                     & 
		\minitab[l]{\mstacktwo{\LOR{{\abseta < 1.37}}{{1.52 < \abseta < 2.47}}}{\pT > \unit{10}{\GeV}}}                         & 
		\minitab[c]{\mstacktwo{\metele}{\sumetele}}                   								&
		all $e^{\pm}$ passing medium re\-con\-struc\-tion quality and kinematic selections                                            \tabularnewline\hline
	\LCC{1}{\myrefstep{ref:gam}}    & \photon											& 
		\minitab[l]{\mstacktwo{\LOR{{\abseta < 1.37}}{{1.52 < \abseta < 2.47}}}{\pT > \unit{25}{\GeV}}}                         &
		\minitab[c]{\mstacktwo{\metgam}{\sumetgam}}                         							&
		all \photon passing tight quality and kinematic selections in reconstruction, and with\-out signal overlap with (\ref{ref:epm}) \tabularnewline\hline
	\LCC{1}{\myrefstep{ref:tau}}	& \tauhad       										& 
		\minitab[l]{\mstacktwo{\LOR{{\abseta < 1.37}}{{1.52 < \abseta < 2.47}}}{\pT > \unit{20}{\GeV}}}                      	&
		\minitab[c]{\mstacktwo{\mettau}{\sumettau}}         									&
		all \tauhad passing medium reconstruction quality and kinematic selections, and without signal overlap with (\ref{ref:epm}) and (\ref{ref:gam}) \tabularnewline\hline
	\LCC{1}{\myrefstep{ref:mu}}     & \muon 											& 
		\minitab[l]{\mstacktwo{\abseta < 2.7}{\pT > \unit{10}{\GeV}}}                                      		 	&  
		\minitab[c]{\mstacktwo{\metmuo}{\sumetmuo}}                                                                             & 
		all \muon passing medium quality and kinematic selections in reconstruction; for the discussion of the \muon--jet overlap removal see \secRef{subsec:muon-jet} \tabularnewline\hline
	\LCC{1}{\myrefstep{ref:jet}}    & \jet												& 
 		\minitab[l]{\tstackfive{\mstacktwo{\abseta < 4.5}{\pT > \unit{60}{\GeV}}}
                                       {\hrulefill\ \text{or\ } \hrulefill }
                                       {\mstacktwo{2.4 < \abseta < 4.5}{\unit{20}{\GeV} < \pT < \unit{60}{\GeV}}}
                                       {\hrulefill\ \text{or\ } \hrulefill }
                                       {\mstackthree{\abseta < 2.4}{\unit{20}{\GeV} < \pT < \unit{60}{\GeV}}{\JVT > 0.59}}}             &                                                
 		\minitab[c]{\mstacktwo{\metjet}{\sumetjet}}				                                                &
 		all jets passing re\-con\-struc\-tion quality (jet cleaning) and kinematic selections, and without signal o\-ver\-lap\tnote{$\dagger$} with  (\ref{ref:epm})--(\ref{ref:tau}); for the dedica\-ted overlap re\-mo\-val stra\-te\-gy with \muon from (\ref{ref:mu}) see \secRef{subsec:muon-jet}
		\tabularnewline\hline
        \LCC{1}{\myrefstep{ref:track}}     										         	& 
        		\ID \text{\ track}    											        & 
       		\minitab[l]{ 	$\pT > \unit{400}{\MeV}$ 					\\
       				$\absdzero < \unit{1.5}{\mm}$ 					\\
       				$\abszzsth < \unit{1.5}{\mm}$					\\
        		                 	$\Delta R(\track,\text{\electron-/\photon cluster}) > 0.05$ \\
        		                 	$\Delta R(\track,\tauhad) > 0.2$                }                                       &
   		\mstacktwo{\metsft}{\sumetsft}										                &					
		all \ID\ tracks from the hard-scatter vertex pass\-ing reconstruction quality and kinematic selections, and not associated with any particle from (\ref{ref:epm}), 
                (\ref{ref:tau}) or (\ref{ref:mu}), or ghost-associated with a \jet from (\ref{ref:jet}) 
 		\tabularnewline	\hline \hline
\end{tabular}

\begin{tablenotes}
\footnotesize
\item[$\dagger$]While for single reconstructed particles no overlap is accepted at all, jets with a signal overlap fraction $\of < 50$\%  can still contribute their associated tracks to \metsft if those pass the selections for $\mathbb{P} = (\ref{ref:track})$, as discussed in \secRef{subsec:jet-sel}. The definition of \of is given in \eqRef{eq:of}.
\end{tablenotes}

\end{threeparttable}

\end{table}

\subsection{\texorpdfstring{\met}{Etmiss} terms}
\label{subsec:etmiss-definition}

Particle and jet selections in a given analysis should be reflected in \met and \sumet for a consistent interpretation of a given event. 
Each reconstructed particle and jet has its own dedicated calibration translating the detector signals into a fully corrected four-momentum. 
This means that e.g. re\-ject\-ing certain electrons in a given analysis can change both \met and \sumet, if the corresponding calorimeter signal is included and calibrated as a jet or a significant part of a jet. 
This also means that systematic uncertainties for the different particles can be consistently propagated to \met.
The applied selections are presented in \secRef{subsec:object-selection}, and summarised in \tabRef{tab:contrib:default}.

In \ATLAS the flexibility needed to recalculate \met and \sumet under changing analysis requirements for the same event is implemented using dedicated variables corresponding to specific object contributions. 
In this approach the full \metvec is the vectorial sum of missing transverse momentum terms \metvecj{p}, with $p \in \{\electron,\photon,\taupm,\muon,\text{jet}\}$ reconstructed from the $\pTvec = (p_{x},p_{y})$ of accepted particles and jets, and the corresponding soft term \metvecj{\text{soft}} from the soft-event signals introduced in \secRef{subsec:etmiss-basics} and further specified in \secRef{subsec:etmiss-softterm}. 
This yields\footnote{In this formula the notion of \textit{selected}, which is only applicable to electrons and muons, means that the choice of reconstructed particles is purely given by a set of criteria similar to those given in \secMultiRef{subsec:electron-sel}{and}{subsec:muon-sel}, respectively, with possible modifications imposed by a given analysis. The notion of \textit{accepted} indicates a modification of the set of initially selected objects imposed by the signal ambiguity resolution.}  
\begin{align} \label{eq:etmiss_terms}
	\metvec = \underbracket{
			                   \underbracket{\vphantom{- \sum_{\substack{ \text{accepted} \\ \text{photons}}}} - \sum_{\substack{ \text{selected}  \\ \text{electrons}  }} \pTvecj{\electron}}_{\metvele}
		                           \underbracket{                                                                                                   - \sum_{\substack{ \text{accepted} \\ \text{photons}   }} \pTvecj{\photon}}_{\metvgam}
		                           \underbracket{\vphantom{- \sum_{\substack{ \text{accepted} \\ \text{photons}}}} - \sum_{\substack{ \text{accepted} \\ \tauleps             }} \pTvecj{\tauhad}}_{\metvtau}
		                           \underbracket{\vphantom{- \sum_{\substack{ \text{accepted} \\ \text{photons}}}} - \sum_{\substack{ \text{selected}  \\ \text{muons}     }} \pTvecj{\muon}}_{\metvmuo}
		                           \underbracket{\vphantom{- \sum_{\substack{ \text{accepted} \\ \text{photons}}}} - \sum_{\substack{ \text{accepted} \\ \text{jets}         }} \pTvecj{\jet}}_{\metvjet}
                                          }_{\text{hard term}}
                        \underbracket{
                                           \underbracket{\vphantom{- \sum_{\substack{ \text{accepted} \\ \text{photons}}}} - \sum_{\substack{ \text{unused}    \\ \text{tracks}     }} \pTvecj{\track}}_{\vphantom{\metvtau}\metvsft}
                                          }_{\text{soft term}}\,.
\end{align}
The \met and \metphi observables can be constructed according to \eqMultiRef{eq:metbasics:met}{and}{eq:metbasics:phi}, respectively, 
for the overall missing transverse momentum (from \metvec) as well as for each individual term indicated in \eqRef{eq:etmiss_terms}.   
In the priority-ordered reconstruction sequence for \met, contributions are defined by a combination of 
analysis-dependent selections and a possible rejection due to the applied signal ambiguity resolution.
The muon and electron contributions are typically not subjected to the signal overlap resolution and are thus exclusively defined by the selection requirements. 
Unused tracks in \eqRef{eq:etmiss_terms} refers to those tracks associated with the \hs vertex but not with any hard object.
Neutral particle signals from the calorimeter suffer from significant contributions from \pu and are not included in the soft term.  
 
Correspondingly, \sumet is calculated from the scalar sums of the transverse momenta of hard objects entering the \met reconstruction and the soft term,
\begin{align}
	\sumet = \underbracket{\sum_{\substack{\text{selected} \\ \text{electrons}}} \pTj{\electron} 
	+ \sum_{\substack{\text{accepted} \\ \text{photons}}} \pTj{\photon} 
	+ \sum_{\substack{\text{accepted} \\ \tauleps}} \pTj{\taupm} 
	+ \sum_{\substack{\text{selected} \\ \text{muons}}} \pTj{\muon}
	+ \sum_{\substack{\text{accepted} \\ \text{jets}}} \pTj{\text{jet}}}_{\text{hard term}} 
	+ \underbracket{\vphantom{\sum_{\substack{\text{accepted} \\ \text{photons}}} \pTj{\photon}} \sum_{\substack{\text{unused} \\ \text{tracks}}} \pTj{\text{track}}}_{\text{soft term}}\,.
\label{eq:sumet}
\end{align}

The hard term in both \met and \sumet is characterised by little dependence on \pu, as it includes only fully calibrated objects, where the calibration includes a \pu correction and objects tagged as originating from \pu are removed. 
The particular choice of using only tracks from the hard-scatter vertex for the soft term strongly suppresses \pu contributions to this term as well. 
The observed residual \pu dependencies are discussed with the performance results shown in \secRef{sec:perf}.

\subsection{Object selection}
\label{subsec:object-selection}
The following selections are applied to reconstructed particles and jets used for the performance evaluations presented in \secMultiRef{sec:perf}{to}{sec:etmiss-variants}.  
Generally, these selections require refinements to achieve optimal \met reconstruction performance in the context of a given physics analysis, 
and the selections performed in this study are an example set of criteria.

\subsubsection{Electron selection}
\label{subsec:electron-sel}

Reconstructed electrons are selected on the basis of their shower shapes in the calorimeter and how well their calorimeter cell clusters are matched to \ID tracks \cite{Aad:2014nim}. 
Both are evaluated in a combined likelihood-based approach \cite{ATLAS-CONF-2014-032}. 
Electrons with at least medium reconstruction quality are selected. 
They are calibrated using the default calibration given in \citRef{Aad:2014nim}. To be considered for \met reconstruction,
electrons passing the reconstruction quality requirements are in addition required to have 
$\pT > \unit{10}{\GeV}$ and $\abseta < 1.37$ or $1.52 < \abseta < 2.47$,  to avoid the transition region between the central and \ec 
electromagnetic calorimeters.
Any energy deposit by electrons within $1.37 < \abseta < 1.52$ is likely reconstructed as a jet and enters the \met reconstruction as such, 
if this jet meets the corresponding selection criteria discussed in \secRef{subsec:jet-sel}. 

\subsubsection{Photon selection}
\label{subsec:photon-sel}

The identification and reconstruction of photons exploits the distinctive evolution of their electromagnetic showers in the 
calorimeters \cite{Aaboud:2016yuq}. Photons are selected and calibrated using the tight selection criteria given in \citRef{Aad:2014nim}.
In addition to the reconstruction quality requirements, photons must have $\pT > \unit{25}{\GeV}$ and $\abseta < 1.37$ or $1.52 < \abseta < 2.37$ to be 
included in the \met reconstruction.
Similarly to electrons, photons emitted within $1.37 < \abseta < 1.52$ may contribute to \met as a jet.

\subsubsection{\taulep selection}
\label{subsec:tau-sel}

Hadronically decaying \tauleps are reconstructed from narrow jets with low associated track multiplicities~\cite{Aad:2015unr}. 
Candidates must pass the medium quality selection given in \citRef{Aad:2014rga}, and in addition have 
$\pT > \unit{20}{\GeV}$ and $\abseta < 1.37$ or $1.52 < \abseta < 2.47$. 
Any \taulep not satisfying these $\tau$-identification criteria may
contribute to \met when passing the jet selection.

\subsubsection{Muon selection}
\label{subsec:muon-sel}

Muons are reconstructed within $\abseta < 2.5$ employing a combined \MS and \ID track fit. 
Outside of the \ID coverage, muons are reconstructed within $2.5 < \abseta < 2.7$ from a track fit to \MS track segments alone.   
Muons are further selected for \met reconstruction by requiring the medium reconstruction quality defined in \citRef{Aad:2016jkr},
$\pT > \unit{10}{\GeV}$, and an association with the \hs vertex for those within $\abseta < 2.5$.

\subsubsection{Jet selection}
\label{subsec:jet-sel}

Jets are reconstructed from clusters of topologically connected calorimeter cells (\topos), described in \citRef{Aad:2016upy}.
The \topos are calibrated at the electromagnetic (\EM) energy scale.\footnote{On this scale the energy deposited in the calorimeter by electrons and photons is represented well. The hadron signal at the \EM scale is not corrected for the non-compensating signal features of the \ATLAS calorimeters.} 
The \antikt algorithm~\cite{Cacciari:2008gp}, as provided by the \FJ toolkit~\cite{Cacciari:2011ma}, is employed with a radius parameter $R = 0.4$ to form jets from these \topos.
The jets are fully calibrated using the \EMJES scheme \cite{Aaboud:2017jcu} including a correction for \pu \cite{Aad:2015ina}.
They are required to have $\pT > \unit{20}{\GeV}$ after the full calibration. 
The jet contribution to \met and \sumet is primarily defined by the signal ambiguity resolution. 

Jets not rejected at that stage are further filtered using a tagging algorithm to select \hs jets (\QI{jet vertex tagging}) \cite{Aad:2015ina}.
This algorithm provides the jet vertex tagger variable \JVT, ranging from 0 (\pu-like) to 1 (\hs-like), for each jet with matched tracks.\footnote{In the calculation of \JVT the total amount of \pT carried by tracks from the \hs vertex matched to the given jet is related to the total amount of \pT carried by all matched tracks, among other inputs, to tag jets from the \hs interaction.} 
The matching of tracks with jets is done by \emph{ghost association}, where tracks are clustered as \emph{ghost particles} into the jet, as described in \citRef{Aad:2016nrq} and based on the approach outlined in \citRef{Cacciari:2008gn}. 

The overlap resolution can result in a partial overlap of the jet with an electron or photon, in terms of the fraction of common signals 
contributing to the respective reconstructed energy. 
This is measured by the ratio \of of the electron(photon) energy $E_{e(\gamma)}^{\EM}$ to the jet energy $E_{\jet}^{\EM}$,
\begin{align} 
	\of = \dfrac{E_{e(\gamma)}^{\EM}}{E_{\jet}^{\EM}}\,,
	\label{eq:of}
\end{align}
with both energies calibrated at the \EM scale. 
In the case of $\of \leq 50$\%, the jet is included in \met reconstruction, with its \pT scaled by $1-\of$.
For $\of > 50$\%, only the tracks associated with the jet, excluding the track(s) associated with the overlapping particle if any, contribute to the soft term as discussed in \secRef{subsec:etmiss-softterm}.

Jets not rejected by the signal ambiguity resolution and with $\pT > \unit{20}{\GeV}$ and $\abseta > 2.4$, or with $\pT \ge \unit{60}{\GeV}$ and $\abseta < 4.5$, are always accepted for \met reconstruction.
Jets reconstructed with $\unit{20}{\GeV} < \pT < \unit{60}{\GeV}$ and $\abseta < 2.4$ are only accepted if they are tagged by $\JVT > 0.59$.
In both cases, the jet \pT thresholds are applied to the jet \pT before applying the \of correction.
Additional configurations for selecting jets used in \met reconstruction are discussed in \appRef{app:jet}, together with the effect of the variation of these selection criteria on the \met reconstruction performance. 

\subsubsection{Muon overlap with jets}\label{subsec:muon-jet}

Jets overlapping with a reconstructed muon affect the \met reconstruction in a manner that depends on their origin. 
If these jets represent a significant (\emph{catastrophic}) energy loss along the path of the muon through the calorimeter, 
or if they are \pu jets tagged by \JVT as originating from the \hs interaction due to the muon \ID track, they need to be rejected for \met reconstruction. 
On the other hand, jets reconstructed from final state radiation (\FSR) off the muon need to be included into \met reconstruction.

In all cases, the muon--jet overlap is determined by ghost-associating the muon with the jet.
For this, each muon enters the jet clustering as ghost particle with infinitesimal small momentum, together with the \EM-scale \topos from the calorimeter.
If a given ghost particle becomes part of a jet, the corresponding muon is considered overlapping with this jet. 
This procedure is very similar to the track associations with jets mentioned in \secRef{subsec:jet-sel}. 

Tagging jets using \JVT efficiently retains those from the \hs vertex for \met reconstruction and rejects jets generated by \pu.
A muon overlapping with a \pu jet can lead to a mis-tag, because the \ID track from the muon represents a significant amount of \pT from the \hs vertex and thus increases \JVT.
As a consequence of this fake tag, the \pu jet \pT contributes to \met, and thus degrades both the \met response and resolution due to the stochastic nature of its contribution.

A jet that is reconstructed from a catastrophic energy loss of a muon tends to be tagged as a \hs jet as well.
This jet is reconstructed from \topos in close proximity to the extrapolated trajectory of the \ID track associated with the muon bend in the axial magnetic field.
Inclusion of such a jet into \met reconstruction leads to double-counting of the transverse momentum associated with the muon energy loss, 
as the fully reconstructed muon \pT is already corrected for this effect. 

To reject contributions from \pu jets and jets reconstructed from muon energy loss, the following selection criteria are applied: 
\begin{itemize}
	\item{$\pTmuonid/\pTjetid > 0.8$} -- the transverse momentum of the \ID track associated with the muon (\pTmuonid) represents a significant fraction of the transverse momentum \pTjetid, the sum of the transverse momenta of all \ID tracks associated with the jet;
	\item{$\pTjet/\pTmuonid < 2$} -- the overall transverse momentum \pTjet of the jet is not too large compared to \pTmuonid;
	\item{$\NtrkPV < 5$} -- the total number of tracks \NtrkPV associated with the jet and emerging from the \hs vertex is small.
\end{itemize}
All jets with overlapping muons meeting these criteria are understood to be either from \pu or a catastrophic muon energy loss and are rejected for \met reconstruction. 
The muons are retained for the \met reconstruction.

Another consideration for muon contributions to \met is \FSR.
Muons can radiate hard photons at small angles, which are typically not reconstructed as such because of the nearby muon \ID track violating 
photon isolation requirements. 
They are also not reconstructed as electrons, 
due to the mismatch between the \ID track momentum and the energy measured by the calorimeter.
Most likely the calorimeter signal generated by the \FSR photon is reconstructed as a jet, with the muon \ID track associated.
As the transverse momentum carried by the \FSR photon is not recovered in muon reconstruction, jets representing this photon need to be included in the \met reconstruction.
Such jets are characterised by the following selections, which are highly indicative of a photon in the \ATLAS calorimeter:
\begin{itemize}
	\item{$\NtrkPV < 3$} -- the jet has low charged-particle content, indicated by a very small number of tracks from the \hs vertex;
	\item{$\femc > 0.9$} -- the jet energy \ejet is largely deposited in the electromagnetic calorimeter (\EMC), as expected for photons and measured by the corresponding energy fraction $\femc = \ejetemc/\ejet$;
	\item{$\pTjetPS > \unit{2.5}{\GeV}$} -- the transverse momentum contribution \pTjetPS from \presampler signals to \pTjet indicates an early starting point for the shower;
	\item{$\jetWidth < 0.1$} -- the jet is narrow, with a width \jetWidth comparable to a dense electromagnetic shower; \jetWidth is reconstructed according to 
	\begin{align*}
		\jetWidth = \dfrac{\sum_{i} \Delta R_{i} \pTi{i}}{\sum_{i}\pTi{i}}\,,
	\end{align*}
	where $\Delta R_{i}=\sqrt{(\Delta\eta_{i})^{2}+(\Delta\phi_{i})^{2}}$ is the angular distance of \topo $i$ from the jet axis, and $\pTi{i}$ is the transverse momentum of this cluster;
	\item{$\pTjettrk/\pTmuonid > 0.8$} -- the transverse momentum \pTjettrk carried by all tracks associated with the jet is close to \pTmuonid.
\end{itemize}  
Jets are accepted for \met reconstruction when consistent with an \FSR photon defined by the ensemble of these selection criteria, with their energy scale set to the EM scale, to improve the calibration.

\subsection{\texorpdfstring{\met}{Etmiss} soft term}
\label{subsec:etmiss-softterm}
The soft term introduced in \secRef{subsec:etmiss-definition} is exclusively reconstructed from \ID tracks from the \hs vertex, thus only using the \pT-flow from soft charged particles. 
It is an important contribution to \met for the improvement of both the \met scale and resolution, in particular in final states with a low hard-object multiplicity. 
In this case it is indicative of (hadronic) recoil, comprising the event components not otherwise represented by reconstructed and calibrated particles or jets.

The more inclusive reconstruction of the \met soft term including signals from soft neutral particles uses calorimeter \topos.
The reconstruction performance using the calorimeter-based \mettcsft is inferior to the track-only-based \metsft, mostly due to a larger residual dependence on \pu. 
More details of the \topo-based \mettcsft reconstruction are discussed in \secRef{subsec:etmiss-calo}.

\subsubsection{Track and vertex selection}
\label{subsec:track-sel}

Hits in the \ID are used to reconstruct tracks pointing to a particular collision vertex ~\cite{Aaboud:2017all}.
Both the tracks and vertices need to pass basic quality requirements to be accepted. 
Each event typically has a number $\NPV > 1$ of reconstructed primary vertices.

Tracks are required to have $\pT > \unit{400}{\MeV}$ and $\abseta < 2.5$, in addition to the reconstruction quality requirements given in \citRef{ATLAS-CONF-2015-017}.
Vertices are constructed from at least two tracks passing selections on the transverse (longitudinal) impact parameter $\absdzero < \unit{1.5}{\mm}$ ($\abszzsth < \unit{1.5}{\mm}$) relative to the vertex candidate.
These tracks must also pass requirements on the number of hits in the \ID.
The hard-scatter vertex is identified as described in \appRef{app:nomenclature}.

\subsubsection{Track soft term}\label{subsec:etmiss-softterm-tst}

The track sample contributing to \metsft for each reconstructed event is collected from high-quality tracks emerging from the \hs vertex 
but not associated with any electron, \taulep, muon, or jet
contributing to \met reconstruction. 
The applied signal-overlap resolution removes
\begin{itemize}
\item \ID tracks with $\Delta R(\text{track,electron/photon cluster}) < 0.05$;
\item \ID tracks with $\Delta R(\text{track,\taulep}) < 0.2$;
\item \ID tracks associated with muons;
\item \ID tracks ghost--associated with fully or partially contributing jets.
\end{itemize}
\ID tracks from the \hs vertex that are associated with jets rejected by the overlap removal or are associated with jets that are likely from \pu, as tagged by the \JVT procedure discussed in \secRef{subsec:jet-sel}, contribute to \metsft. 

Since only reconstructed tracks associated with the \hs vertex are used, the track-based \metsft is largely insensitive to \pu effects.
It does not include contributions from any soft neutral particles, including those produced by the \hs interaction.

\section{Data and simulation samples}
\label{sec:samples}
The determination of the \met reconstruction performance uses selected 
final states without ($\mettrue = 0$) and with genuine missing transverse momentum from neutrinos ($\mettrue = \pTnu$). 
Samples with $\mettrue = 0$ are composed of leptonic \Zboson boson decays (\Zee and \Zmm) collected by a trigger and event selection 
that do not depend on the particular \pu conditions, since both the electron and muon triggers as well as the corresponding reconstructed kinematic variables are only negligibly affected by \pu. 
Also using lepton triggers, samples with neutrinos were collected from \Wen and \Wmn decays. 
In addition, samples with neutrinos and higher hard-object multiplicity were collected from top-quark pair ($\ttbar\,$) production with at least either the $t$ or the \tbar decaying semi-leptonically.

\subsection{Data samples}
\label{sec:datasample}
The data sample used corresponds to a total integrated luminosity of \ilum, collected  with a proton bunch-crossing interval of \unit{25}{\ns}.
Only high-quality data with a well-functioning calorimeter, inner detector and muon spectrometer are analysed.
The data-quality criteria are applied, which reduce the impact of instrumental noise and out-of-time calorimeter deposits from cosmic-ray and beam backgrounds.

\subsection{Monte Carlo samples}
\label{sec:mc-samples}
The \Zll and \Wln samples were generated using \POWHEGBOX~\cite{Frixione:2007vw} (version v1r2856) employing a matrix element calculation at next-to-leading order (NLO) in perturbative QCD.
To generate the particle final state, the (parton-level) matrix element output was interfaced to \PYTHIAEight~\cite{Pythia8a},\footnote{Version 8.186 was used for all final states generated with \PYTHIAEight.}
which generated the parton shower (PS) and the underlying event (UE) using the AZNLO tuned parameter set~\cite{Aad:2014xaa}.
Parton distribution functions (PDFs) were taken from the CTEQ6L1 PDF set~\cite{Pumplin:2002vw}.

The \ttbar-production sample was generated with a \POWHEG NLO kernel (version v2r3026) interfaced to \PYTHIASix~\cite{Sjostrand:2006za} (version 6.428) with the Perugia2012 set of tuned parameters~\cite{Skands:2010ak} for the PS and UE generation. 
The CT10 NLO PDF set~\cite{Lai:2010vv} was employed. 
The resummation of soft-gluon terms in the next-to-next-to-leading-logarithmic (NNLL) approximation with \TOPPPV~\cite{Czakon:2011xx} was included.

Additional processes contributing to the \Zll and \Wln final state samples are the production of \dibosons, single top quarks, and \multijets.
\Dibosons were generated using \SHERPA \cite{Gleisberg:2008ta,Gleisberg:2008fv,Hoeche:2009rj,Schumann:2007mg} version v2.1.1 employing the CT10 PDF set. 
Single top quarks were generated using \POWHEG version v1r2556 with the CT10 PDF set for the $t$-channel production and \POWHEG version v1r2819 for the $s$-channel and the associated top quark  ($\Wboson t$) production, all interfaced to the PS and UE from the same \PYTHIASix configuration used for \ttbar production. 
\Multijet events were generated using \PYTHIAEight with the \NNPDFLO PDF set \cite{Ball:2012cx} and the A14 set of tuned PS and UE parameters described in \citRef{ATL-PHYS-PUB-2014-021}.

Minimum bias (\MB) events were generated using \PYTHIAEight with the MSTW2008LO PDF set \cite{Martin:2009iq} and the A2 tuned parameter set~\cite{ATL-PHYS-PUB-2011-014} for PS and UE.
These \MB events were used to model \pu, as discussed in \secRef{sec:samples:pu}.

For the determination of the systematic uncertainties in \met reconstruction, an alternative inclusive sample of \Zmm events was generated 
using the \textsc{MadGraph\_aMC@NLO} (version v2.2.2) matrix element generator~\cite{Alwall:2014hca}  employing the CTEQ6L1 PDF set. 
Both PS and UE were generated using \PYTHIAEight with the \NNPDFLO PDF set and the A14 set of tuned parameters.

The \MC-generated events were processed with the \GEANTFour software toolkit \cite{Agostinelli:2002hh}, which simulates the propagation of the generated stable particles\footnote{In \ATLAS stable particles are those with an expected laboratory lifetime $\tau$ corresponding to $c\tau > \unit{10}{\mm}$.}  through the ATLAS detector and their interactions with the detector material \cite{Aad:2010ah}.

\subsection{\PU}
\label{sec:samples:pu}
The calorimeter signals are affected by \pu and the short bunch-crossing period at the \LHC.
In 2015, an average of about 13 \pu collisions per bunch crossing was observed. 
The dominant contribution of the additional \pp collisions to the detector signals of the recorded event arises from a diffuse emission of soft particles superimposed to the 
\hs interaction final state (\ipu).
In addition, the \LAr calorimeter signals are sensitive to signal remnants from up to 24 previous bunch crossings and one following bunch crossing (\opu), as discussed in 
\citMultiRefLabel{} \cite{Aad:2016upy,Aad:2010ai}.
Both types of \pu affect signals contributing to  \met. 

The \ipu activity is measured by the number of reconstructed primary collision vertices \NPV. 
The \opu is proportional to the number of collisions per bunch crossing $\mu$, measured as an average over time periods of up to two minutes by integrated signals from the luminosity detectors in \ATLAS \cite{Aad:2013ucp}. 

To model \ipu in \MC simulations, a number of generated \pu collisions was drawn from a Poisson distribution around the value of $\mu$ recorded in data. 
The collisions were randomly collected from the \MB sample discussed in \secRef{sec:mc-samples}. 
The particles emerging from them were overlaid onto the particle-level final state of the generated hard-scatter interaction and converted into detector signals before event reconstruction.
The event reconstruction then proceeds as for data. 

Similar to the \LHC proton-beam structure, events in \MC simulations are organised in bunch trains, where the structure in terms of bunch-crossing interval and gaps between trains is taken into account to model the effects of \opu. 
The fully reconstructed events in \MC simulation samples are finally weighted such that the distribution of the number of overlaid collisions over the whole sample corresponds to the $\mu$ distribution observed in data.    

The effect of \pu on the signal in the \Tile calorimeter is reduced due to its location behind the electromagnetic calorimeter and its fast time response \cite{Aad:2010af}.
Reconstructed \ID and \MS tracks are largely unaffected by \pu.

\section{Event selection}
\label{sec:event-selection}
\subsection{\texorpdfstring{\Zmm}{Zmm} event selection}
\label{subsec:event-selection-zmumu}
The \Zmm final state is 
ideal 
for the evaluation of \met reconstruction performance, since it can be selected with a high signal-to-background ratio and the \Zboson kinematics can be measured with high precision, even in the presence of \pu.
Neutrinos are produced only through very rare heavy-flavour decays in the hadronic recoil. 
This channel can therefore be considered to have no genuine missing transverse momentum.
Thus, the scale and resolution for the reconstructed \met 
are indicative of the reconstruction quality and reflect limitations introduced by both the detector and the ambiguity resolution procedure. 
The well-defined expectation value $\mettrue = 0$ allows the reconstruction quality to be determined in both data and \MC simulations.
The reconstructed \met in this final state is also sensitive to the effectiveness of the muon--jet overlap resolution, which can be explored in this
low-multiplicity environment in both data and \MC simulations, with a well-defined \met.
  
Events must pass one of three high-level muon triggers with different \pTmuon thresholds and isolation requirements.
The isolation is determined by the ratio of the scalar sum of \pT of reconstructed tracks other than the muon track itself, in a cone of size $\deltaR = 0.2$ around the muon track (\pTcone), to \pTmuon. 
The individual triggers require (1) $\pTmuon > \unit{20}{\GeV}$ and $\pTcone/\pTmuon < 0.12$, or (2) $\pTmuon > \unit{24}{\GeV}$ and $\pTcone/\pTmuon < 0.06$, or (3) $\pTmuon > \unit{50}{\GeV}$ without isolation requirement.

The offline selection of \Zmm events requires exactly two muons, each selected as defined in \secRef{subsec:muon-sel}, 
with the additional criteria that (1) the muons must have opposite charge, (2) $\pTmuon > \unit{25}{\GeV}$, 
and (3) the reconstructed invariant mass \mmm of the dimuon system is consistent with the mass \mZ of the \Zboson boson, $|\mmm - \mZ| < \unit{25}{\GeV}$.

\subsection{\texorpdfstring{\Wen}{Wen} event selection}
\label{subsec:event-selection-wenu}
Events with \Wen or \Wmn in the final state provide a well-defined topology with neutrinos produced in the hard-scatter interaction.
In combination with \Zmm, the effectiveness of signal ambiguity resolution and lepton energy reconstruction for both the electrons and muons can be observed. 
The \Wen events in particular provide a good metric with $\mettrue = \pTnu > 0$ to evaluate and validate the scale, resolution and direction (azimuth) of the reconstructed \met,
as the \met reconstruction is sensitive to the electron--jet overlap resolution performance. 
This metric is only available in \MC simulations where \pTnu is known.
Candidate \Wen events are required to pass the high-level electron trigger with $\pT > \unit{17}{\GeV}$.
Electron candidates are selected according to criteria described in \secRef{subsec:electron-sel}.
Only events containing exactly one electron are considered.

Further selections using \met  and the reconstructed transverse mass \mT, given by 
\begin{equation*}
	\mT = \sqrt{ 2\pTj{e} \met (1-\cos{\deltaPhi})}\,,
\end{equation*}
are applied to reduce the \multijet background with one jet emulating an isolated electron from the \Wboson boson. 
Here \met is calculated as presented in \secRef{sec:etmiss-reco}. The transverse momentum of the electron is denoted by \pTj{e}, and
\deltaPhi is the distance between \phimiss and the azimuth of the electron.
Selected events are required to have $\met > \unit{25}{\GeV}$ and $\mT > \unit{50}{\GeV}$.

\subsection{\texorpdfstring{\ttbar}{ttbar} event selection}
\label{subsec:event-selection-ttbar}
Events with \ttbar in the final state allow the evaluation of the \met performance in interactions with a large jet multiplicity.
Electrons and muons used to define these samples are reconstructed as discussed in \secRef{subsec:electron-sel} and \secRef{subsec:muon-sel}, respectively, and are required to have $\pT > \unit{25}{\GeV}$.

The final \ttbar sample is selected by imposing additional requirements.
Each event must have exactly one electron and no muons passing the selections described above. 
In addition, at least four jets reconstructed by the \antikt algorithm with $R = 0.4$ and selected following the description in \secRef{subsec:jet-sel} are required. 
At least one of the jets needs to be $b$-tagged using the tagger configuration for a 77\% efficiency working point described in \citRef{Aad:2015ydr}. 
All jets are required to be at an angular distance of $\deltaR > 0.4$ from the electron.

\section{Performance of \met reconstruction in data and Monte Carlo simulation}
\label{sec:perf}

Unlike for fully reconstructed and calibrated particles and jets, 
and in the case of the precise reconstruction of charged particle kinematics provided by \ID tracks,
\met reconstruction yields a non-linear response, especially in regions of phase space where the observation bias discussed in \secRef{subsec:etmiss-basics} dominates the reconstructed \met.
In addition, the \met resolution functions are characterised by a high level of complexity, due to the composite character of the observable.
Objects with different \pT-resolutions contribute, and the \met composition can fluctuate significantly for events from the same final state.
Due to the dependence of the \met response on the resolution, both performance characteristics change as a function of the total event activity and are affected by \pu.  
There is no universal way of mitigating these effects, due to the 
inability to validate in data a stable and universal calibration reference for \met.

The \met reconstruction performance is therefore assessed by comparing a set of reconstructed \met-related observables in data and \MC simulations 
for the same final-state selection, with the same object and event selections applied. 
Systematic uncertainties in the \met response and resolution are derived from these comparisons and are used to quantify the level of understanding of the data 
from the physics models. 
The quality of the detector simulation is independently determined for all reconstructed jets, particles and \ID tracks, 
and can thus be propagated to the overall \met uncertainty for any given event. 
Both the distributions of observables as well as their average behaviour with respect to relevant scales measuring the overall kinematic activity of the  
\hs event or the \pu activity are compared. 
To focus on distribution shapes rather than statistical differences in these comparisons, the overall distribution of a given observable obtained from
\MC simulations is normalised to the integral of the corresponding distribution in data. 
  
As the reconstructed final state can be produced by different physics processes, the individual process contributions in \MC simulations are scaled according to the \xsect of the process. 
This approach is taken to both show the contribution of a given process to the overall distribution, 
and to identify possible inadequate modelling arising from any individual process, or a subset of processes, by its effect on the overall shape of the
\MC distribution. 

Inclusive event samples considered for the \met performance evaluation are obtained by applying selections according to \secRef{subsec:event-selection-zmumu}  
for a final state without genuine \met (\Zmm), and according to \secRef{subsec:event-selection-wenu} for a final state with genuine \met (\Wen). 
From these, specific exclusive samples are extracted by applying conditions on the number of jets reconstructed. 
In particular, \emph{zero jet} ($\Njet = 0$) samples without any jet with $\pT > \unit{20}{\GeV}$ (fully calibrated) and $|\eta|<4.9$ are useful for exclusively studying the performance of the soft term. 
Samples with events selected on the basis of a non-zero number of reconstructed jets with $\pT > \unit{20}{\GeV}$ are useful for evaluating the contribution of jets to \met. 
While the \pT response of jets is fully calibrated and provides a better measurement of the overall event \pT-flow, the \pT resolution for jets is affected by \pu and can introduce a detrimental effect on \met reconstruction performance. 
 
Missing transverse momentum and its related observables presented in \secRef{subsec:etmiss-basics} are reconstructed for the performance evaluations shown in the 
following sections using a standard reconstruction configuration. 
This configuration implements the signal ambiguity resolution in the \met reconstruction sequence discussed in \secRef{subsec:etmiss-basics}.  
It employs the hard-object selections defined in \secMultiRef{subsec:electron-sel}{to}{subsec:muon-sel}, with jets selected according to the prescriptions given in \secRef{subsec:jet-sel}.
The overlap resolution strategy for jets and muons described in \secRef{subsec:muon-jet} is applied. 
The soft term is formed from \ID tracks according to \secRef{subsec:etmiss-softterm}. 

\subsection{\texorpdfstring{\met}{Etmiss} modelling in Monte Carlo simulations}
\label{subsec:etmiss-distributions}

\begin{figure}[t!]\centering
\subfigure[]{\includegraphics[width=\fighalfwidth]{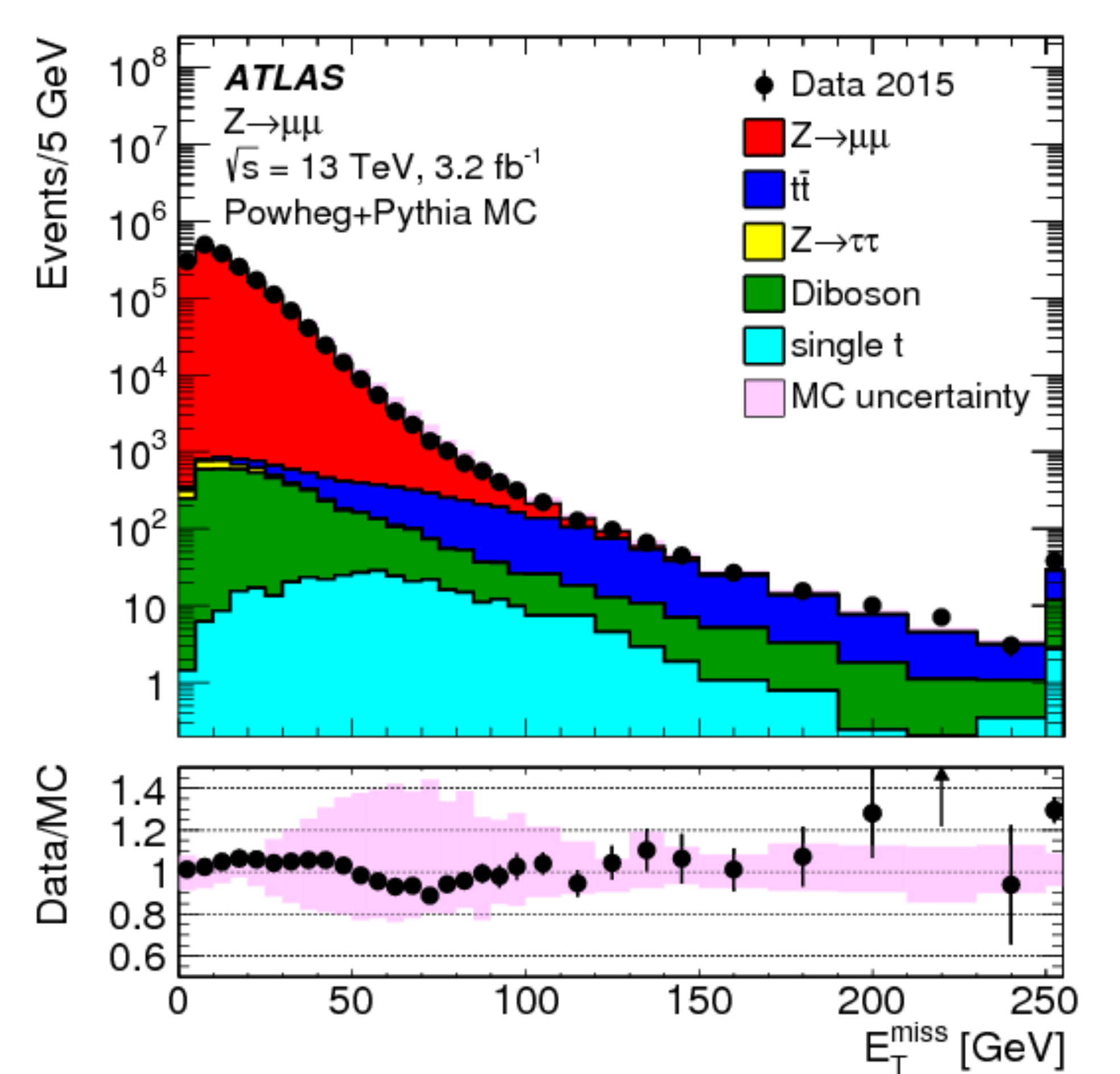}\label{fig:perf:zll:met}}\quad
\subfigure[]{\includegraphics[width=\fighalfwidth]{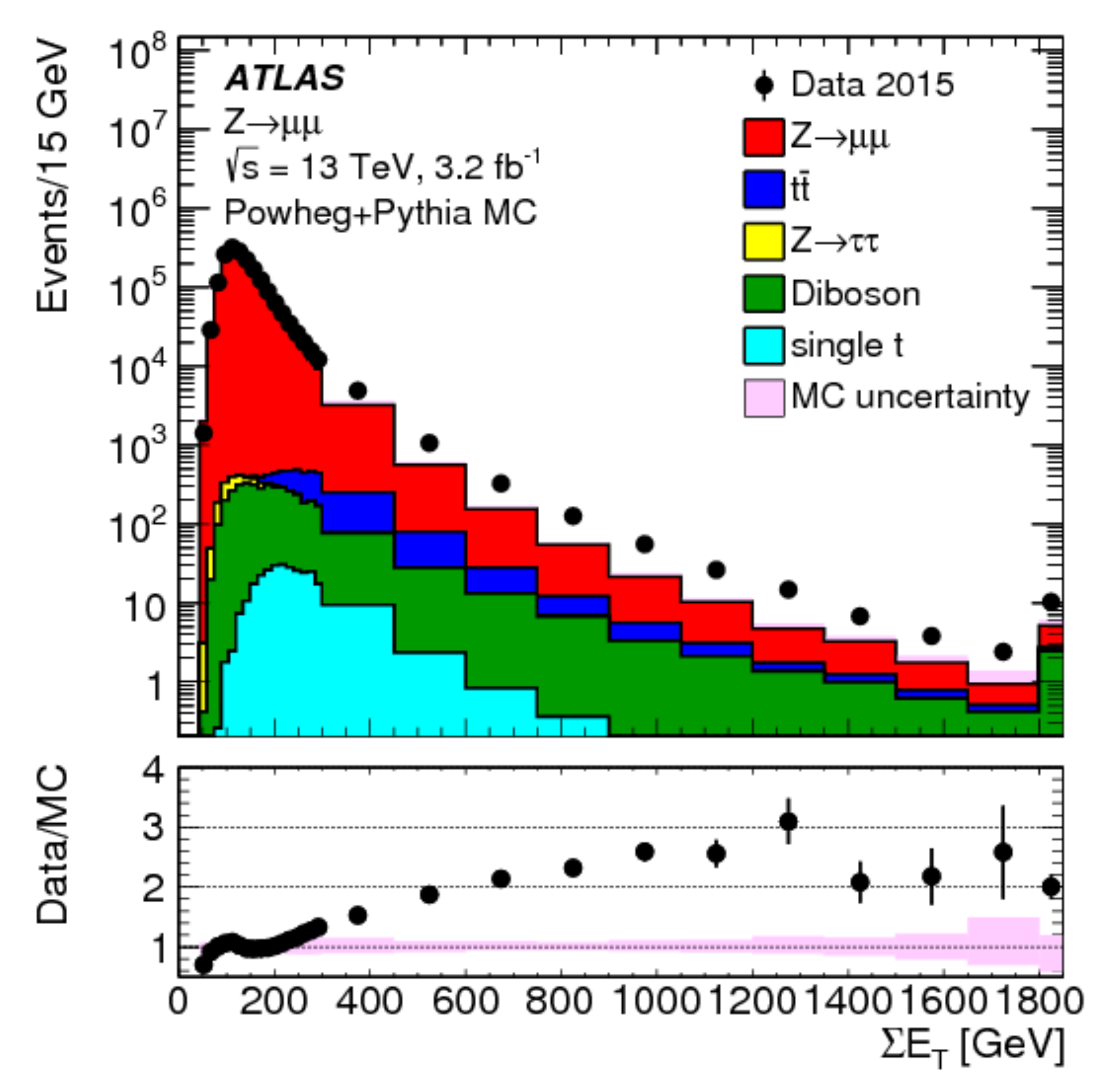}\label{fig:perf:zll:sumet}}\qquad
\subfigure[]{\includegraphics[width=\fighalfwidth]{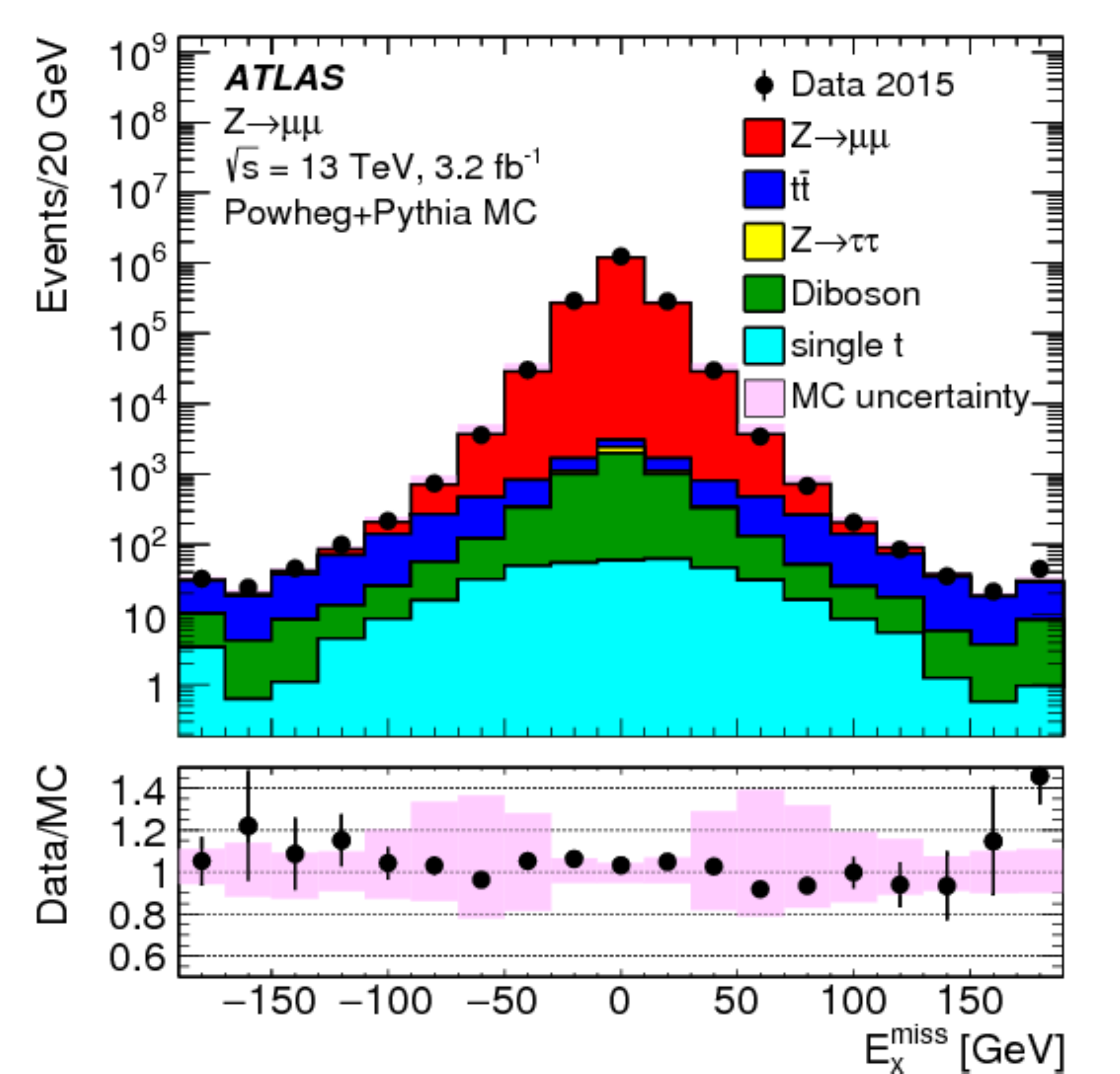}\label{fig:perf:zll:metx}}\quad
\subfigure[]{\includegraphics[width=\fighalfwidth]{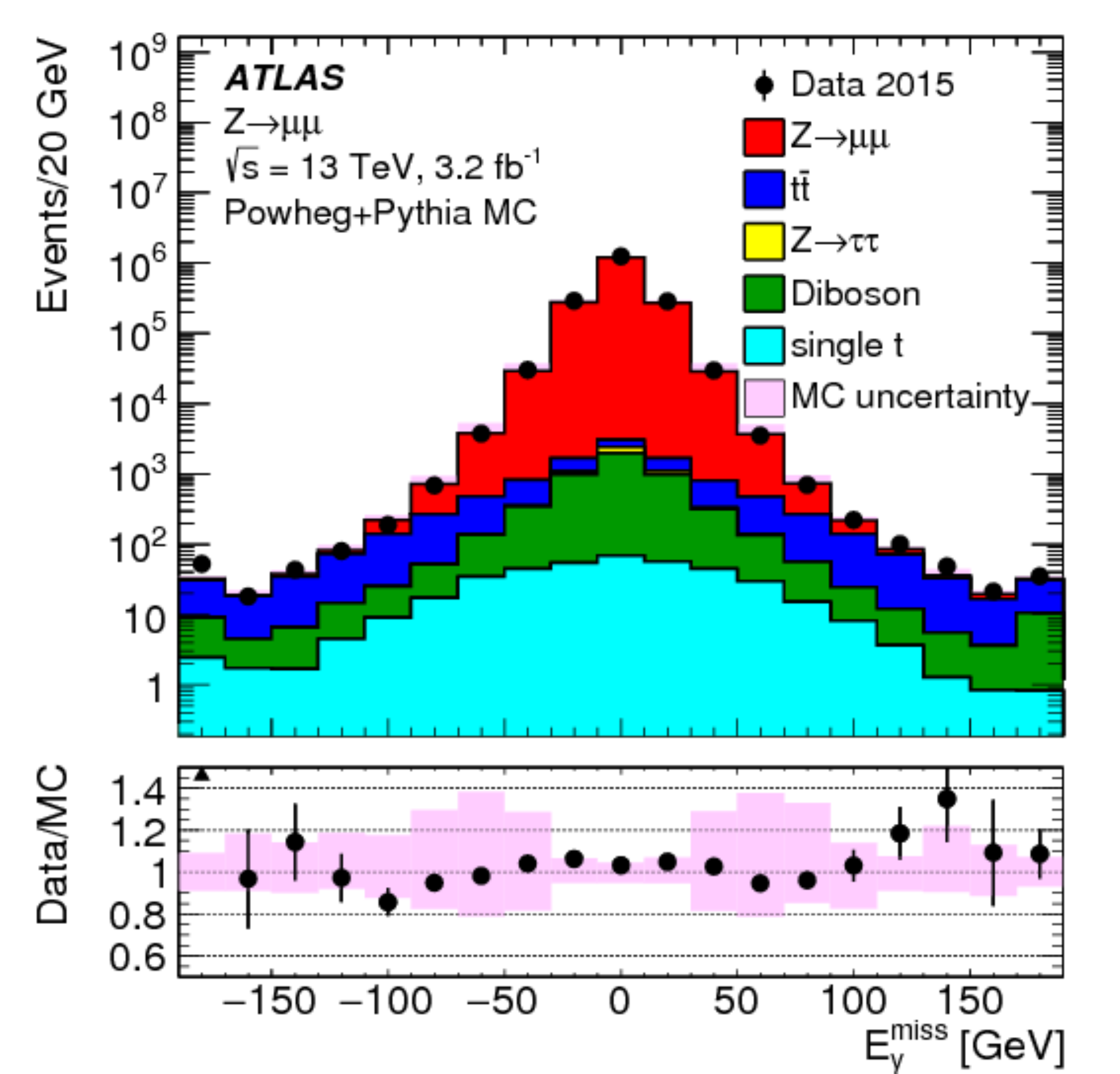}\label{fig:perf:zll:mety}}\qquad
\caption[]{Distributions of \subref{fig:perf:zll:met} \met, \subref{fig:perf:zll:sumet} \sumet, \subref{fig:perf:zll:metx} \metx and \subref{fig:perf:zll:mety} \mety for an inclusive sample of \Zmm events extracted from data and compared to \MC simulations including all relevant backgrounds. The shaded areas indicate the total uncertainty for \MC simulations, including the overall statistical uncertainty combined with systematic uncertainties from the \pT scale and resolution which are contributed by muons, jets, and the soft term. The last bin of each distribution includes the overflow, and the first bin contains the underflow in \subref{fig:perf:zll:metx} and \subref{fig:perf:zll:mety}. The respective ratios between data and \MC simulations are shown below the distributions, with the shaded areas showing the total uncertainties for \MC simulations.}
\label{fig:perf:zll}
\end{figure}

\begin{figure}[t!]\centering
\subfigure[]{\includegraphics[width=\fighalfwidth]{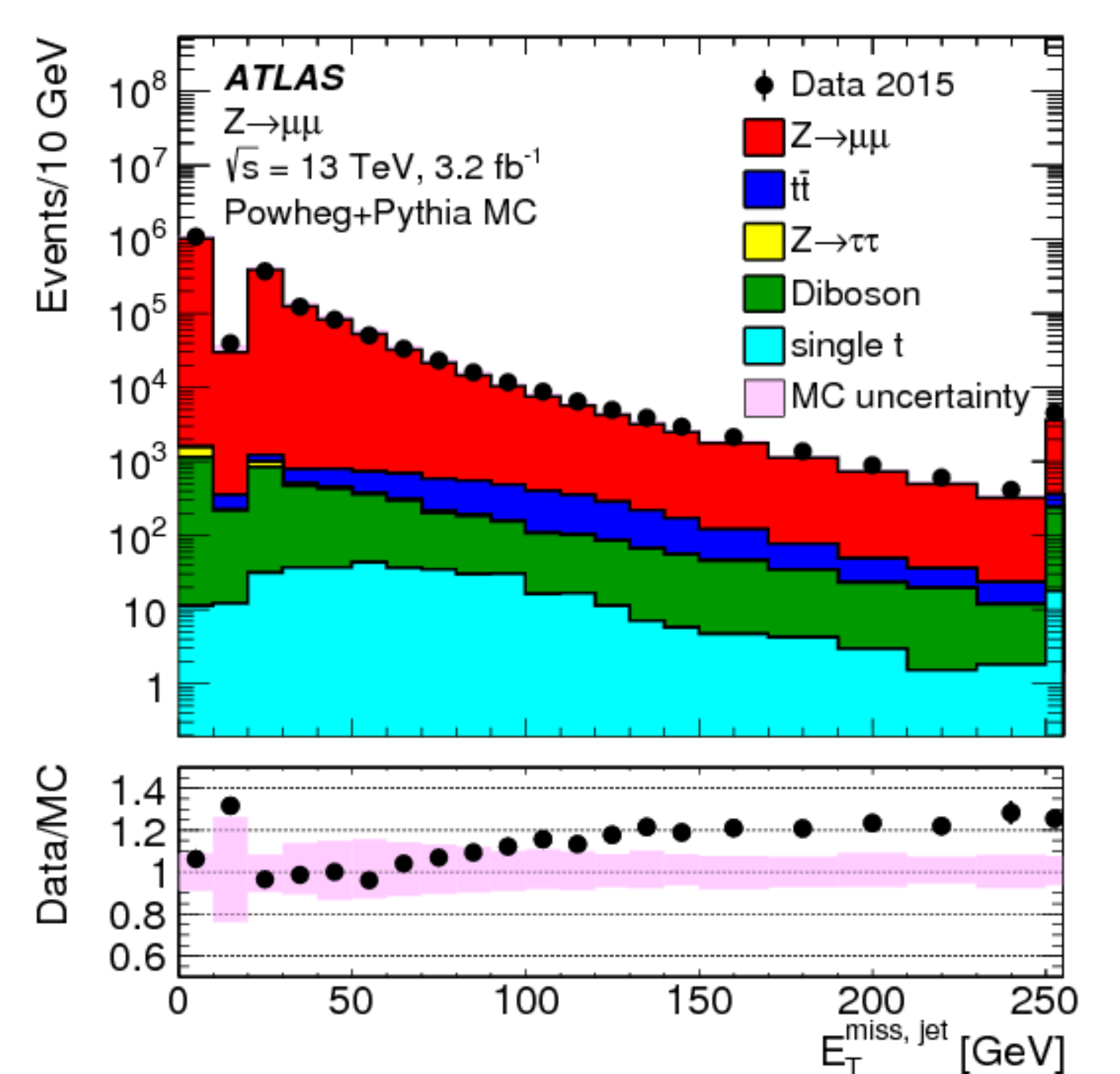}\label{fig:perf:zll:terms:met:jet}}\quad
\subfigure[]{\includegraphics[width=\fighalfwidth]{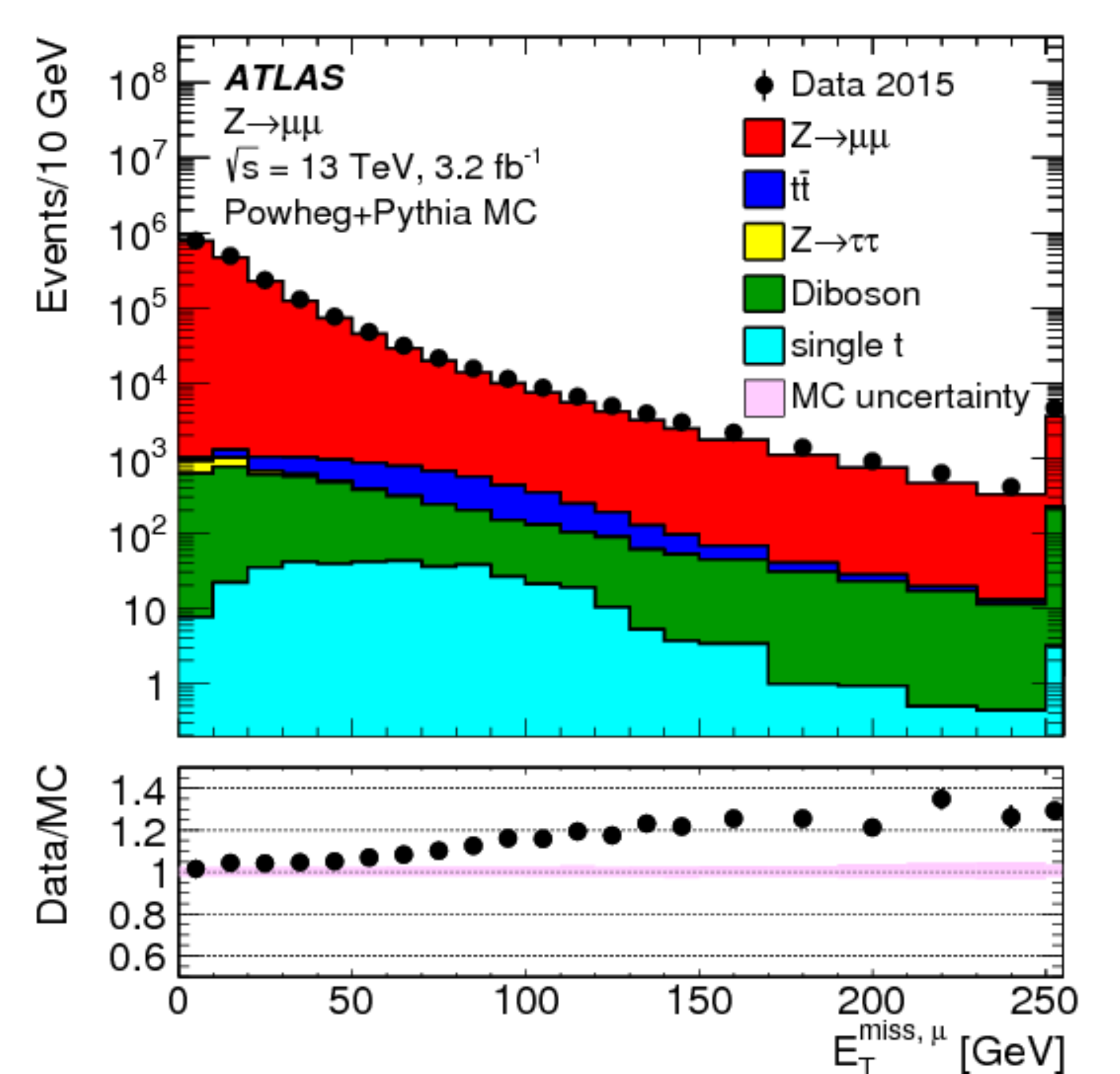}\label{fig:perf:zll:terms:met:muon}}\qquad
\subfigure[]{\includegraphics[width=\fighalfwidth]{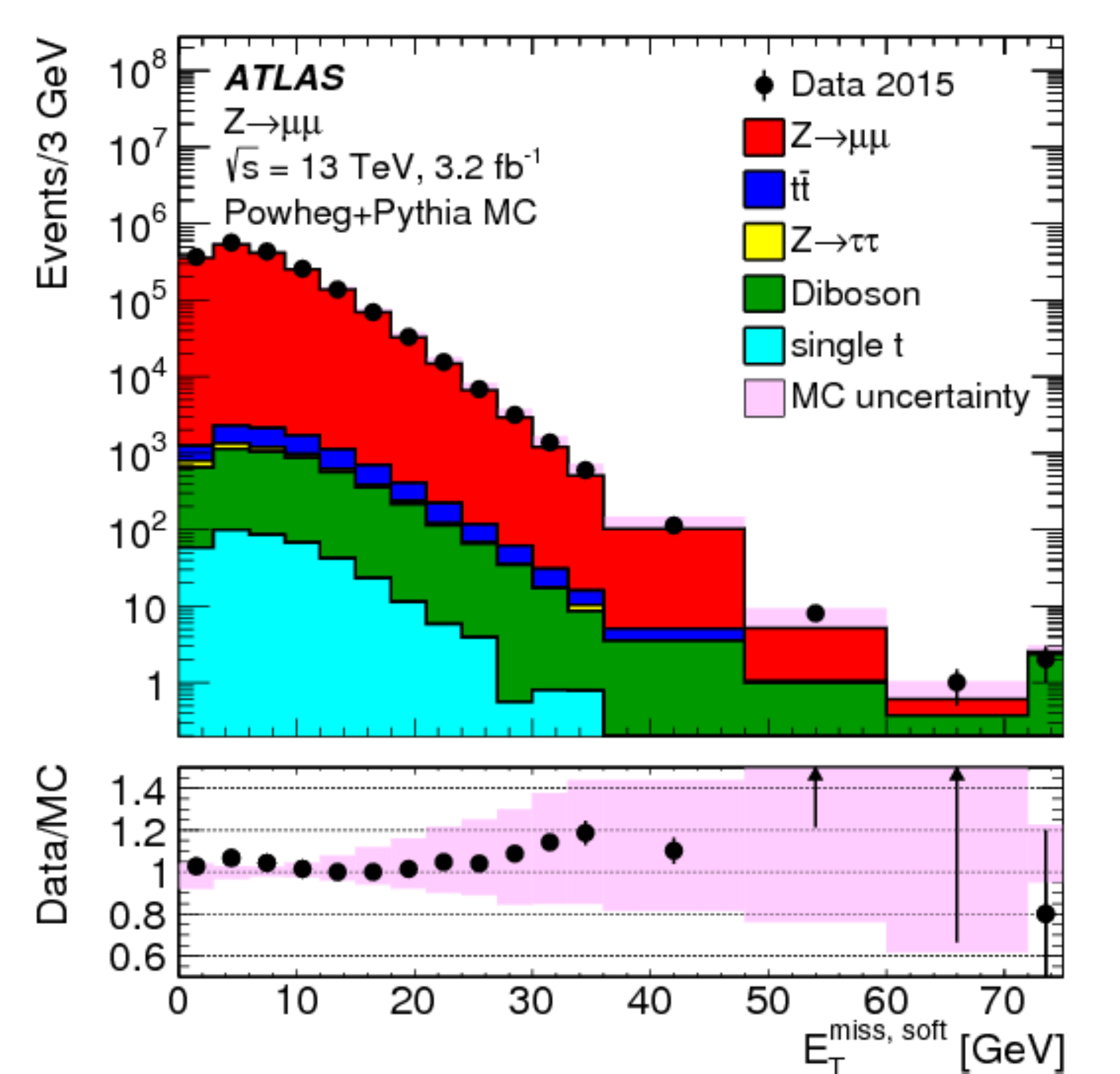}\label{fig:perf:zll:terms:met:soft}}\quad
\caption{Distributions of \subref{fig:perf:zll:terms:met:jet} the jet term \metjet, \subref{fig:perf:zll:terms:met:muon} the muon term \metmuo, and \subref{fig:perf:zll:terms:met:soft} the soft  term \metsft for the inclusive samples of \Zmm events in data, compared to \MC simulations including all relevant backgrounds. 
The shaded areas indicate the total uncertainty from \MC simulations, including  the overall statistical uncertainty combined with the respective systematic uncertainties from \subref{fig:perf:zll:terms:met:jet} the jet, \subref{fig:perf:zll:terms:met:muon} the muon, and \subref{fig:perf:zll:terms:met:soft} the soft term.
The last bin of each distribution includes the overflow entries. The respective ratios between data and \MC simulations are shown below the distributions, with the shaded areas showing the corresponding total uncertainties from \MC simulations.}
\label{fig:perf:zll:terms:met}
\end{figure}

The quality of the \MC modelling of \metx, \mety, \met and \sumet, reconstructed as given in \eqMultiRefLabel~(\ref{eq:metbasics:comp}), (\ref{eq:metbasics:met}) and (\ref{eq:metbasics:sumet}),
is evaluated for an inclusive sample of \Zmm events by comparing the distributions of these observables to data. 
The results are presented in \figRef{fig:perf:zll}. 
The data and \MC simulations agree within 20\% for the bulk of the \met distribution shown in \subfigRef{fig:perf:zll}{fig:perf:zll:met}, 
with larger differences not accommodated by the total (systematic and statistical) uncertainties of the distributions for high \met. 
These differences suggest a mismodelling in \ttbar events, the dominant background in the tail regime~\cite{ATLAS-CONF-2015-065}.
The \sumet distributions compared between data and \MC simulations in \subfigRef{fig:perf:zll}{fig:perf:zll:sumet} show discrepancies significantly larger than the overall uncertainties for $\unit{200}{\GeV} < \sumet < \unit{1.2}{\TeV}$. 
These reflect the level of mismodelling of the final state mostly in terms of hard-object composition in \MC simulations. 
The \metx and \mety spectra shown in \subfigMultiRef{fig:perf:zll}{fig:perf:zll:metx}{and}{fig:perf:zll}{fig:perf:zll:mety}, respectively, 
show good agreement between data and \MC simulations for the bulk of the distributions within $|\metxy|<\unit{100}{\GeV}$, 
with larger differences observed outside of this range still mostly within the uncertainties.  

\begin{figure}[t!]\centering
\subfigure[]{\includegraphics[width=\fighalfwidth]{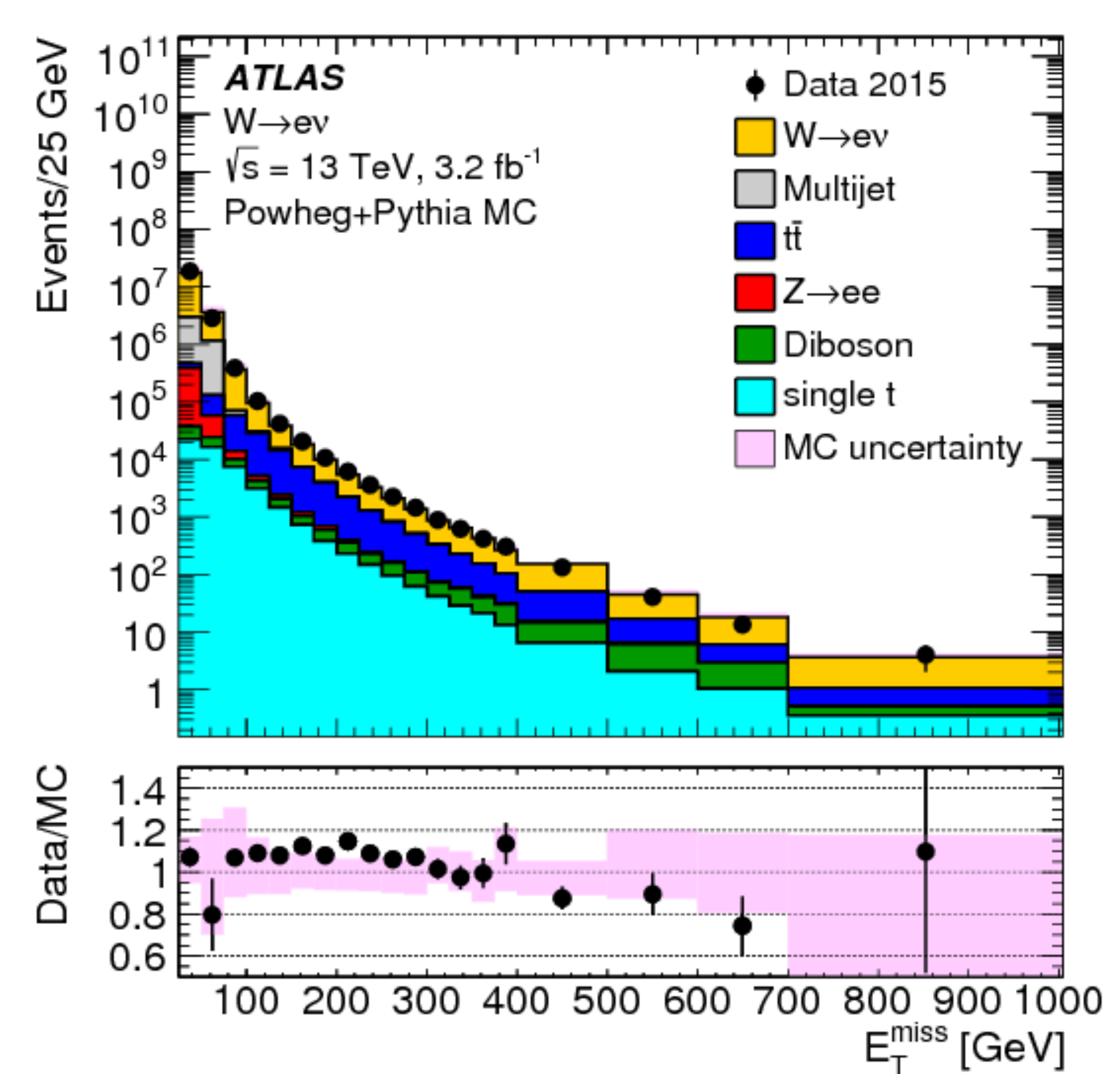}\label{fig:perf:wen:met:total}}\quad
\subfigure[]{\includegraphics[width=\fighalfwidth]{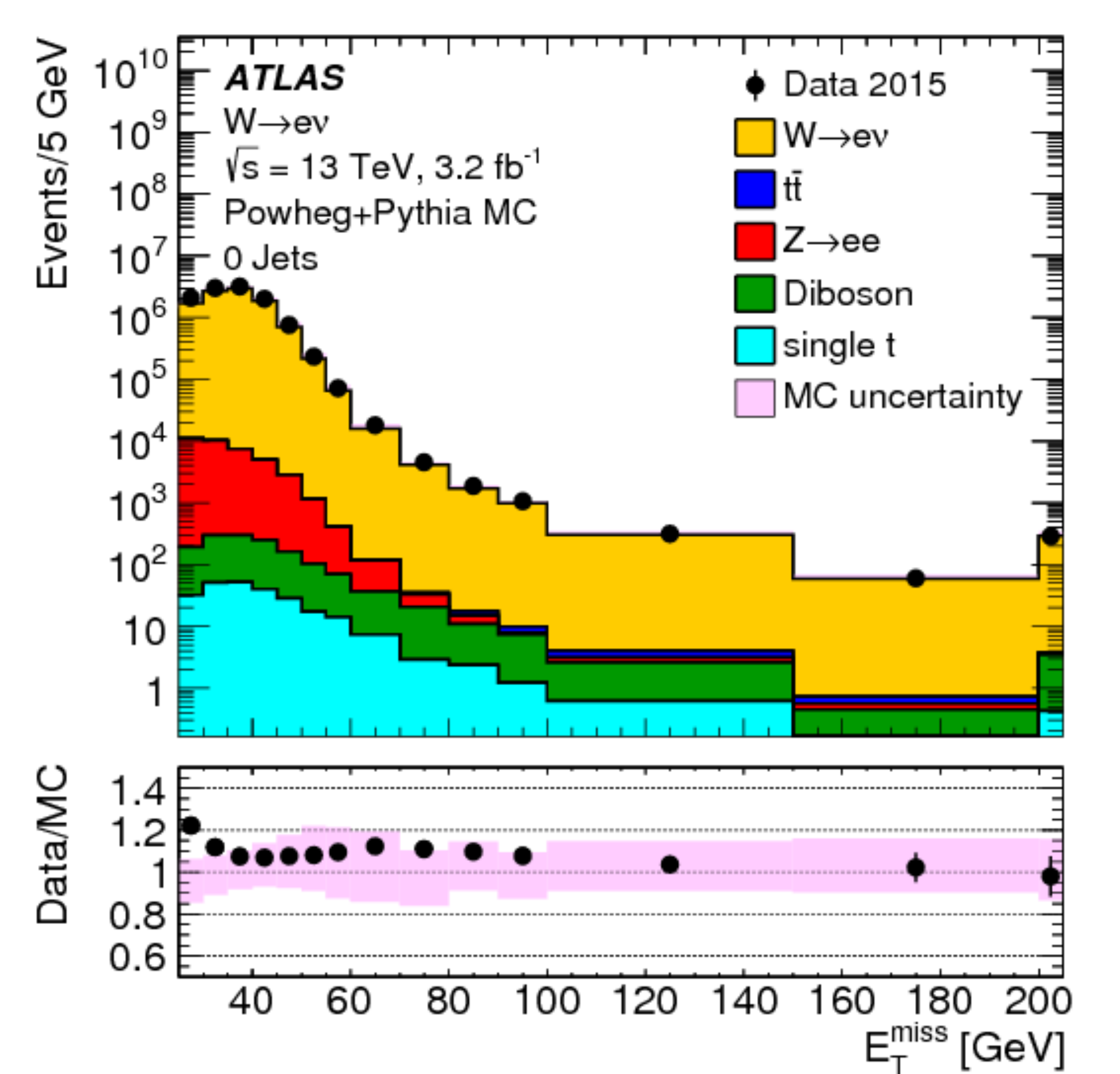}\label{fig:perf:wen:met:total0j}}\qquad
\subfigure[]{\includegraphics[width=\fighalfwidth]{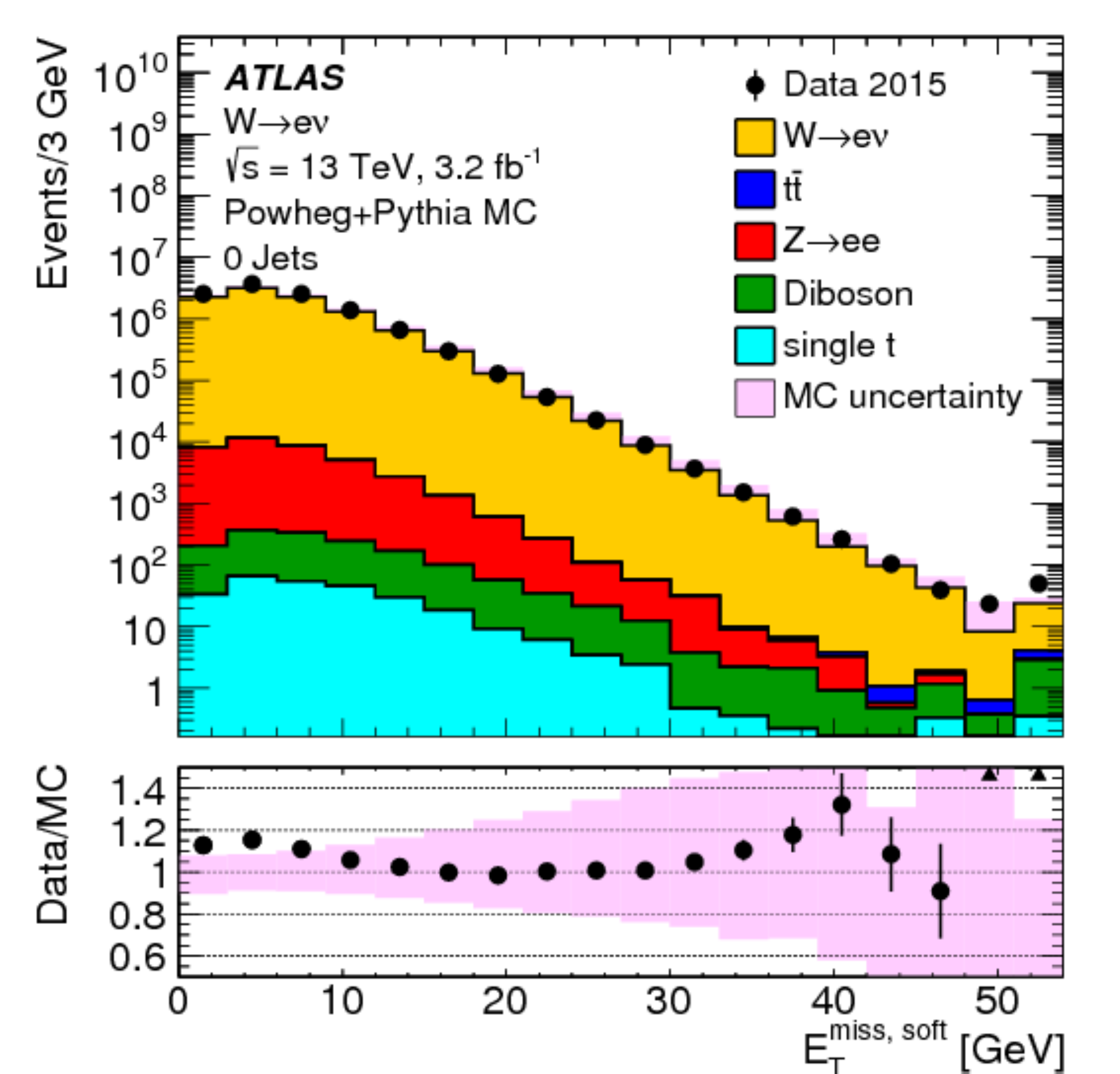}\label{fig:perf:wen:met:soft0j}}\quad
\caption{Distributions of the total \met in \subref{fig:perf:wen:met:total} the inclusive case and \subref{fig:perf:wen:met:total0j} the \nojets case, as well as \subref{fig:perf:wen:met:soft0j} the soft term \metsft reconstructed in \nojets events with \Wen in data. The expectation from \MC simulation is superimposed and includes all relevant background final states passing the event selection. 
The inclusive \met distribution from \MC simulations contains a small contribution from \multijet final states at low \met, which is absent for the \nojets selection.  
The shaded areas indicate the total uncertainty for \MC simulations, including the overall statistical uncertainty combined with systematic uncertainties comprising contributions from the electron, jet, and the soft term.
The last bins contain the respective overflows. The respective ratios between data and \MC simulations are shown below the distributions, with the shaded areas indicating the total uncertainties for \MC simulations.}
\label{fig:perf:wen:met}
\end{figure}

The distributions of individual contributions to \met from jets (\metjet), muons (\metmuo), and the soft term (\metsft), as defined in \eqRef{eq:etmiss_terms}, 
are compared between data and \MC simulations for the same inclusive \Zmm sample in \figRef{fig:perf:zll:terms:met}.  
Agreement between data and \MC simulations for \metjet in \subfigRef{fig:perf:zll:terms:met}{fig:perf:zll:terms:met:jet} is of the order of $\pm 20$\% and within the total uncertainties for $\metjet \lesssim \unit{120}{\GeV}$, 
but beyond those for higher \metjet.
A similar observation holds for \metmuo in \subfigRef{fig:perf:zll:terms:met}{fig:perf:zll:terms:met:muon}, where data and \MC simulations agree within the uncertainties for low \metmuo but significantly beyond them for larger \metmuo.
Agreement between data and \MC simulations is better for the soft term \metsft, with differences up to 10\% for $\metsft \lesssim \unit{30}{\GeV}$, as seen in \subfigRef{fig:perf:zll:terms:met}{fig:perf:zll:terms:met:soft}.
Larger differences for larger \metsft are still found to be within the uncertainties.
 
The peak around $\metjet = \unit{20}{\GeV}$ indicates the onset of single-jet events at the threshold $\pT = \unit{20}{\GeV}$ for jets contributing to \metjet. 
Larger values of \metjet arise from events with one or more high-\pT jets balancing the \pT of the \Zboson boson.     

For the \Wen sample with genuine missing transverse momentum given by \pTnu, both the total reconstructed \met and the soft term are compared between data and \MC simulations in \figRef{fig:perf:wen:met}. 
The level of agreement between the \met distributions for data and \MC simulations shown in \subfigRef{fig:perf:wen:met}{fig:perf:wen:met:total} for the inclusive event sample is at $\pm20\%$, similar to that observed for the \Zmm sample in \subfigRef{fig:perf:zll}{fig:perf:zll:met}, except that for this final state it is found to be within the total uncertainties of the measurement.
The differences between the \met distributions observed with the exclusive \nojets sample shown in \subfigRef{fig:perf:wen:met}{fig:perf:wen:met:total0j} are well below 20\%, but show a trend to larger discrepancies for decreasing $\met \lesssim \unit{40}{\GeV}$.  
This trend is due to the missing background contribution in \MC simulations from \multijet final states.
The extraction of this contribution is very inefficient and only possible with large statistical uncertainties.
Even very large \MC samples of \multijet final states provide very few events with only one jet that is accidentally reconstructed as an electron, and with the amount of \met required in the \Wen selection described in \secRef{subsec:event-selection-wenu}.
The comparison of the \metsft distributions from data and \MC simulations shown in  \subfigRef{fig:perf:wen:met}{fig:perf:wen:met:soft0j} yields agreement well within the 
uncertainties, for $\metsft \gtrsim \unit{10}{\GeV}$. 
The rising deficiencies observed in the \MC distribution for decreasing $\metsft \lesssim \unit{10}{\GeV}$ are expected to be related to the missing \multijet contribution.

\subsection{\texorpdfstring{\met}{Etmiss} response and resolution}
\label{subsec:etmiss-recoperf}
The response in the context of \met reconstruction is determined by the deviation of the observed \met from the expectation value for a given final state.
This deviation sets the scale for the observed \met.
If this deviation is independent of the genuine missing transverse momentum, or any other hard \pT indicative of the overall hard-scatter activity, the \met response is linear.
In this case, a constant bias in the reconstructed \met is still possible due to detector inefficiencies and coverage (acceptance) limitations. 

Final states balanced in transverse momentum are expected to show a non-linear \met response at low event activity, 
as the response in this case suffers from the observation bias in \met reconstruction discussed in \secRef{subsec:etmiss-basics}.
With increasing momentum transfers in the hard-scatter interaction, the \met response becomes increasingly dominated by a well-measured hadronic recoil and thus more linear.  
In the case of final states with genuine missing transverse momentum,  the \met response is only linear once \mettrue exceeds the observation bias. 
These features are discussed in \secRef{subsec:etmiss-scale:method} and explored in \secRef{subsec:etmiss-scale:results}.

Contributions to the fluctuations in the \met measurement arise from (1) the limitations in the detector acceptance not allowing the reconstruction of the complete
transverse momentum flow from the hard interaction, (2) the irreducible intrinsic signal fluctuations in the detector response, and from (3) the additional response fluctuations 
due to \pu.   
In particular (1) introduces fluctuations driven by the large variations of the particle composition of the final state with respect to their types, momenta and directions.     
The limited detector coverage of $\abseta < 4.9$ for all particles, together with the need to suppress the \pu-induced signal fluctuations as much as possible, 
restricts the contribution of particles to \met to the reconstructed and accepted \electron, \photon, \tauhad and \muon, 
and those being part of a reconstructed and accepted jet.
In addition, the \pT-flow of not explicitly reconstructed charged particles emerging from the \hs vertex is represented by \ID tracks 
contributing to \metsft given in \eqMultiRef{eq:etmiss_terms}{and}{eq:sumet}, but only in the phase space defined by the selections given in 
\secRef{subsec:track-sel}. 
All other charged and neutral particles
do not contribute to \met reconstruction.  

Like for the \met response, resolution-related aspects of \met reconstruction are understood from \datatomc-simulations comparisons.  
The scales used for the corresponding evaluations are the overall event activity represented by \sumet, and the \pu activity measured by \NPV. 
The measurement of the \met resolution is discussed in \secRef{subsec:etmiss-resolution:method} and results are presented in \secRef{subsec:etmiss-resolution:results}.

\subsubsection{\met scale determination}\label{subsec:etmiss-scale:method}

In events with \Zmm decays, the transverse momentum of the \Zboson boson (\ptZ) is an indicator of the hardness of the  interaction.
It provides a useful scale for the evaluation of the \met response for this final state without genuine missing transverse momentum.
The direction of the corresponding \Zboson boson transverse momentum vector \ptZvec defines an axis \Az in the transverse plane of the collision, which is reconstructed from the 
\pTvec of the decay products by
\begin{equation}
	\Az =\frac{\pTvecj{\muon^{+}}+\pTvecj{\muon^{-}}}{\left|\pTvecj{\muon^{+}}+\pTvecj{\muon^{-}}\right|} = \frac{\ptZvec}{\ptZ}\,.
	\label{eq:az}
\end{equation}
The magnitude of the component of \metvec parallel to \Az is
\begin{equation}
	\projpar = \metaz\,.
	\label{eq:projpar}
\end{equation}
This projection is sensitive to any limitation in \met reconstruction, in particular with respect to the contribution from the hadronic recoil against \ptZvec, 
both in terms of response and resolution. 
Because it can be determined both for data and \MC simulations, it provides an important tool for the validation of the \met response and the associated systematic uncertainties. 

The expectation value for a balanced interaction producing a \Zboson boson against a hadronic recoil is $\EXV{\projpar} = 0$. 
Any observed deviation from this value represents a bias in the \met reconstruction. 
For $\projpar < 0$, the reconstructed hadronic activity recoiling against \ptZvec is too small, while for  $\projpar > 0$ too much hadronic recoil is reconstructed.
The evolution of \projpar as a function of the hardness of the \Zboson boson production can be measured by evaluating the mean \AVE{\projpar} in bins of the hard-scatter scale $\pThard = \ptZ$.

In addition to measuring the \met response in data and \MC simulation without genuine \met, 
its linearity can be determined using samples of final states with genuine \met in \MC simulations. 
This is done by evaluating the relative deviation \metdl of the reconstructed \met from the expected $\mettrue > 0$ as a function of \mettrue,
\begin{equation}
	\metdl(\mettrue) = \dfrac{\met - \mettrue}{\mettrue}\,.
	\label{eq:devlin}
\end{equation}

\subsubsection{Measuring the \met response}\label{subsec:etmiss-scale:results}

\begin{figure}[t!] \centering
\subfigure[]{\includegraphics[width=\fighalfwidth]{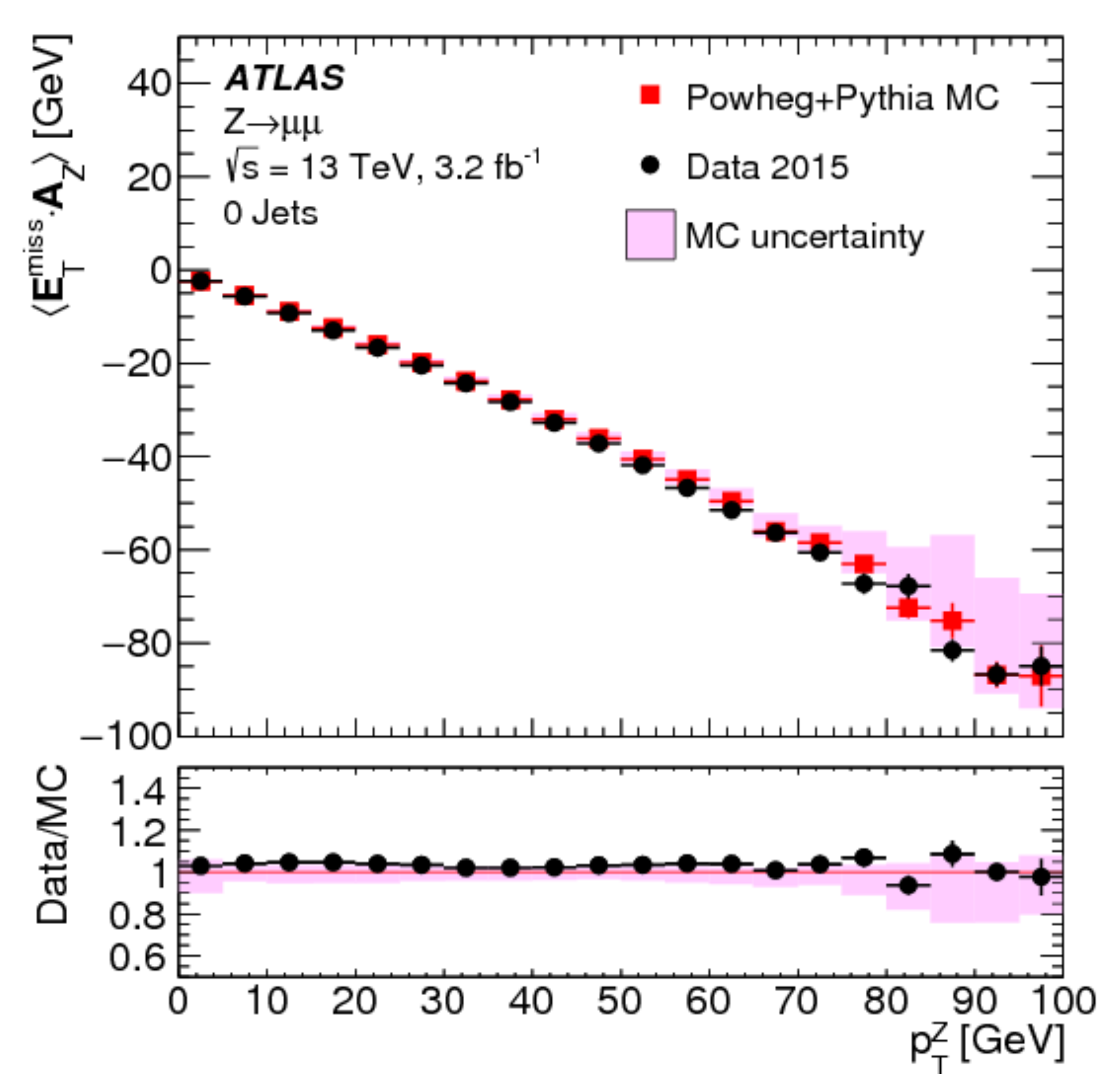}\label{fig:perf:zmm:nojets:scale}}\quad
\subfigure[]{\includegraphics[width=\fighalfwidth]{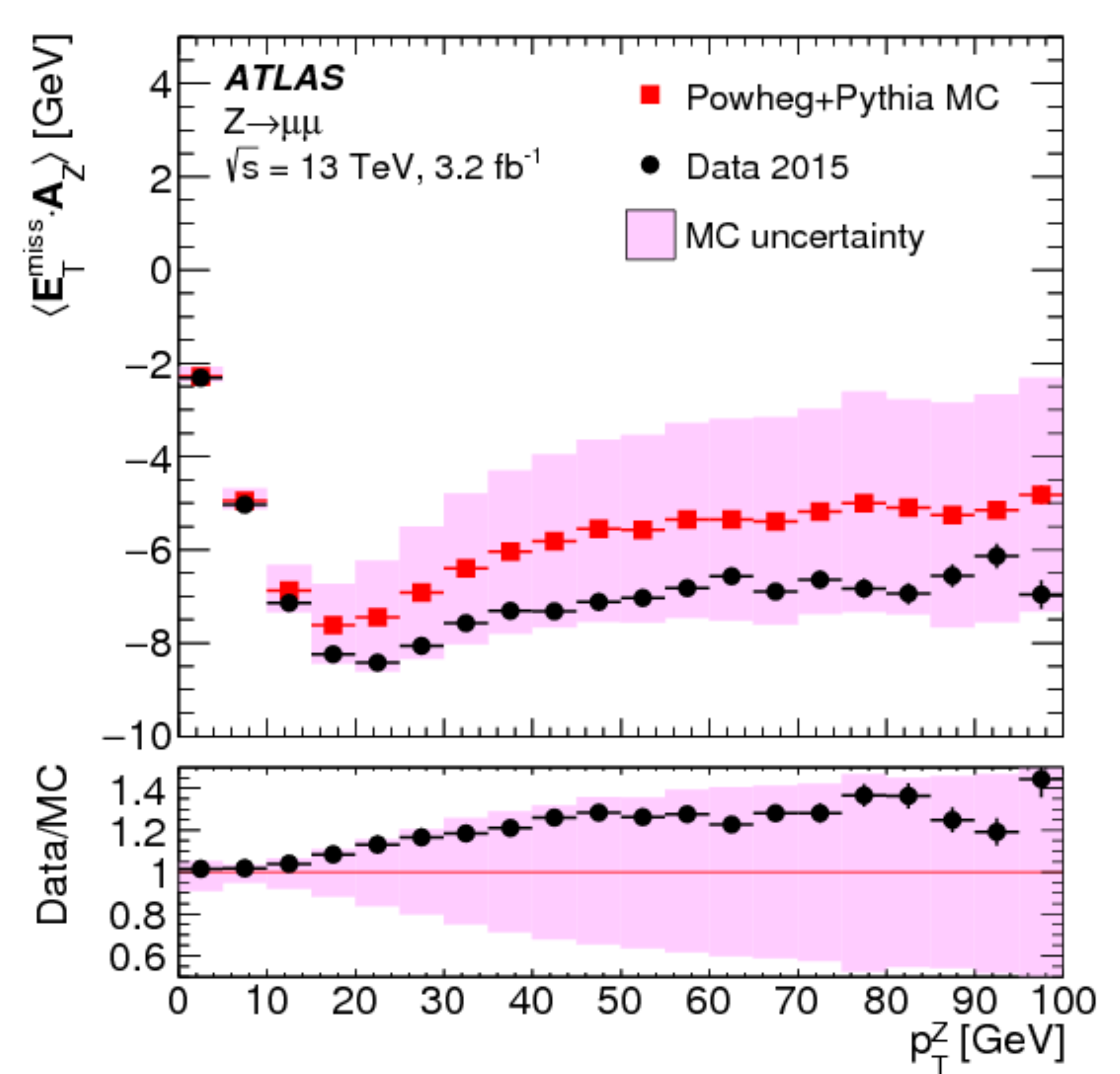}\label{fig:perf:zmm:incl:scale}}\qquad
\caption{The average projection of \metvec onto the direction \Az of the \Zboson boson's transverse momentum vector \ptZvec, as given in \eqRef{eq:projpar}, is shown as a function of $\ptZ = |\ptZvec|$ in \Zmm events from \subref{fig:perf:zmm:nojets:scale} the \nojets sample  and from \subref{fig:perf:zmm:incl:scale} the inclusive sample. In both cases data are compared to \MC simulations. The ratio of the averages from data and \MC simulations are shown below the plots. 
The shaded areas indicate the overall statistical uncertainty combined with systematic uncertainties comprising contributions from the muon and soft-term systematic uncertainties in \subref{fig:perf:zmm:nojets:scale}, and including the additional jet systematic uncertainties in \subref{fig:perf:zmm:incl:scale}, for \MC simulations.}
\label{fig:perf:zmm:scale}
\end{figure}

\FigRef{fig:perf:zmm:scale} shows $\AVE{\projpar}$ as a function of \ptZ for the \nojets and the inclusive \Zmm sample, respectively.
\MC simulations compare well with the data for \nojets, but show larger deviations up to 30\% for the inclusive selection.
Nevertheless, these differences are still found to be within the total uncertainty of the measurement. 

The steep decrease of  $\AVE{\projpar}$ with increasing \ptZ in the \nojets sample seen in \subfigRef{fig:perf:zmm:scale}{fig:perf:zmm:nojets:scale} reflects 
the inherent underestimation of the soft term, as in this case the hadronic recoil is exclusively represented by \ID tracks with $\pT > \unit{400}{\MeV}$ within $\abseta < 2.5$. 
It thus does not contain any signal from (1) neutral particles, (2) charged particles produced with $\abseta > 2.5$, and (3) charged particles produced  within $\abseta < 2.5$  but with \pT below threshold, rejected by the track quality requirements, or not represented by a track at all due to insuffcient signals in the \ID (e.g., lack of hits for track fitting).

In the case of the inclusive sample shown in \subfigRef{fig:perf:zmm:scale}{fig:perf:zmm:incl:scale}, the \met response is recovered better as \ptZ increases, 
since an increasing number of events enter the sample with a reconstructed recoil containing fully calibrated jets. 
These provide a more complete representation of the hadronic transverse momentum flow.
The residual offsets in \AVE{\projpar} of about \unit{8}{\GeV} in data and \unit{6}{\GeV} in \MC simulations observed for $\ptZ \gtrsim \unit{40}{\GeV}$ in 
\subfigRef{fig:perf:zmm:scale}{fig:perf:zmm:incl:scale} agree within the uncertainties of this measurement.  

\begin{figure}[t!] \centering
\includegraphics[width=\fighalfwidth]{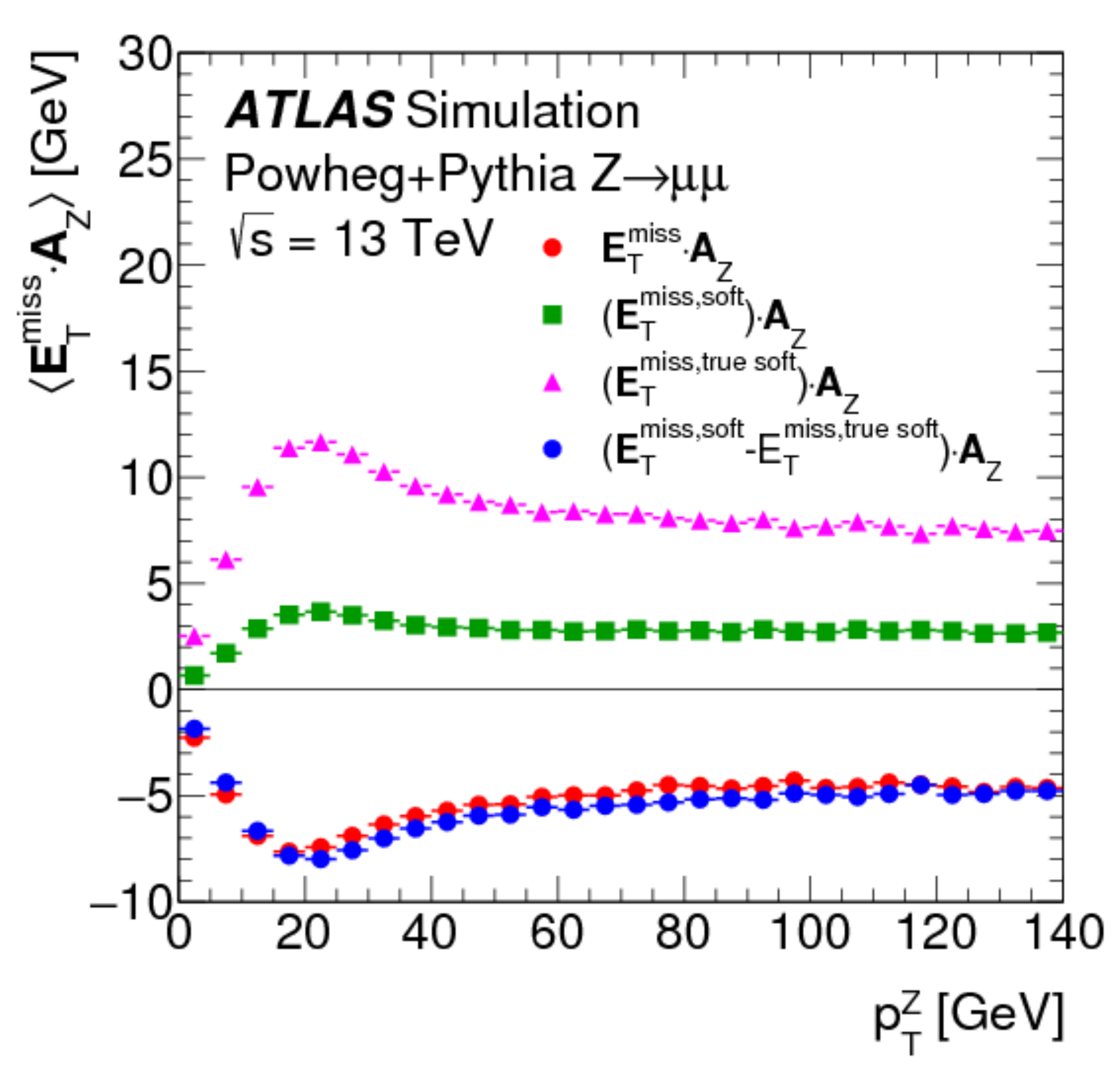}
\caption{The average projection of \metvec onto the direction \Az of the \Zboson boson's transverse momentum vector \ptZvec, as given in \eqRef{eq:projpar}, is shown as a function of $\ptZ = |\ptZvec|$ in \Zmm events from the inclusive \MC sample. The average projection of the soft term and the true soft term are also shown, 
to demonstrate the source of the deviation from zero.}
\label{fig:perf:zmm:scale:true}
\end{figure}

The persistent bias in \AVE{\projpar} is further explored in \figRef{fig:perf:zmm:scale:true}, which compares variations of \AVE{\projpar} respectively using the full \metvec, the soft-term contribution \metvsft only, the hard-term contribution $\metvec-\metvsft$, and the true soft term \metvecj{\text{true\ soft}} only, as a function of \ptZ, for the \Zmm sample from \MC simulations. 
In particular the difference between the projections using \metvecj{\text{true\ soft}} and \metvsft indicates the lack of reconstructed hadronic response, 
when $\metvsft = \metvecj{\text{true\ soft}}$ is expected for a fully measured recoil. 
The parallel projection using only the soft terms is larger than zero for all \ptZ due to the missing \Zboson-boson contribution to \metvec given by $-\ptZvec$.

\begin{figure}[t!]\centering
\begin{center}
\includegraphics[width=\fighalfwidth]{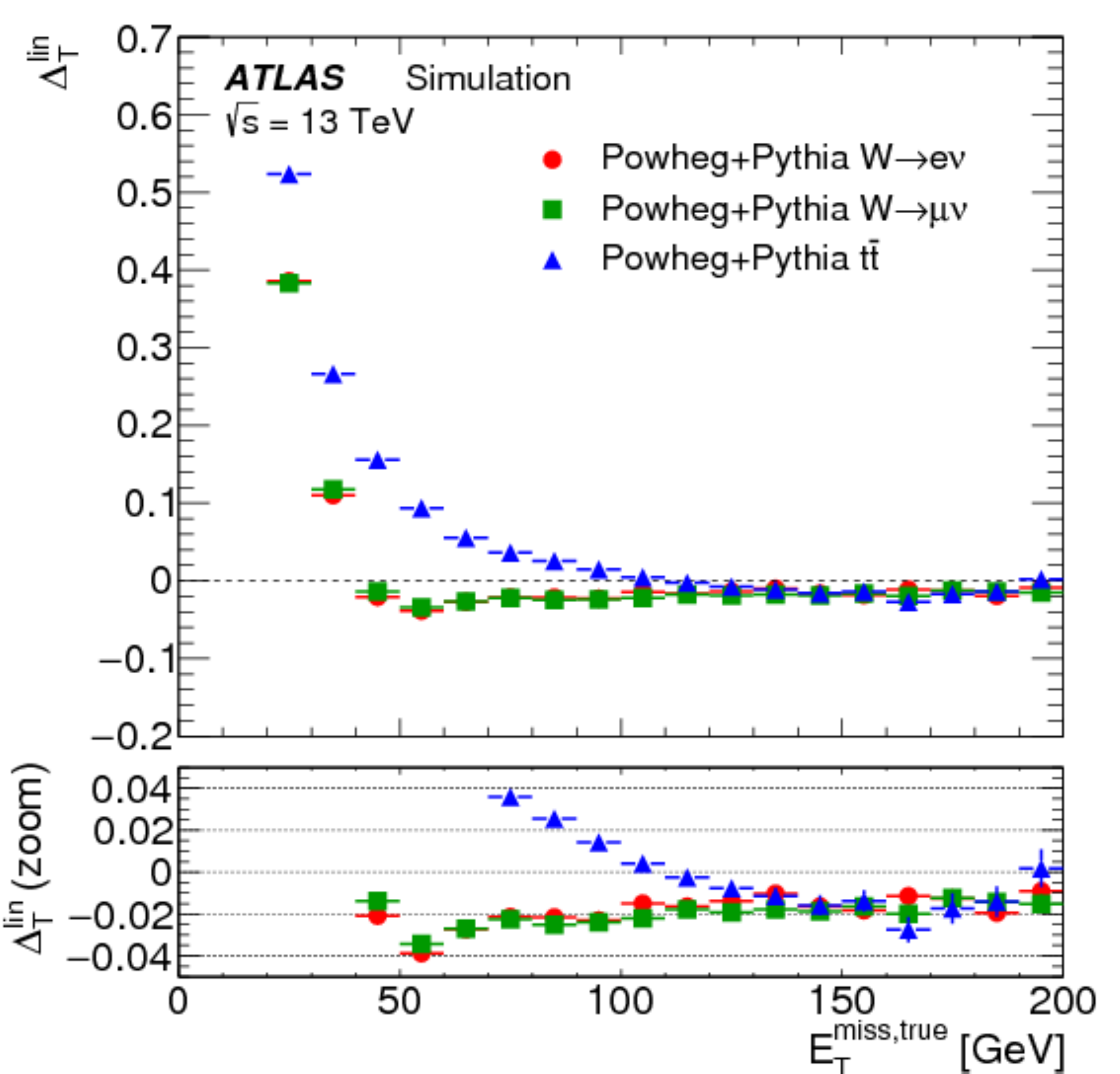}
\end{center}
\caption{The deviation of the \met response from linearity, measured as a function of the expected \mettrue by \metdl in \eqRef{eq:devlin}, in \Wen, \Wmn, and \ttbar final states in \MC simulations. The lower plot shows a zoomed-in view on the \metdl dependence on \mettrue with a highly suppressed ordinate.}
\label{fig:perf:linearity}
\end{figure}

The deviation from linearity in \met reconstruction, measured by \metdl given in \eqRef{eq:devlin}, is shown as a function of \mettrue for \MC simulations of \Wen, \Wmn and \ttbar production in \figRef{fig:perf:linearity}.
The observed $\metdl > 0$ at low \mettrue indicates an overestimation of \mettrue by the reconstructed \met due to the observation biases arising from the finite \met resolution, as discussed in \secRef{subsec:etmiss-basics}.  
This bias overcompensates the lack of reconstructed \pT-flow from the incompletely measured hadronic recoil in \Wen and \Wmn events for $\mettrue \lesssim \unit{40}{\GeV}$ with an increasing non-linearity observed with decreasing \mettrue.
For $\mettrue \gtrsim \unit{70}{\GeV}$ the \met response is directly proportional to \mettrue, with the reconstructed recoil being approximately 2\% too small. 
The \Wen and \Wmn final states show very similar $\metdl(\mettrue)$, thus indicating the universality of the recoil reconstruction 
and the independence on the lepton flavour of the reconstructed \met 
in a low-multiplicity final state with $\mettrue > 0$.

In \ttbar final-state reconstruction, resolution effects tend to dominate \metdl at $\mettrue \lesssim \unit{120}{\GeV}$. 
Compared to the \Wen and \Wmn final states, a significantly poorer \met resolution is observed in this kinematic region, due to the presence of at least four jets with relatively low \pT and high sensitivity to \pu-induced fluctuations in each event of the \ttbar sample. 
For $\mettrue > \unit{120}{\GeV}$, $\metdl(\mettrue) \approx 2$\% indicates a proportional \met response with a systematic shift similar to the one observed in inclusive \Wboson-boson production.

\subsubsection{Determination of the \met resolution}\label{subsec:etmiss-resolution:method}

The \met resolution is determined by the width of the combined distribution of the differences between the measured \metxy and the  components of the true missing transverse momentum vector $\metvtrue = (\metxj{\true},\metyj{\true})$. 
The width is measured in terms of the \RMS{}, with
\begin{align}
	\metresxy = \left\{\begin{array}{ll}
					\RMS{\metxy - \mettruexy} & \text{\Wen or \ttbar sample\ } (\mettrue > 0) \\[1.5mm]
					\RMS{\metxy}                    & \text{\Zmm sample\ } (\mettrue = 0)
				\end{array}\right.\,.
	\label{eq:metreso}
\end{align} 
This metric does not capture all of the effects driving the fluctuations in \met reconstruction, such as biases between individual \met terms or the behaviour of outliers, 
but it is an appropriate general measure of how well \met represents \mettrue. 

Using the \Zmm sample  allows direct comparisons of \metresxy between data and \MC simulations, as $\mettrue = 0$ in this case.
The resolution in final states with genuine \met is determined with \MC simulations alone. 
For \Wen and \ttbar final states, $\mettruexy = \pxyj{\nu}$ is used.

\subsubsection{\met resolution measurements}\label{subsec:etmiss-resolution:results}

\begin{figure}[t!] \centering
\subfigure[]{\includegraphics[width=\fighalfwidth]{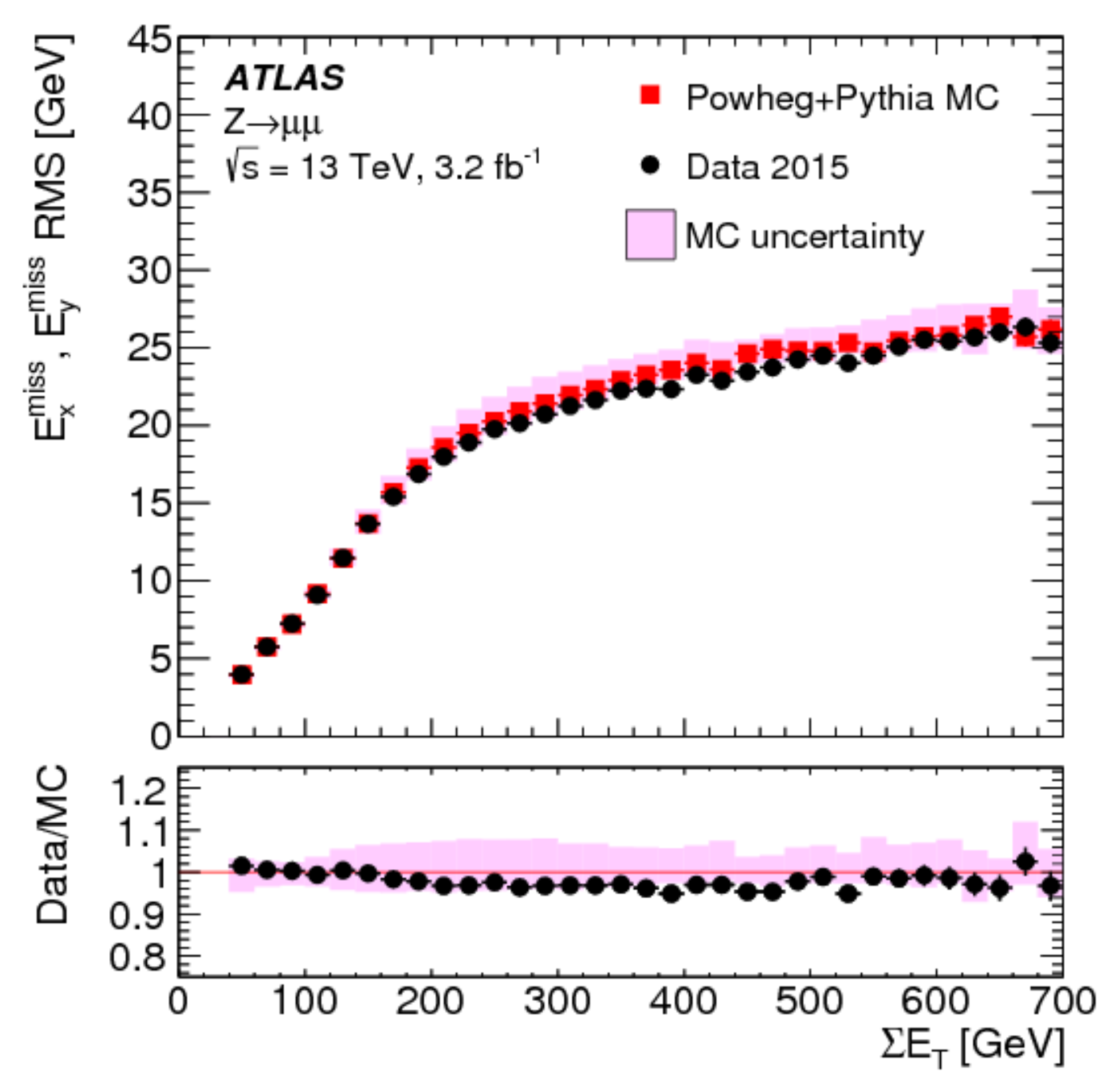}\label{fig:perf:zmm:reso:sumet}}\quad
\subfigure[]{\includegraphics[width=\fighalfwidth]{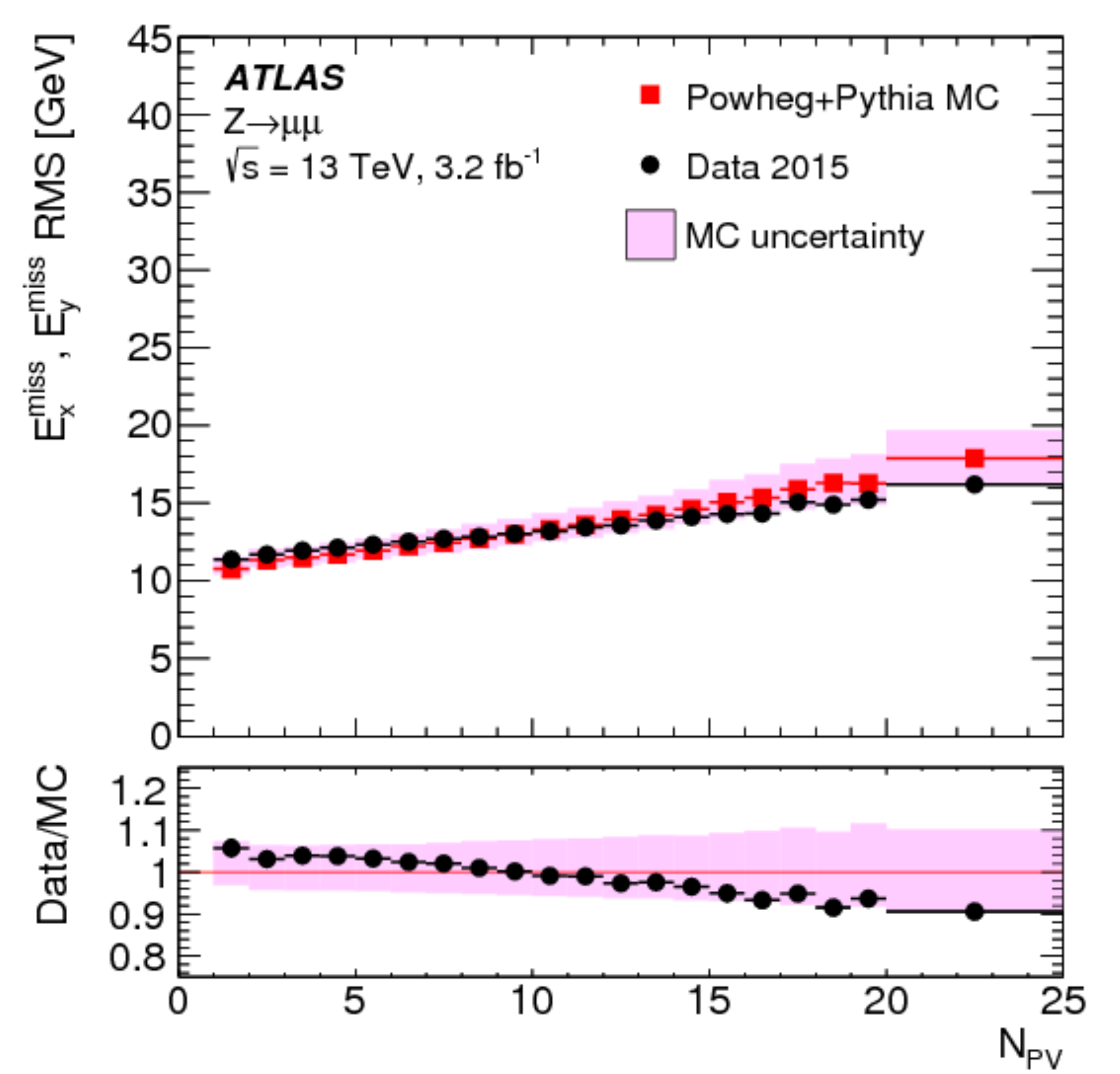}\label{fig:perf:zmm:reso:npv}}\qquad
\caption{The RMS width of the \metxy distributions \subref{fig:perf:zmm:reso:sumet} in bins of \sumet  and \subref{fig:perf:zmm:reso:npv} in bins of the number of primary vertices in an inclusive sample of \Zmm events. Predictions from \MC simulations are overlaid on the data points, and the ratios are shown below the respective plot. The shaded bands indicate the combined statistical and systematic uncertainties of the resolution measurements. }
\label{fig:perf:zmm:reso}
\end{figure}
 
\begin{figure}[t!]\centering
\subfigure[]{\includegraphics[width=\fighalfwidth]{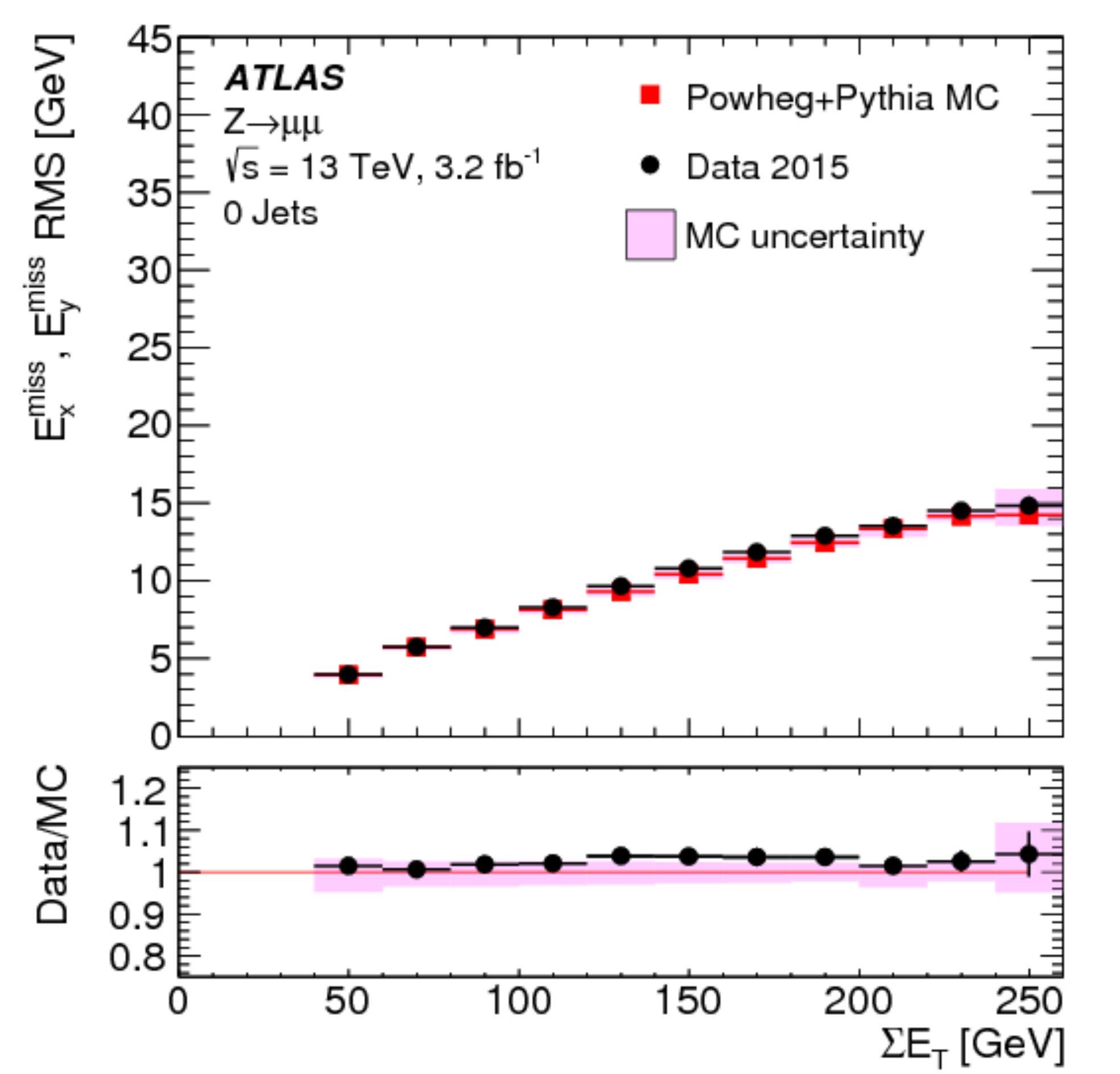}\label{fig:perf:zmm:nojets:reso:sumet}}\quad
\subfigure[]{\includegraphics[width=\fighalfwidth]{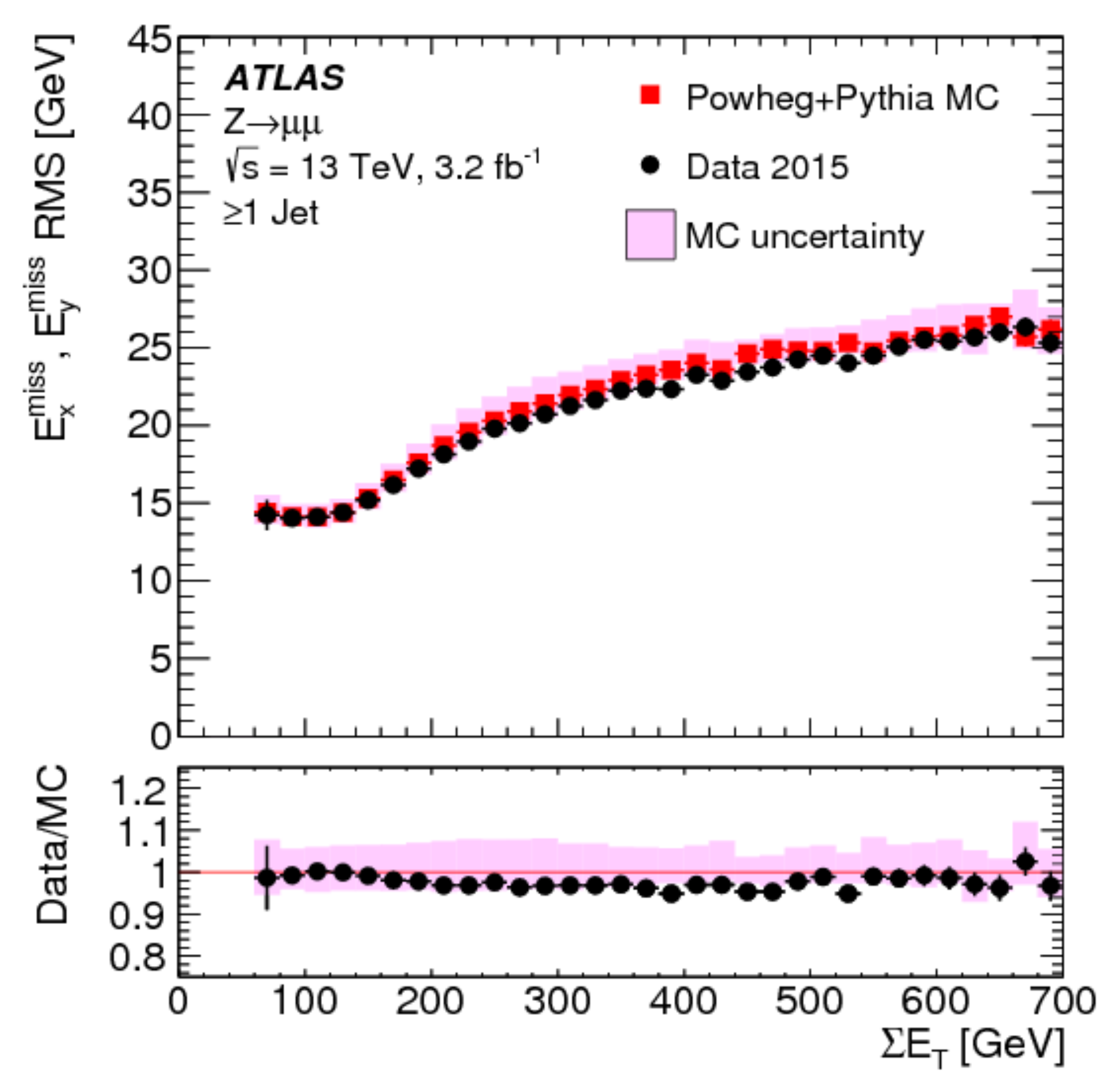}\label{fig:perf:zmm:jets:reso:sumet}}\qquad
\subfigure[]{\includegraphics[width=\fighalfwidth]{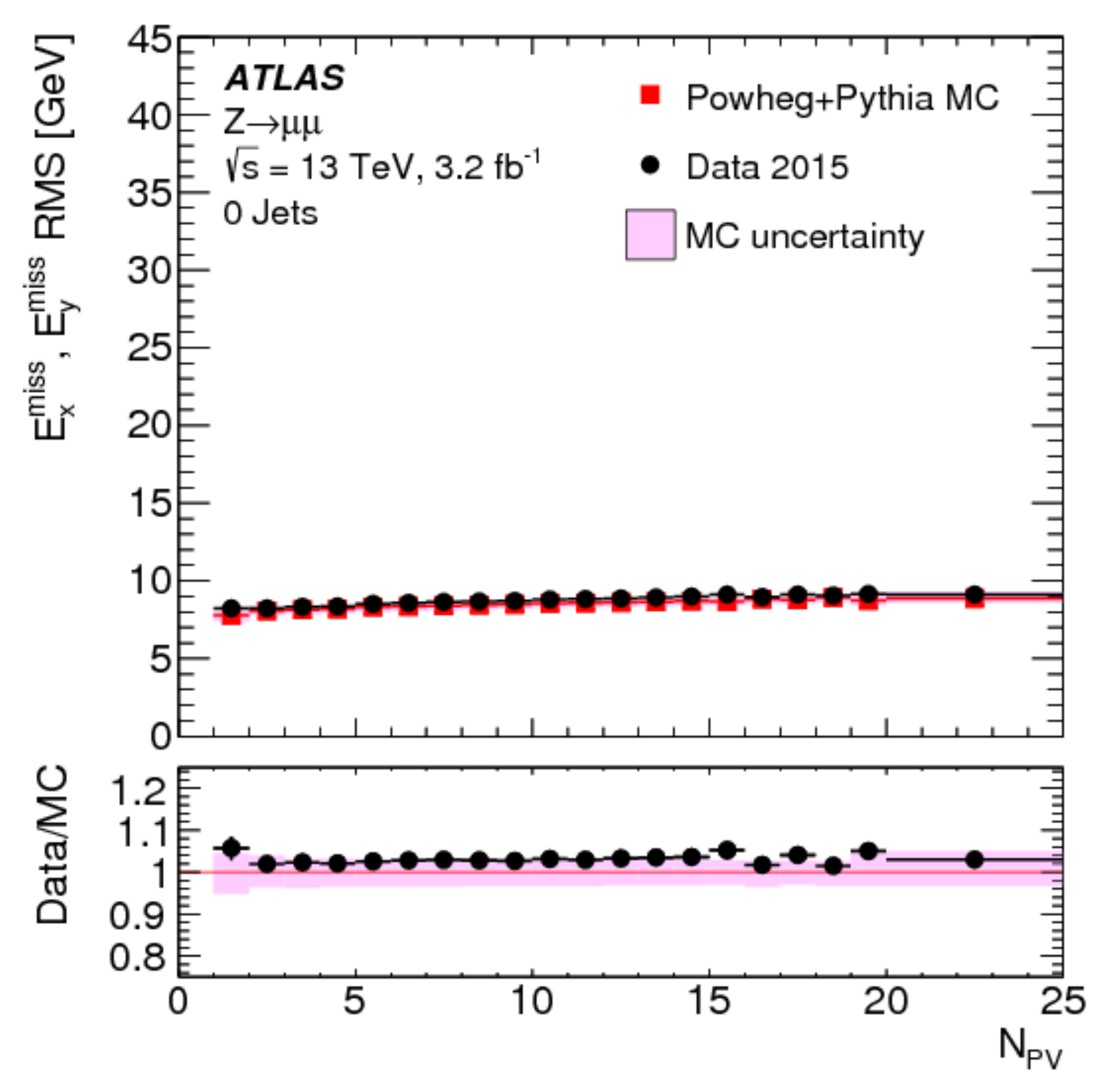}\label{fig:perf:zmm:nojets:reso:npv}}\quad
\subfigure[]{\includegraphics[width=\fighalfwidth]{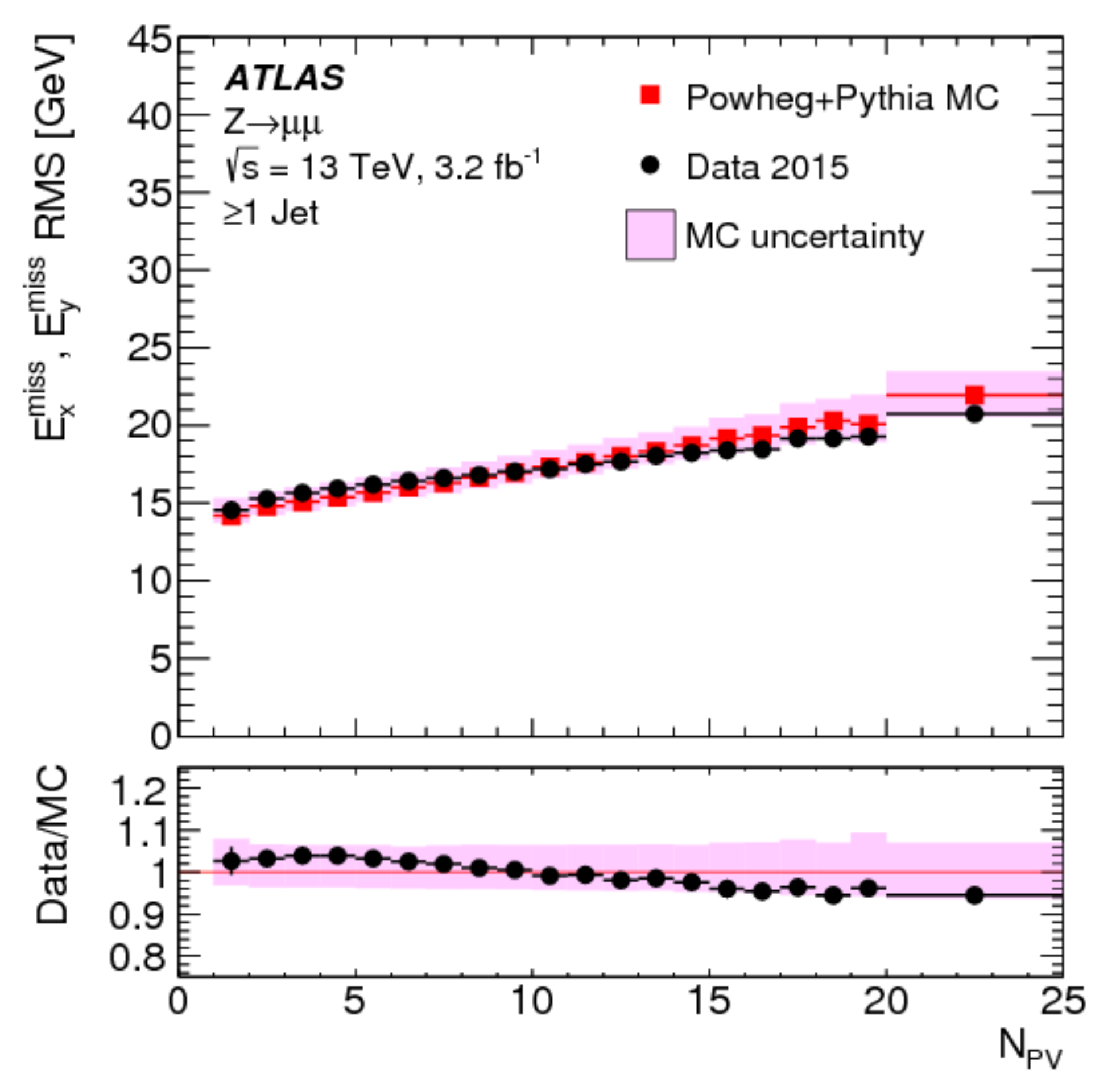}\label{fig:perf:zmm:jets:reso:npv}}\qquad
\caption{The \met resolution \metresxy determined for \subref{fig:perf:zmm:nojets:reso:sumet} an exclusive \Zmm sample without jets with $\pT > \unit{20}{\GeV}$ ($\Njet = 0$) and for \subref{fig:perf:zmm:jets:reso:sumet} an exclusive sample with at least one jet above this threshold ($\Njet \geq 1$), as a function of \sumet in data and \MC simulations. The dependence of \metresxy on the \pu activity, as measured by \NPV, for these two samples is shown in \subref{fig:perf:zmm:nojets:reso:npv} and \subref{fig:perf:zmm:jets:reso:npv}, respectively. The shaded bands indicate the combined statistical and systematic uncertainties associated with the measurement.}
\label{fig:perf:zmm:excl:reso}
\end{figure} 
The \met resolution measured by \metresxy is evaluated as a function of the event activity measured by \sumet given in \eqRef{eq:sumet}.
For the inclusive \Zmm sample, \subfigRef{fig:perf:zmm:reso}{fig:perf:zmm:reso:sumet} shows \metresxy quickly rising from less than \unit{5}{\GeV} to about \unit{10}{\GeV}
with increasing \sumet within $\unit{50}{\GeV} \leq \sumet < \unit{70}{\GeV}$.\footnote{This lower boundary of this range is given by the muon selection with $\pTmuon > \unit{25}{\GeV}$, as described in \secRef{subsec:event-selection-zmumu}, assuming no other \hs vertex tracks, i.e. $\metsft = 0$.  The upper boundary indicates the lower limit of \sumet to accommodate at least one jet with $\pTjet > \unit{20}{\GeV}$ in addition the two muons (for the jet selection see \secRef{subsec:jet-sel}).}
This is due to the fact that in this range the two muons are the dominant hard objects contributing, with a \pT resolution proportional to $(\pTmuon)^{2}$.
A convolution of the muon resolution with a small contribution from \metsft is possible for $\sumet > \unit{50}{\GeV}$. 
This component is on average about 60\% of \ptZ, and subject to the stochastic fluctuations further discussed below.

The increase of $\Zmm + 1$ jet topologies in the \Zmm sample leads to an additional source of fluctuations affecting $\metresxy(\sumet)$ for $\unit{70}{\GeV} < \sumet \lesssim \unit{180}{\GeV}$.   
In general the \Zmm sample collected for this study covers $\ptZ \lesssim \unit{140}{\GeV}$ with relevant statistics.
At this limit it is expected that the hadronic recoil contains two reconstructed jets, with the onset of this contribution at \sumet of about \unit{180}{\GeV}.
The corresponding change of the dominant final state composition for $\sumet > \unit{180}{\GeV}$ leads to a change of shape of $\metresxy(\sumet)$, as the transverse momentum of the individual jets rises and the number of contributing jets slowly increases.
The expected $\metresxy(\sumet) \propto \sqrt{\sumet}$ scaling driven by the jet-\pT resolution~\cite{PERF-2011-04} therefore dominates \metresxy at these higher \sumet. 
The \MC predictions for $\metresxy(\sumet)$ agree with the data within a few percent and well within the total uncertainties of this measurement.
A tendency for slightly poorer resolution in \MC simulations is observed, in particular for $\sumet > \unit{200}{\GeV}$.

Any contribution from \pu to \metresxy is expected to be associated with the jets. 
While dedicated corrections applied to the jets largely suppress \pu contributions in the jet response, residual irreducible fluctuations introduced into the calorimeter signals by \pu lead to a degradation of the jet energy resolution and thus poorer resolution in the jet-\pT measurement.  
The dependence of \metresxy on the \pu activity measured by \NPV is shown in \subfigRef{fig:perf:zmm:reso}{fig:perf:zmm:reso:npv}. 
Data show a less steep slope of $\metresxy(\NPV)$ than \MC simulations, but with about 10\% worse resolution in the low \pu region of $\NPV \lesssim 5$.
The resolution in data is better than in \MC simulations by about 10\% for the region of higher \pu activity at $\NPV \approx 20$.   

The differences between data and \MC simulations seen in $\metresxy(\sumetevt)$ for the inclusive \Zmm sample can be further analysed by splitting the sample according to the value of \Njet. 
\SubfigRef{fig:perf:zmm:excl:reso}{fig:perf:zmm:nojets:reso:sumet} shows the dependence of \metresxy on \sumetevt for \Zmm events with $\Njet = 0$. 
The dominant source of fluctuations other than the muon-\pT resolution is in this case introduced by the incomplete reconstruction of the hadronic recoil.
These fluctuations increase with increasing \ptZ, which in turn means higher overall event activity measured by \sumetevt. 
For this sample \metresxy in data compares well to \MC simulations, at a level of a few percent, without any observed dependence on \sumetevt.

The exclusive $\Njet \geq 1$ samples extracted from \Zmm data and \MC simulations show the expected $\metresxy \propto \sqrt{\sumetevt}$ scaling in \subfigRef{fig:perf:zmm:excl:reso}{fig:perf:zmm:jets:reso:sumet}. 
The resolution in data is well represented by \MC simulations, at the level of a few percent.
The slightly better resolution observed in data with increasing \sumetevt
follows the trend observed in \subfigRef{fig:perf:zmm:reso}{fig:perf:zmm:reso:sumet}. 
The similar trends are expected as this kinematic region is largely affected by the jet contribution.

The dependence of \metresxy on \NPV  shown in \subfigRef{fig:perf:zmm:excl:reso}{fig:perf:zmm:nojets:reso:npv} indicates that the \met resolution is basically independent of \pu, for the $\Njet = 0$ sample. 
This is expected from the exclusive \met composition comprising the (track-based) \metmuo and \metsft terms only.
Data and \MC simulations compare well within a few percent, and without any observable dependence on \NPV.
\SubfigRef{fig:perf:zmm:excl:reso}{fig:perf:zmm:jets:reso:npv} shows the \NPV dependence of \metresxy for the $\Njet \geq 1$ sample. 
Comparing this result to \subfigRef{fig:perf:zmm:reso}{fig:perf:zmm:reso:npv} confirms that all \pu dependence of the \met resolution is arising from the jet term. 
Both trend and magnitude of the \datamc comparison follow the observation from the inclusive analysis. 

\subsubsection{\met resolution in final states with neutrinos}\label{subsec:etmiss-resolution:neutrino}

\begin{figure}[t!]
\begin{center}
\subfigure{\includegraphics[width=\fighalfwidth]{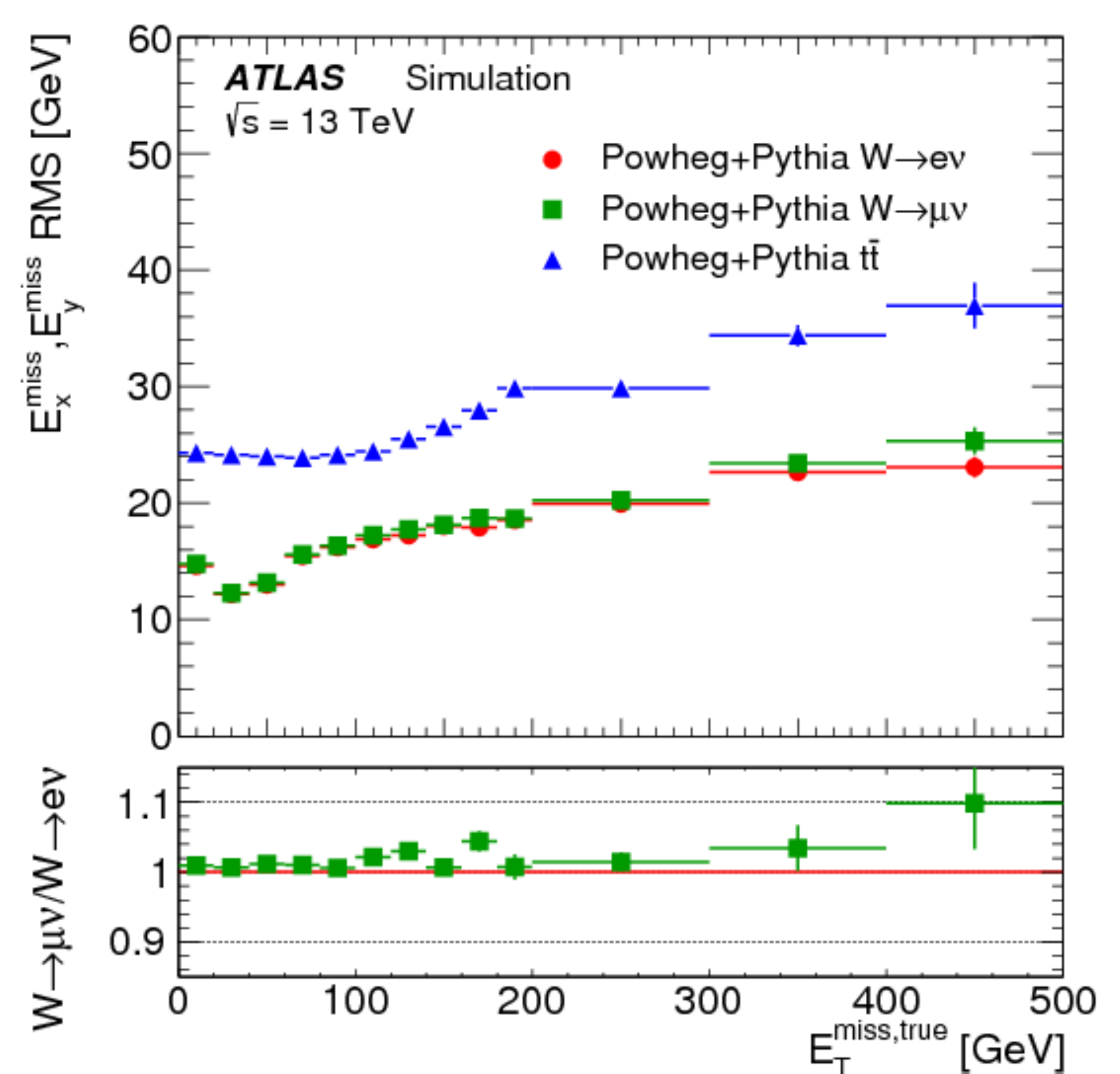}}
\end{center}
\caption{The \met resolution measured by \metresxy as a function of the true missing transverse momentum \mettrue for the \Wen, \Wmn, and \ttbar samples from \MC simulations.}
\label{fig:perf:mtrue:reso}
\end{figure}

The \met resolution for final states with $\mettrue > 0$ is measured by \metresxy according to \eqRef{eq:metreso} and evaluated using dedicated inclusive \Wen and \Wmn samples from \MC simulations, and the inclusive \ttbar \MC sample defined in \secRef{subsec:event-selection-ttbar}.
For these samples, \metresxy can be determined as a function of $\mettrue = \pTnu$.
The dedicated \Wen and \Wmn samples are obtained with an event selection based on the description in \secRef{subsec:event-selection-wenu}, but omitting both the \met-based and the \mT-based selections. 

\FigRef{fig:perf:mtrue:reso} shows \metresxy evaluated as a function of \mettrue for these samples.
The universality of the response to the hadronic recoil observed in \figRef{fig:perf:linearity}, together with the different but subdominant contributions from the \pT resolutions of the electrons and muons,
yield a very similar \met resolution for \Wen and \Wmn final states.
Generally, poorer resolution is observed in \ttbar final states.
The deviation from the expected $\metresxy \propto \sqrt{(\mettrue)}$ scaling behaviour for \Wln at lower \mettrue reflects the kinematic features of the \Wboson boson and its decay. 
Events with low \pTW, and therefore small hadronic recoil, lie predominantly in the region $\unit{25}{\GeV}\lesssim\pTnu\lesssim\unit{50}{\GeV}$. 
Since the hadronic recoil is generally the poorly measured component of an event and the reconstructed \met    
is dominated by the lepton \pT in this region, the \met resolution tends to be better here than for   
events with larger hadronic recoil populating $\pTnu\lesssim\unit{25}{\GeV}$ and $\pTnu\gtrsim\unit{50}{\GeV}$.

\subsection{\texorpdfstring{\met}{Etmiss} tails}
\label{subsec:etmiss-tails}

Large reconstructed \met is an indicator for the production of (potentially new) undetectable particles, but can also be generated by detector problems and/or poor reconstruction of the objects used for its reconstruction. 
Enhanced tails in the distribution of the \metvec components for final states with well-known expectation values for \met are indicative of such inefficiencies.

Non-Gaussian shapes in the distribution arise from a combination of object selection inefficiencies and potentially non-Gaussian resolutions of the \met constituents.
Even for a well-defined final state, event-by-event fluctuations in terms of which particles, jets, and soft tracks enter the \met reconstruction, and with which \pT, lead to deviations from a normally distributed (\metx,\mety) response.

\begin{figure}[t!] \centering
\includegraphics[width=\fighalfwidth]{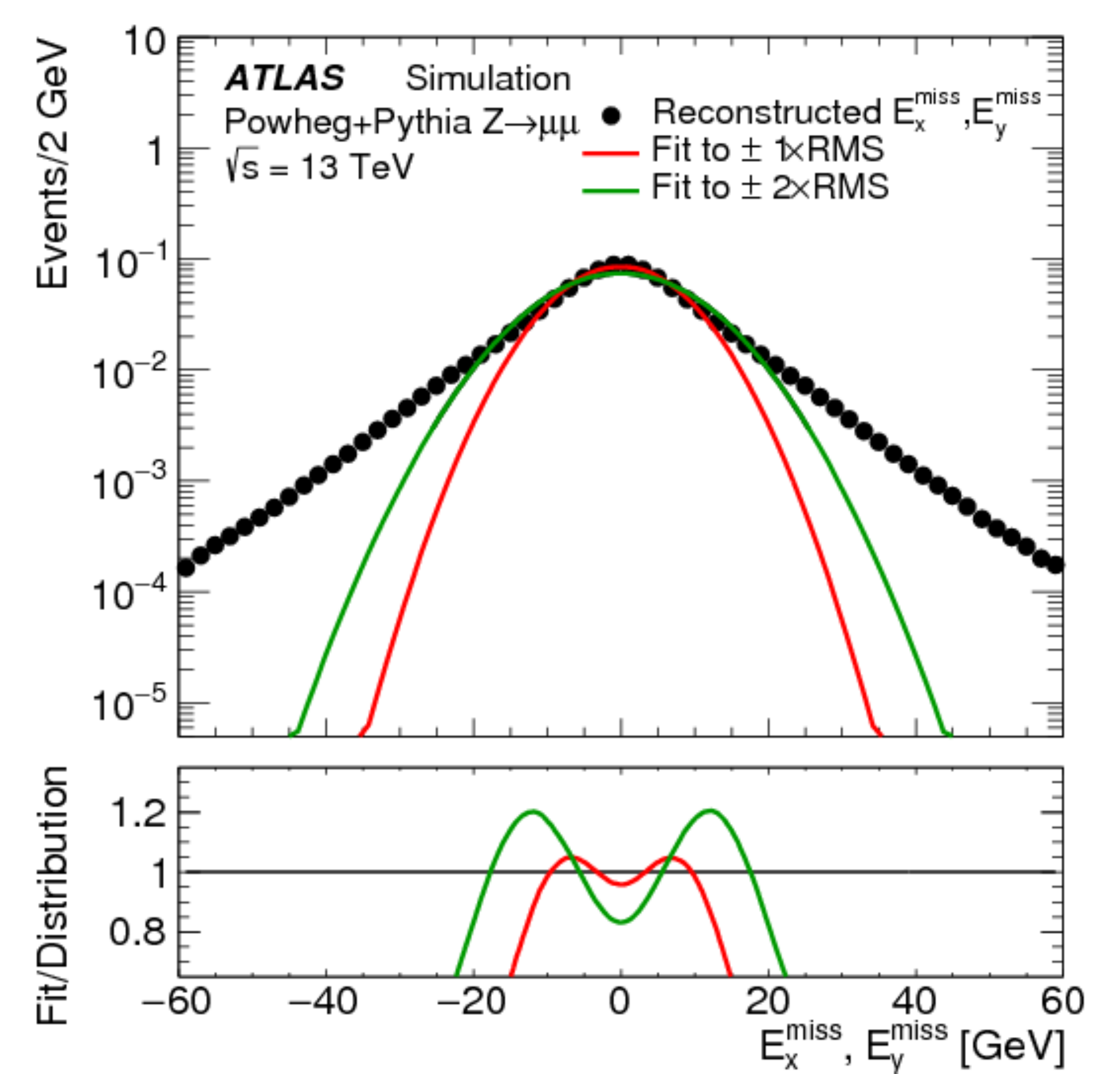}
\caption{The combined distribution of \metx and \mety for an inclusive \Zmm from simulation. Gaussian fits limited to the $\pm 1\times\text{RMS}$ and $\pm 2\times\text{RMS}$ ranges around the centre of the distribution are shown, together with the respective differences between the fitted functions and the actual distribution.}
\label{fig:perf:zmm:gaus}
\end{figure}

\FigRef{fig:perf:zmm:gaus} shows the combined (\metx,\mety) distribution for the inclusive \Zmm sample from \MC simulations. 
To illustrate its symmetric nature and its deviation from a normal distribution in particular with respect to the tails, Gaussian functions are fitted to two limited ranges around the centre of the distribution, $\pm 1\times\text{RMS}$ and $\pm 2\times\text{RMS}$. 
The differences between these functions and the data distribution (lower panel of \figRef{fig:perf:zmm:gaus}) indicate a more peaked shape around the most probable value for \metxy with near exponential slopes.
The result of this comparison supports the choice of \metresxy defined in \eqRef{eq:metreso} in \secRef{subsec:etmiss-resolution:method} for the determination of the \met resolution, rather than using any of the widths measured by fitting Gauss functions in selected ranges of the distribution.

The tails in this shape are reflected in the distribution of \met itself and can be estimated by measuring the fraction of events with $\met > \metj{\text{threshold}}$,
\begin{align}
	f_{\text{tail}} = \dfrac{1}{\mathcal{H}}{\int_{\metj{\text{threshold}}}^{\infty} h(\met) \text{d}\met}\,\text{, with\ } & \mathcal{H} = \int_{0}^{\infty} h(\met) \text{d}\met\,.
	\label{eq:tails}
\end{align}
Here $h(\met)$ is the \met distribution for a given event sample, and \metj{\text{threshold}} is a threshold set to estimate tails.
Any decrease of $f_{\text{tail}}$ at a fixed integral $\mathcal{H}$ indicates an improvement of the \met resolution, and is more sensitive to particular improvements than e.g. \metresxy. 
For example, improving the \metsft reconstruction by rejecting \ID tracks from the \hs vertex with poor reconstruction quality yields a significantly smaller $f_{\text{tail}}$ for the same event sample.   

\begin{figure}[t!]\centering
\subfigure[]{\includegraphics[width=\fighalfwidth]{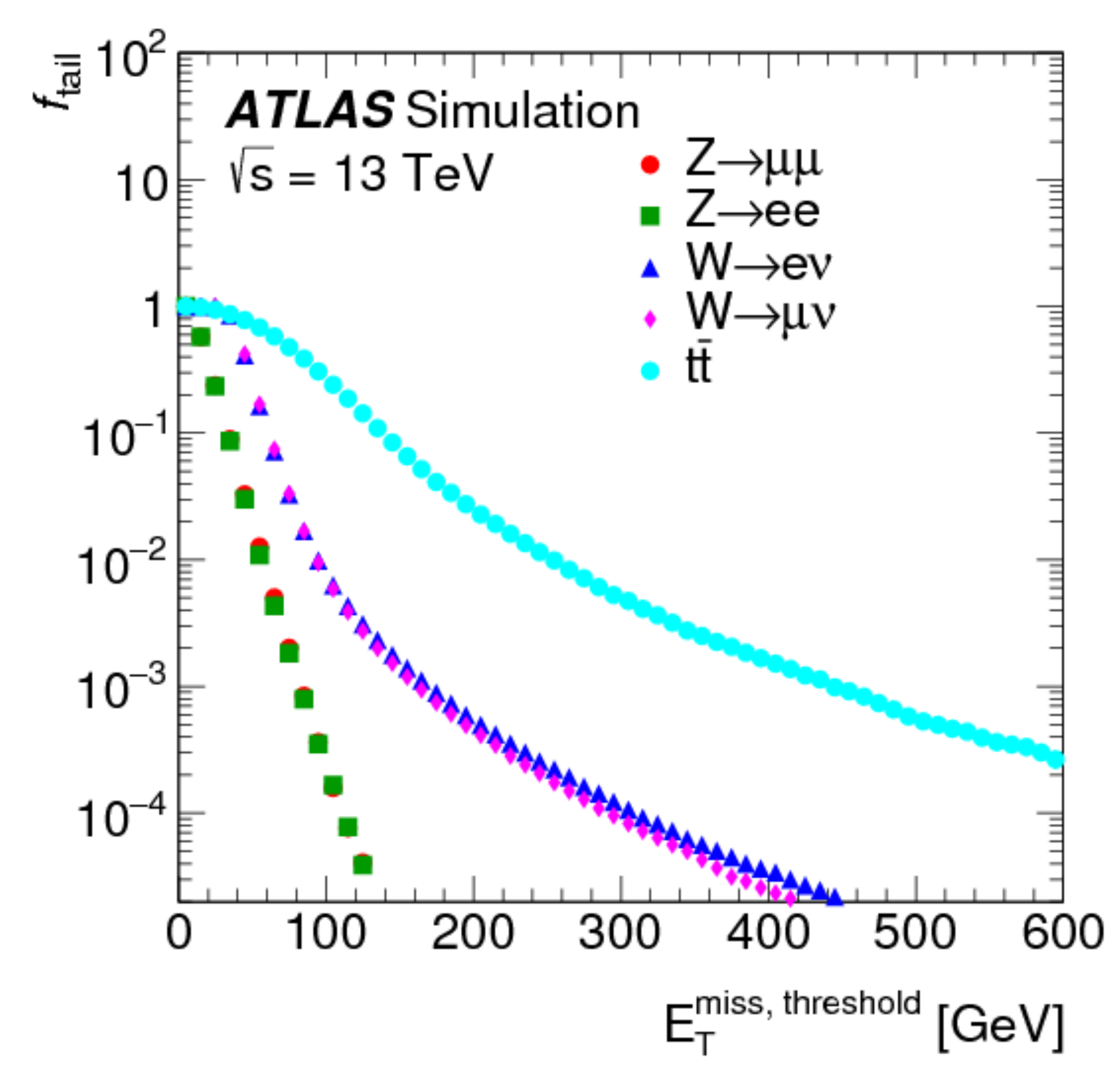}\label{fig:perf:tails:abs}}
\subfigure[]{\includegraphics[width=\fighalfwidth]{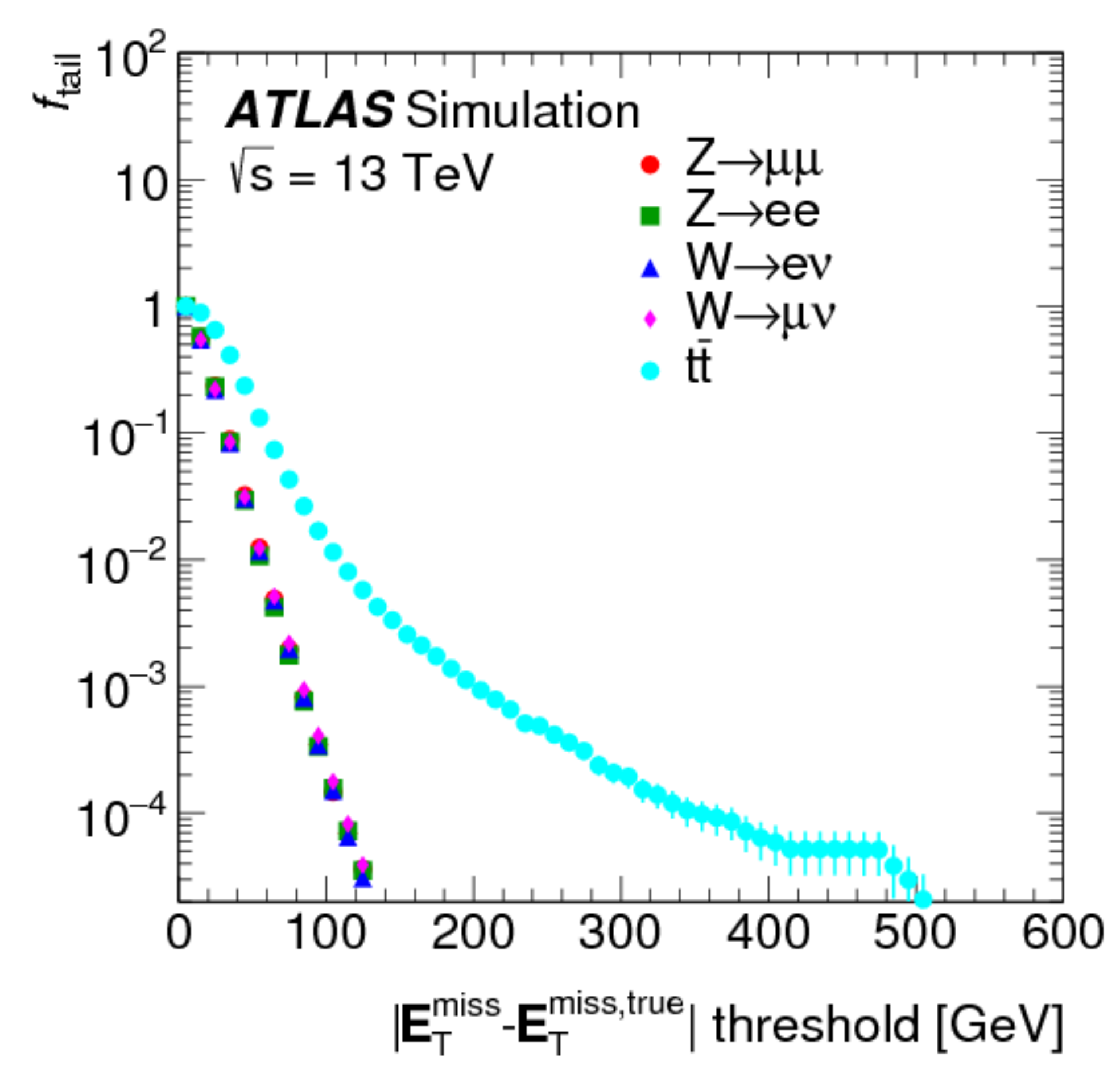}\label{fig:perf:tails:norm}}\\
\subfigure[]{\includegraphics[width=\fighalfwidth]{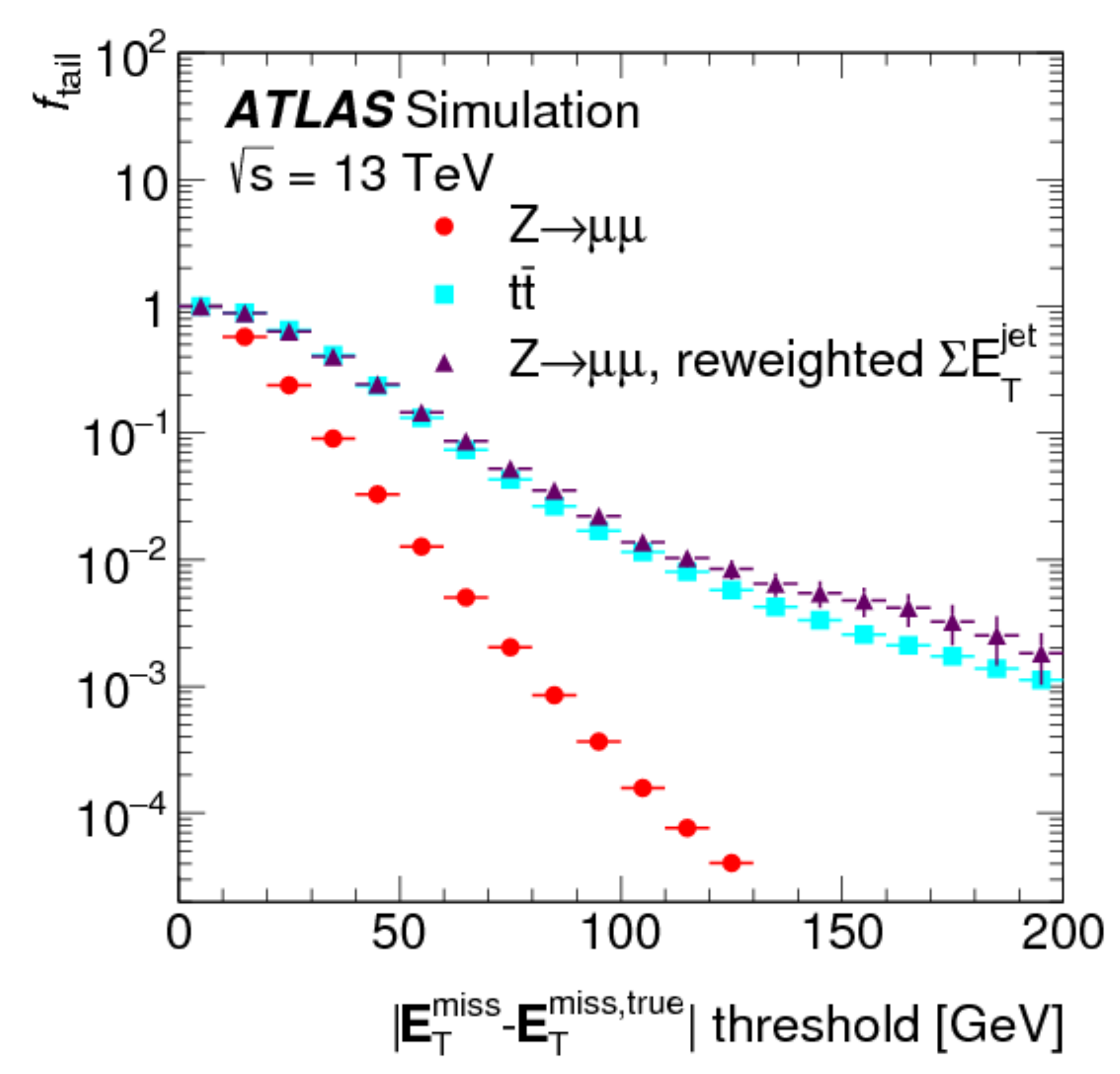}\label{fig:perf:tails:reweight}}
\caption{In \subref{fig:perf:tails:abs} the integral tail fraction $f_{\text{tail}}$ given in \eqRef{eq:tails} is shown as a function of the integration threshold \metj{\text{threshold}}, for \MC simulations of \Zll, \Wln, and \ttbar final states. The tail fraction in terms of a threshold applied to $|\metvec-\metvtrue|$, the distance between the reconstructed (\metvec) and the expected (\metvtrue) vectors, is shown in \subref{fig:perf:tails:norm} for all considered final states. The same fraction is shown in \subref{fig:perf:tails:reweight} for the \met distributions for \Zmm before and after a reweighting following the \sumet distribution for \ttbar is applied, together with $f_{\text{tail}}$ from the \ttbar final state.}  
\label{fig:perf:process_tails}
\end{figure}

The tails in the \met distributions for the final states considered for this study are quantified by the fraction of events above a certain \met threshold using \MC simulations. 
\SubfigRef{fig:perf:process_tails}{fig:perf:tails:abs} shows that the \Zll events ($\ell = \electron$ or $\ell = \muon$) with $\mettrue = 0$ have significantly reduced tails when compared to \Wln and \ttbar with this metric, and that the tails do not depend on the lepton flavour.
A modification of this metric, taking into account \mettrue such that the fraction of events with $|\metvec - \metvtrue|$ above a given threshold is determined, shows the universality of the hadronic recoil in \Zll and \Wln, as can be seen in \subfigRef{fig:perf:process_tails}{fig:perf:tails:norm}. 
 
Another finding of this study is that the tail in the $|\metvec -\metvtrue|$ distribution for the higher \sumetjet \ttbar sample is considerably larger than for the low-\sumetjet samples with \Zll or \Wln final states. 
As can be seen in   \subfigRef{fig:perf:process_tails}{fig:perf:tails:reweight}, the tails are much more consistent between \Zmm and \ttbar samples when the distribution for the \Zmm sample is reweighted such that it follows the same \sumetjet distribution as the \ttbar sample.
The enhanced tails are thus likely introduced by the jet response and multiplicity, which has a residual sensitivity to \pu.

\section{Systematic uncertainties}
\label{sec:uncertainties}
The systematic uncertainties associated with the measurement of \met are provided for the response (\met scale) as well as for the resolution. 
They depend on the composition of the hard terms and on the magnitude of the corresponding soft term.
As the hard-term composition is generally defined by optimisations implemented in the context of a given analysis, the contributions of the \met terms need to be extracted from the scale and resolution uncertainties for the individual contributing objects comprising electrons, photons, muons, \tauleps, and jets. 
In the corresponding propagations, correlations between systematic uncertainties for the same type of object are typically taken into account.
However, it is assumed that systematic uncertainties of the different object types entering \met reconstruction are uncorrelated.
The determination of the \met scale and resolution uncertainties arising from the soft term \metsft is described in this section.

\subsection{Methodology}\label{subsec:systematics:method} 

The extraction of the systematic uncertainties for the reconstructed \met 
is based on \datatomc comparisons of spectra of observables measuring the contribution of \metsft to the overall \met. 

\subsubsection{Observables}\label{subsec:systematics:method:obs}

The vector sum of the transverse momentum vectors of all particles and jets emerging from a hard-scatter interaction (\pTintVec) is given by
\begin{align*}
	\pTintVec = \underbracket{\sum \pTvecj{e} + \sum \pTvecj{\photon} + \sum \pTvecj{\tau} + \sum \pTvecj{\muon} + \sum \pTvecj{\jet}}_{\pTobsVec \text{\ (observable)}} \;\;+\;\; 
	\underbracket{\sum \pTvecj{\nu}}_{\mathclap{\pTinvVec \text{\ (not observable)}}}\,.
\end{align*}
Here \pTvecj{\nu} generally represents the transverse momenta of non-observable particles, which are summed up to form \pTinvVec. 
All other transverse momenta are carried by particles that are observable in principle, and sum up to \pTobsVec.
Momentum conservation dictates $\pTint = |\pTintVec| = 0$.

Due to detector acceptance limitations and inefficiencies in hard-object reconstruction and calibration, and all other effects 
discussed in \secRef{sec:etmiss-reco}, only a proxy (\pThardVec) for the observable-particle contribution \pTobsVec can be measured. 
The reconstructed hard final-state objects entering \met as described in \secRef{subsec:etmiss-definition} are used to measure \pThardVec as
\begin{align*}
	\pThardVec = \sum_{\substack{ \text{contributing} \\ \text{electrons}  }} \pTvecj{\electron}
		          + \sum_{\substack{ \text{contributing} \\ \text{photons}    }} \pTvecj{\photon}
                          + \sum_{\substack{ \text{contributing} \\ \tauleps             }} \pTvecj{\tauhad}
                          + \sum_{\substack{ \text{contributing} \\ \text{muons}      }} \pTvecj{\muon}
                          + \sum_{\substack{ \text{contributing} \\ \text{jets}          }} \pTvecj{\jet}\,.
\end{align*}
The expectation is that $\pThard = |\pThardVec| > 0$ and $\pThardVec \neq \pTobsVec$.  
Adding $\pTsoftVec = -\metvsft$, with \metvsft defined in \eqRef{eq:etmiss_terms}, to \pThardVec yields an improved estimate of the net transverse momentum carried by undetectable particles, as some of the experimental inefficiencies are mitigated.\footnote{As discussed in \secRef{subsec:etmiss-softterm}, the soft term represents only charged particles with $\pT > \unit{400}{\MeV}$ not associated with fully identified and reconstructed particles or jets. Therefore, including \pTsoftVec can only recover a part of the actual soft \pT-flow of the interaction.}

In the \Zmm final state without genuine missing transverse momentum the expectation is that $\metvec = -(\pThardVec + \pTsoftVec) = \mathbf{0}$.
While this expectation does not hold due to the experimental inefficiencies, it nevertheless raises the expectation that for events without jets \pTsoftVec points into the direction of the hadronic recoil, i.e.~opposite to \pThardVec in the transverse-momentum plane. 
The deviation from this expectation is measured in terms of the parallel (\projparhard) and perpendicular (\projperhard) projections of \pTsoftVec onto \pThardVec.
\FigRef{fig:syst:zmm:pthardproj} schematically shows these projections for $\Zboson + 0$-jet and $\Zboson + 1$-jet topologies.   

The average \AVE{\projparhard} in a given bin $k$ of phase space defined by \pThard measures the \metsft response, with $\AVE{\projparhard} = \AVE{\pThard}_{k}$ indicating a perfect response in this bin.
The \met resolution contribution from \metsft reconstruction is measured by two components, the fluctuations in response (\projparvar) and the fluctuations of the (transverse) angular deflection around the \pThardVec axis, measured by \projpervar. 
These fluctuations are expressed in terms of variances, with 
\begin{align*}
	\projparvar = \left\langle(\projparhard)^{2}\right\rangle - \left\langle\projparhard\right\rangle^{2} \qquad\text{and}\qquad
	\projpervar = \left\langle(\projperhard)^{2}\right\rangle\,.
\end{align*} 

\newlength{\figspecwidthfull} \setlength{\figspecwidthfull}{\figfullwidth}
\newlength{\figspecwidthnarrow} \setlength{\figspecwidthnarrow}{0.441\figspecwidthfull}
\newlength{\figspecwidthwide}\setlength{\figspecwidthwide}{\figspecwidthfull}
\begin{figure}[t!]\centering
	\subfigure[$\Zboson + \unit{0}{\jet}$ topology]{\includegraphics[width=\figspecwidthnarrow]{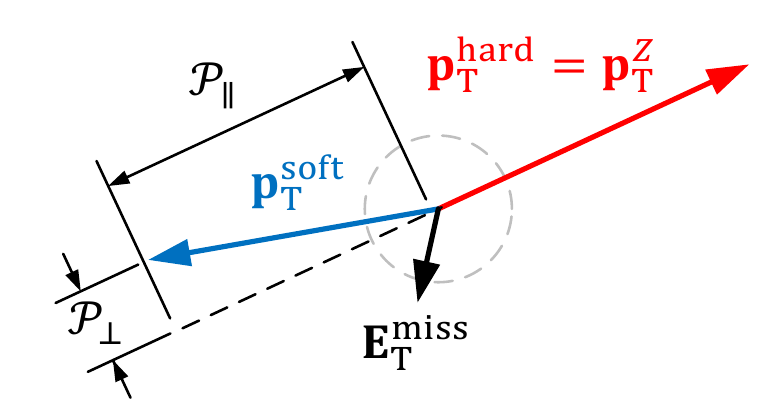}\label{fig:syst:zmm:pthardproj:0jet}}\qquad
	\subfigure[$\Zboson + \unit{1}{\jet}$ topology]{\includegraphics[width=\figspecwidthwide]{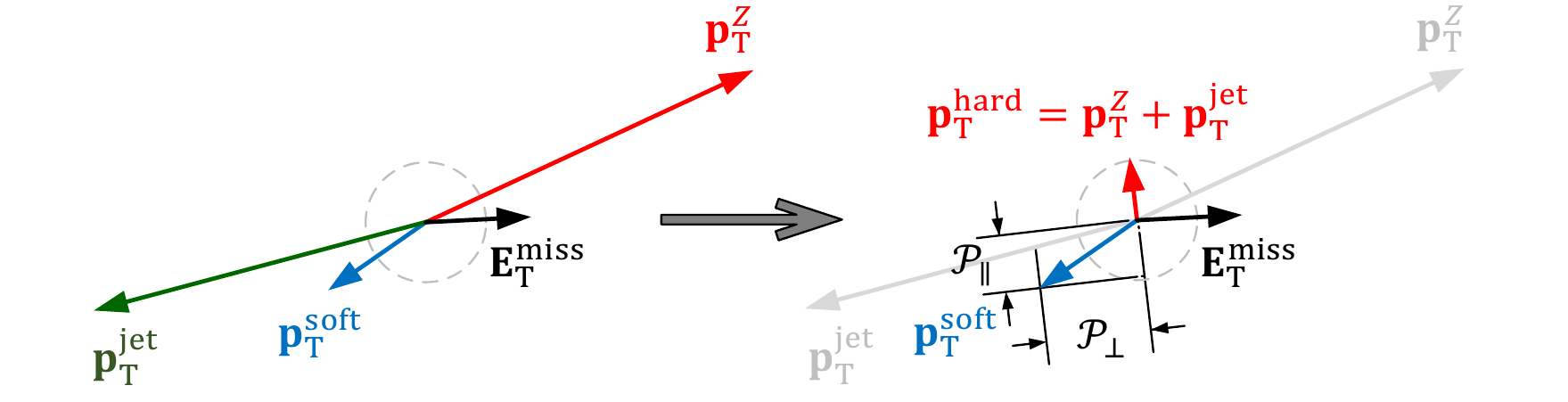}\label{fig:syst:zmm:pthardproj:1jet}}\qquad
\caption{Schematic view of the parallel (\projparhard) and perpendicular (\projperhard) projections of \pTsoftVec on \pThardVec for \Zmm events without genuine \met, for  \subref{fig:syst:zmm:pthardproj:0jet} a final state without  any jets and \subref{fig:syst:zmm:pthardproj:1jet} a final state with one jet. The expectation values for a perfect \met reconstruction are $\EXV{\projparhard} = \ptZ$ for $\Njet = 0$ and $\EXV{\projparhard} = \pThard$ for $\Njet \geq 1$, with $\EXV{\projperhard}= 0$ in all cases.}\label{fig:syst:zmm:pthardproj}
\end{figure}

\subsubsection{Procedures}\label{subsec:systematics:proc}

The extraction of the systematic uncertainties introduced into the \met measurement by the \metsft term is based on \datatomc-simulations comparisons of $\AVE{\projparhard}(\pThard)$ for the response, and of $\projparvar(\pThard)$ and $\projpervar(\pThard)$ for the resolution. 
Alternative \MC samples are considered, with variations of either the event generator or the detector simulation (description and shower models).
For the highest impact of \metsft on \met, the exclusive \Zmm selection with \nojets is the basis for the determination of the systematic uncertainty components for both data and all \MC simulations. 
In this case, the only hard contribution is from the reconstructed \Zboson boson, i.e.~$\pThardVec = \ptZvec$ as shown in \subfigRef{fig:syst:zmm:pthardproj}{fig:syst:zmm:pthardproj:0jet}.

\begin{figure}[t!]\centering
	\subfigure[]{\includegraphics[width=\fighalfwidth]{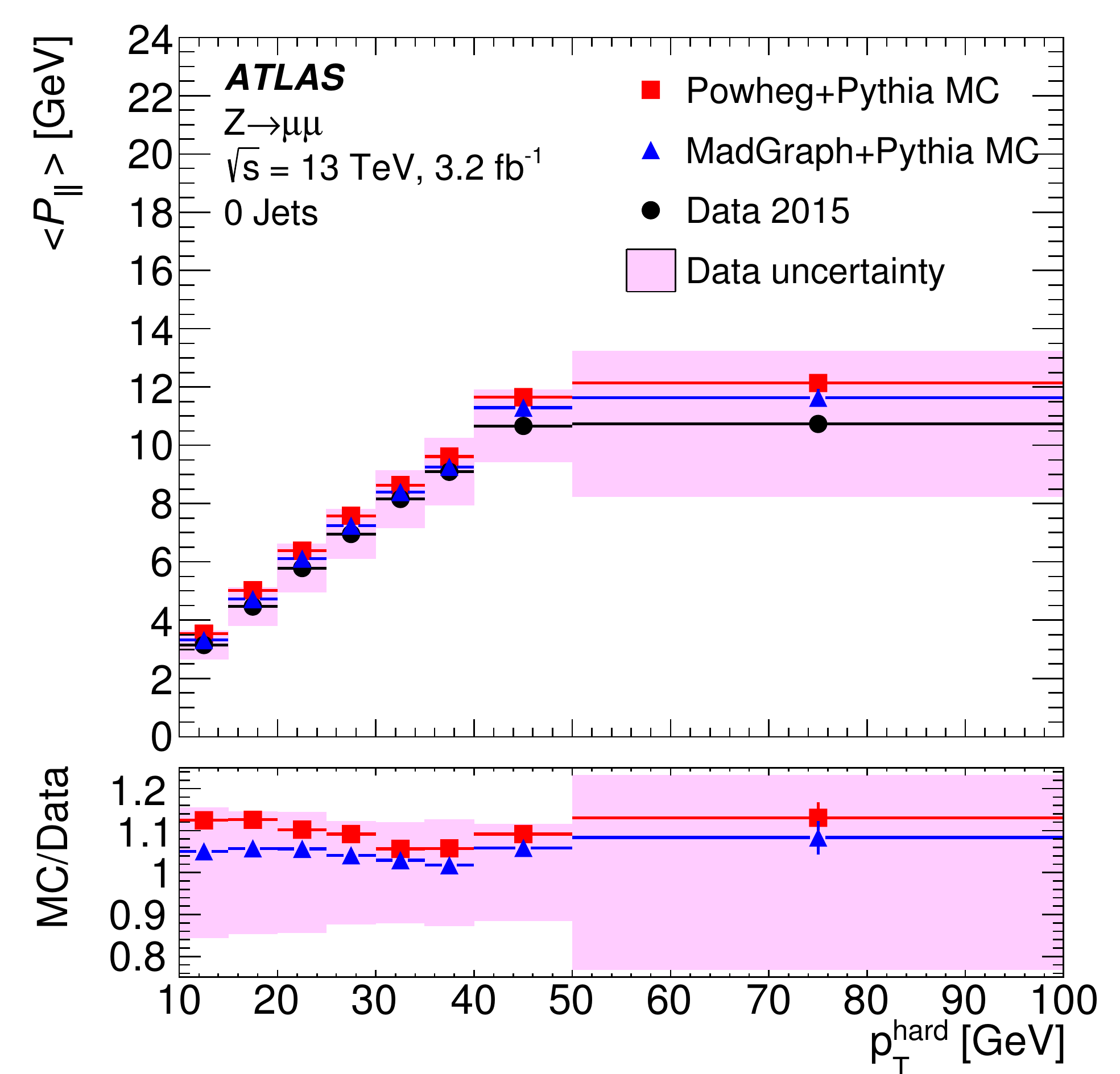}\label{fig:syst:zmm:projpar:mean:comp}}
	\subfigure[]{\includegraphics[width=\fighalfwidth]{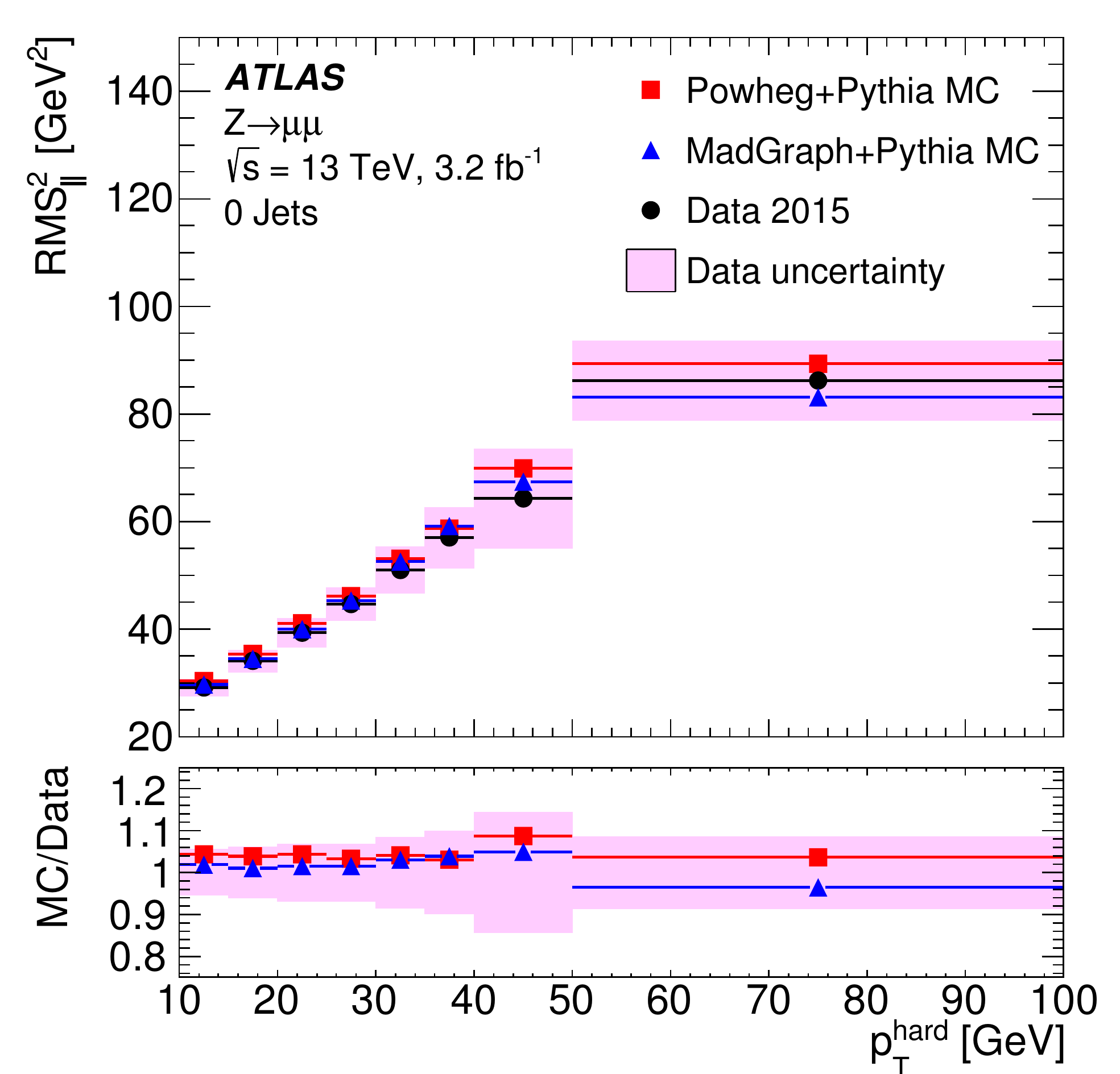}\label{fig:syst:zmm:projpar:rms:comp}}\\
	\subfigure[]{\includegraphics[width=\fighalfwidth]{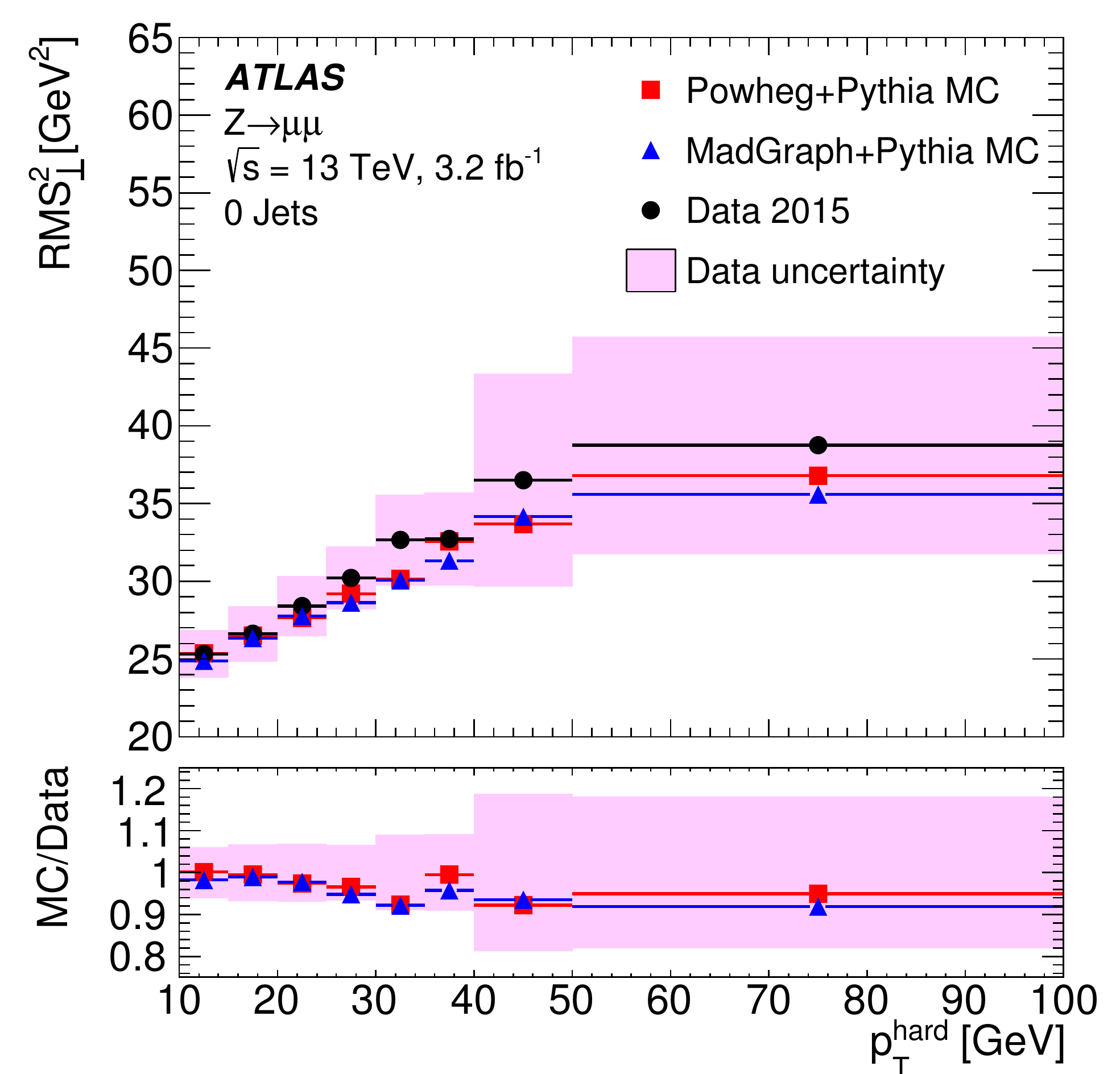}\label{fig:syst:zmm:projper:rms:comp}}
	\caption{The \subref{fig:syst:zmm:projpar:mean:comp} average value of the longitudinal projection \AVE{\projparhard} and the \subref{fig:syst:zmm:projpar:rms:comp} variance \projparvar of the longitudinal projection \projparhard of \pTsoftVec onto \pThardVec for \Zmm event with \nojets, for data and two different \MC simulations, shown as a function of \pThard. The variance \projpervar of the perpendicular projection \projperhard is shown in \subref{fig:syst:zmm:projper:rms:comp} for the same event samples.
	The shaded band indicates the systematic uncertainties derived as described in the text.}
	\label{fig:syst:zmm:projall}
\end{figure}

The uncertainties are determined by comparing \projparhard and \projperhard spectra from data and \MC simulations, in bins of \pThard. 
For \projparhard, 
the smearing of the response and the width both yield
scale and (longitudinal) resolution offsets.
In the case of \projperhard, only smearing of the width is applied to provide transverse resolution offsets.
These fitted offsets, determined for the various \MC configurations, provide the systematic uncertainties with respect to a specific \MC modelling configuration.
In practice, to account for the resolution offsets, Gaussian smearing is applied in simulation to the longitudinal and transverse components of \metvsft 
relative to the direction of 
\pThardVec.
To account for differences in response between data and simulation, the longitudinal component of \metvsft is scaled up and down to give an uncertainty band.

In order to generate the required number of simulated events,
some analyses in \ATLAS may have to use the fast detector simulation \ATLFAST \cite{Aad:2010ah,ATL-PHYS-PUB-2010-013} for the calorimeter response.
It employs parameterisations for electromagnetic and hadronic showers, instead of the explicit simulation of the particle tracking through matter and the energy-loss mechanisms in a detailed detector geometry. 
An additional uncertainty is assigned to effects introduced by \ATLFAST.
This uncertainty contribution only needs to be considered in analyses using this fast simulation, and does not apply for the results presented in this paper. 
In analyses where it is applicable, it is added in quadrature to the standard uncertainties. 

\subsection{Systematic uncertainties in \met response and resolution}\label{subsec:systematics:results}

The result for the systematic uncertainty of the \met scale, determined as discussed in the previous section, is summarised in \figRef{fig:syst:zmm:projall}. 
The average longitudinal projection of \pTsoftVec onto \pThardVec, \AVE{\projparhard}, as a function of \pThard  is shown in \subfigRef{fig:syst:zmm:projall}{fig:syst:zmm:projpar:mean:comp} 
which compares data 
to both the standard \PPEight-based simulations and the alternative \MC simulation employing \MADGRAPH, as described in \secRef{sec:mc-samples}. 
All \MC simulation results are expected to have $\AVE{\projparhard}_{\text{\MC}}$ within the uncertainties of the data. 
The lower panel of \subfigRef{fig:syst:zmm:projall}{fig:syst:zmm:projpar:mean:comp} confirms that the ratio $\AVE{\projparhard}_{\text{\MC}}/\AVE{\projparhard}_{\text{data}}$ 
lies within the systematic uncertainty band over the full \pThard range. 

The systematic uncertainty for the \met resolution is extracted from the variances of the parallel (\projparvar) and perpendicular (\projpervar) projections of \metvec onto \pThardVec defined in \secRef{subsec:systematics:proc}.
\SubfigRef{fig:syst:zmm:projall}{fig:syst:zmm:projpar:rms:comp} shows the \pThard dependence of \projparvar measured for the exclusive \Zmm sample (\nojets) in data and two \MC simulations.
The variances $(\projparvar)_{\text{\MC}}$ calculated for both sets of simulations agree within the systematic uncertainties of $(\projparvar)_{\text{data}}$ with the data, 
as illustrated in the lower panel of the figure, where the ratio $(\projparvar)_{\text{\MC}}/(\projparvar)_{\text{data}}$ is shown as a function of \pThard. 
The results of the evaluation of the variances \projpervar of the perpendicular projections as a function of \pThard are shown in \subfigRef{fig:syst:zmm:projall}{fig:syst:zmm:projper:rms:comp}, 
together with the resulting \pThard dependence of the ratio $(\projpervar)_{\text{\MC}}/(\projpervar)_{\text{data}}$. 
The systematic uncertainties of the data cover all differences to \MC simulations.

\section{Missing transverse momentum reconstruction variants}
\label{sec:etmiss-variants}
\subsection{Calorimeter-based \texorpdfstring{\met}{Etmiss}}
\label{subsec:etmiss-calo}
The \met soft term from the calorimeter \mettcsft is reconstructed from \topos.
As discussed in \citRef{Aad:2016upy}, each \topo provides a basic \EM scale signal as well as a calibrated signal reconstructed using local cell weighting (\LCW), 
and \mettcsft is calculated from \topos calibrated at the LCW scale.
Only \topos with a calibrated energy $\ecluslcw > 0$, not contributing to the reconstruction of the hard objects used to calculate the hard term given in \eqRef{eq:etmiss_terms}, are considered for \mettcsft. 
In addition, \topos that are formed at the same location as the hard object signals are not considered for \mettcsft even if their signals are not directly contributing to the reconstruction of the hard objects.   
The fully reconstructed \met using \mettcsft is \metcalo.  

Compared to the reference \met and \sumet, \metcalo and \sumetcalo have an enhanced dependence on \pu, mostly introduced by the soft term.
To partly compensate for the irreducible contribution of \pT-flow reconstructed from \topos generated by \pu to \metcalo, a modified jet selection and ambiguity resolution is applied in their reconstruction.
The considered jets are reconstructed following the prescription in \secRef{subsec:jet-sel}, and required to have a fully calibrated $\pT > \unit{20}{\GeV}$. 
The contribution of these jets to \metcalo and \sumetcalo, defined in terms of momentum components $(p_{x},p_{y})$, depends on the overlap with already accepted reconstructed particles,
\begin{align}
	(p_{x},p_{y}) = 
		\left\{ 
			\begin{array}{lll}
				(0,0)               & \of \geq 50\%                      & \text{(large overlap)}    \\
				(1-\of)\times(p_{x}^{\jet},p_{y}^{\jet}) & \of < 50\% & \text{(small or no overlap)} \\
			\end{array}
		\right.\,.
	\label{eq:cstof}
\end{align}
The overlap fraction \of is given in \eqRef{eq:of}. 
Jets with $\of \geq 50\%$ are not used at all. 
The \JVT-based tagging of non-\pu jets is omitted.
It is found that this strategy reduces the fluctuations in the \metcalo reconstruction.
The transverse momentum contribution of groups of clusters representing a jet-like \pT-flow e.g.~from \pu in a given direction that are not reconstructed and calibrated as a jet, or do not pass the jet-\pT threshold applied in \met reconstruction, is reduced   
if all jets and jet fragments, including those from \pu, are included.

\subsection{\texorpdfstring{\met}{Etmiss} from tracks}
\label{subsec:etmiss-track}
The reference track-based soft term \metsft is largely insensitive to \pu, as indicated by the dependence of the \met resolution \metresxy on \NPV in the exclusive \Zmm sample  (\nojets) shown in \subfigRef{fig:perf:zmm:reso}{fig:perf:zmm:nojets:reso:npv}.
As discussed in \secRef{subsec:etmiss-resolution:results} and from the comparison of \subfigMultiRef{fig:perf:zmm:reso}{fig:perf:zmm:nojets:reso:npv}{and}{fig:perf:zmm:reso}{fig:perf:zmm:jets:reso:npv}, the \pu dependence of \metresxy  in the inclusive \Zmm sample is largely introduced by the jet contribution.
This contribution suffers from (1) the lack of \pu suppression for forward jets with $\abseta > 2.4$, (2) any inefficiency connected with the \JVT-based tagging, and (3) irreducible \pu-induced fluctuations in the calorimeter jet signals.     
Using a representation of \met employing only reconstructed \ID tracks from the primary vertex increases stability against \pu as long as the tracking and vertex resolution is not affected by it. 
In this representation (\mettrk) all jets and reconstructed particles are ignored, i.e. the \mettrk reconstruction does not include any calorimeter or \MS signals.
The \mettrk resolution is then inherently immune to \pu, while the \mettrk response is low as all neutral \pT-flow in $\abseta < 2.5$ as well as all \pT-flow outside of this region is excluded.

\subsection{Performance evaluations for \texorpdfstring{\met}{Etmiss} variants}
\label{subsec:etmiss-variants-perf}
The main motivation to study \met-reconstruction variants is to improve some combination of the \met resolution, scale, and stability against \pu. 
As with the composition of objects entering \met reconstruction in general, the particular choice of variant used for a given analysis strongly depends on the performance requirements for this analysis. 
The comparison of both the resolution and response of \metcst and \mettrk to the corresponding measurements using the reference \met illustrates their principal features for the \Zmm and \ttbar production final state. 

\begin{figure}[t!]\centering
\includegraphics[width=\fighalfwidth]{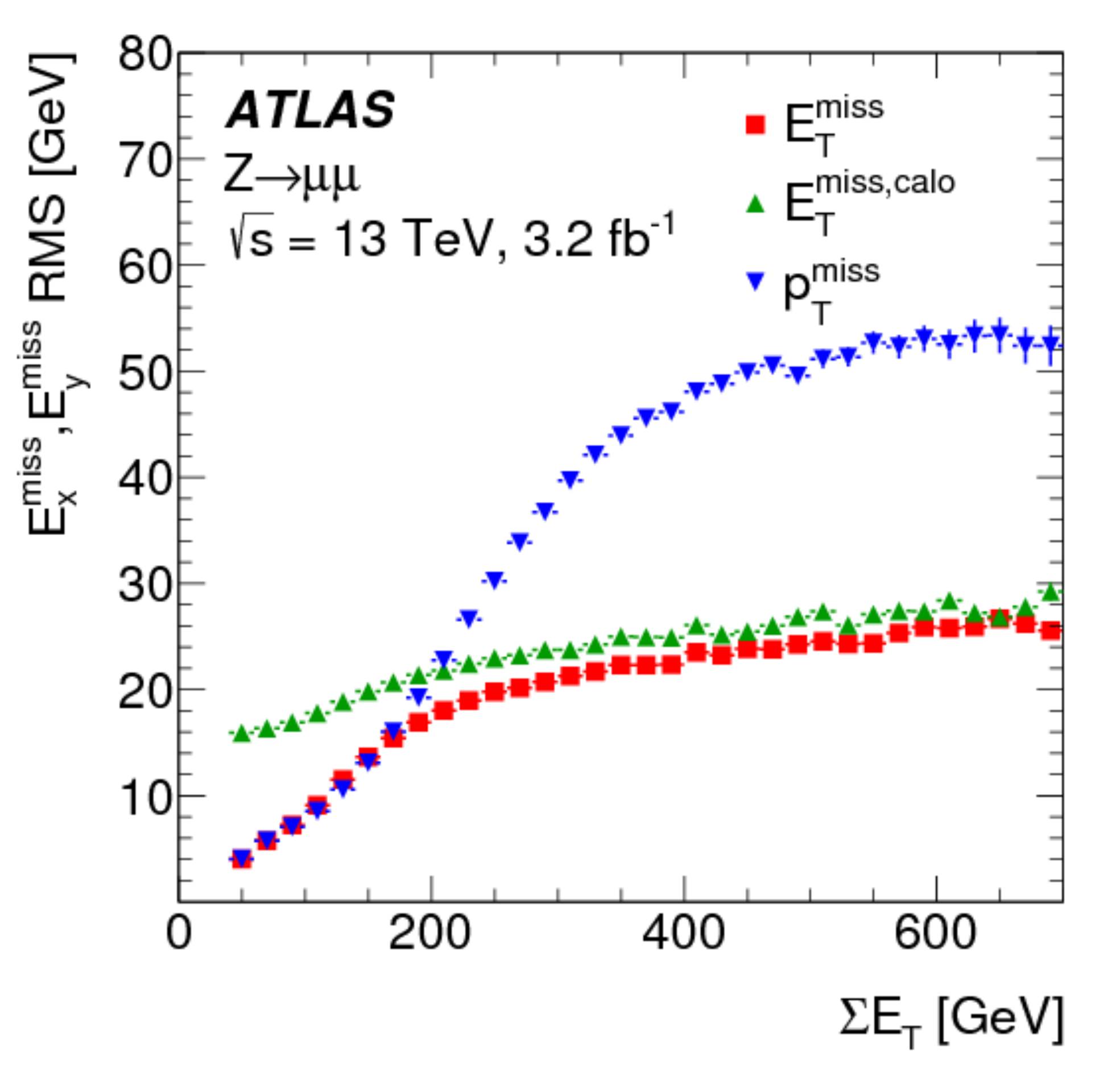}
\caption{Comparison of the reference \met resolution with the resolutions of the track-only-based variant \mettrk described in \secRef{subsec:etmiss-track}, and the reconstruction variant \metcst employing a calorimeter-based soft term, as discussed in \secRef{subsec:etmiss-calo}. The resolutions are determined as described in \secRef{subsec:etmiss-resolution:method} and shown as a function of the \sumet. For consistency, for all three variants, the \sumet value is taken from \met.}
\label{fig:performance:track_resos}
\end{figure}

\begin{figure}[t!]\centering
\subfigure[]{\includegraphics[width=\fighalfwidth]{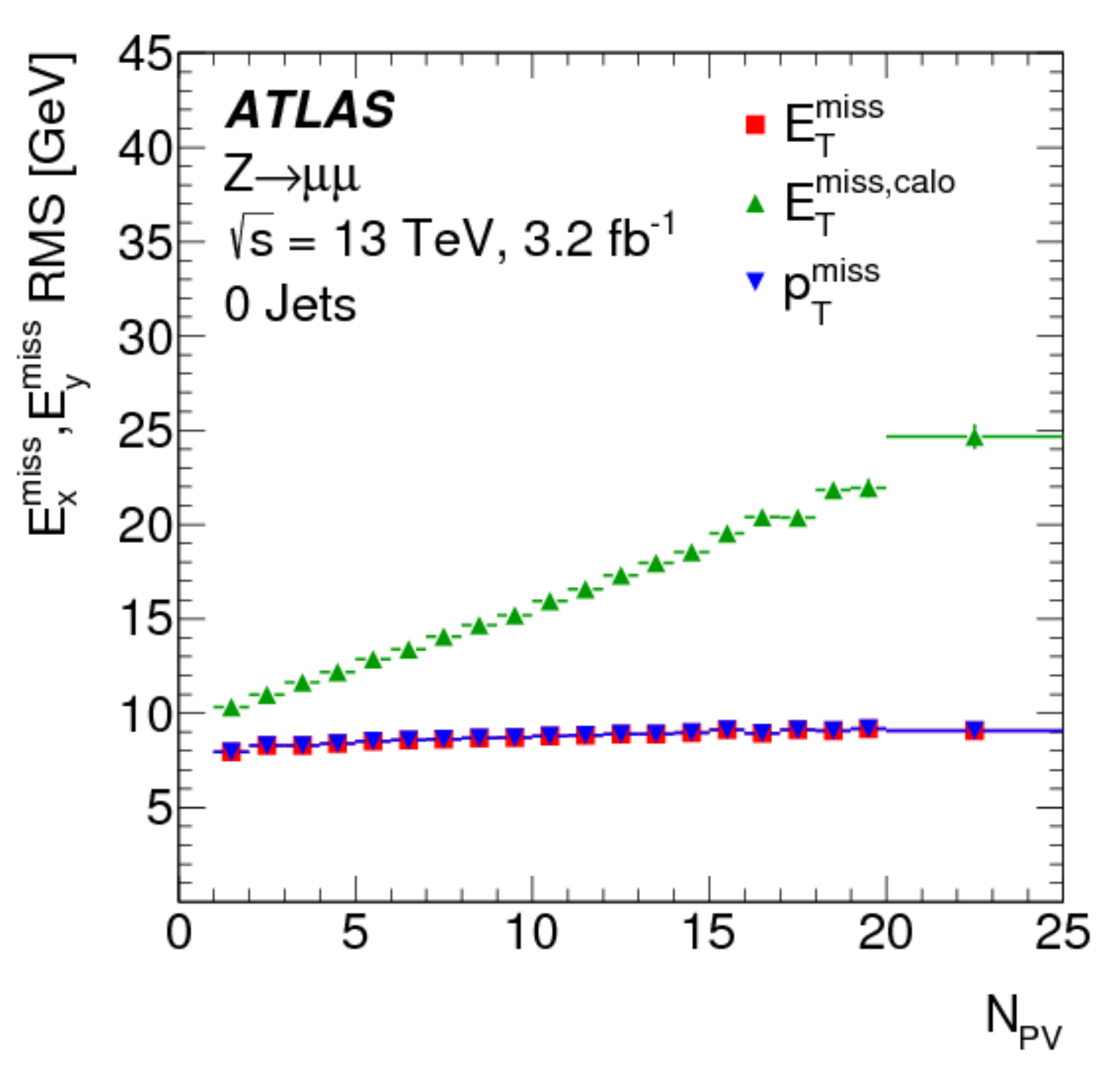}\label{fig:perf:zmm:reso_excl}}\quad
\subfigure[]{\includegraphics[width=\fighalfwidth]{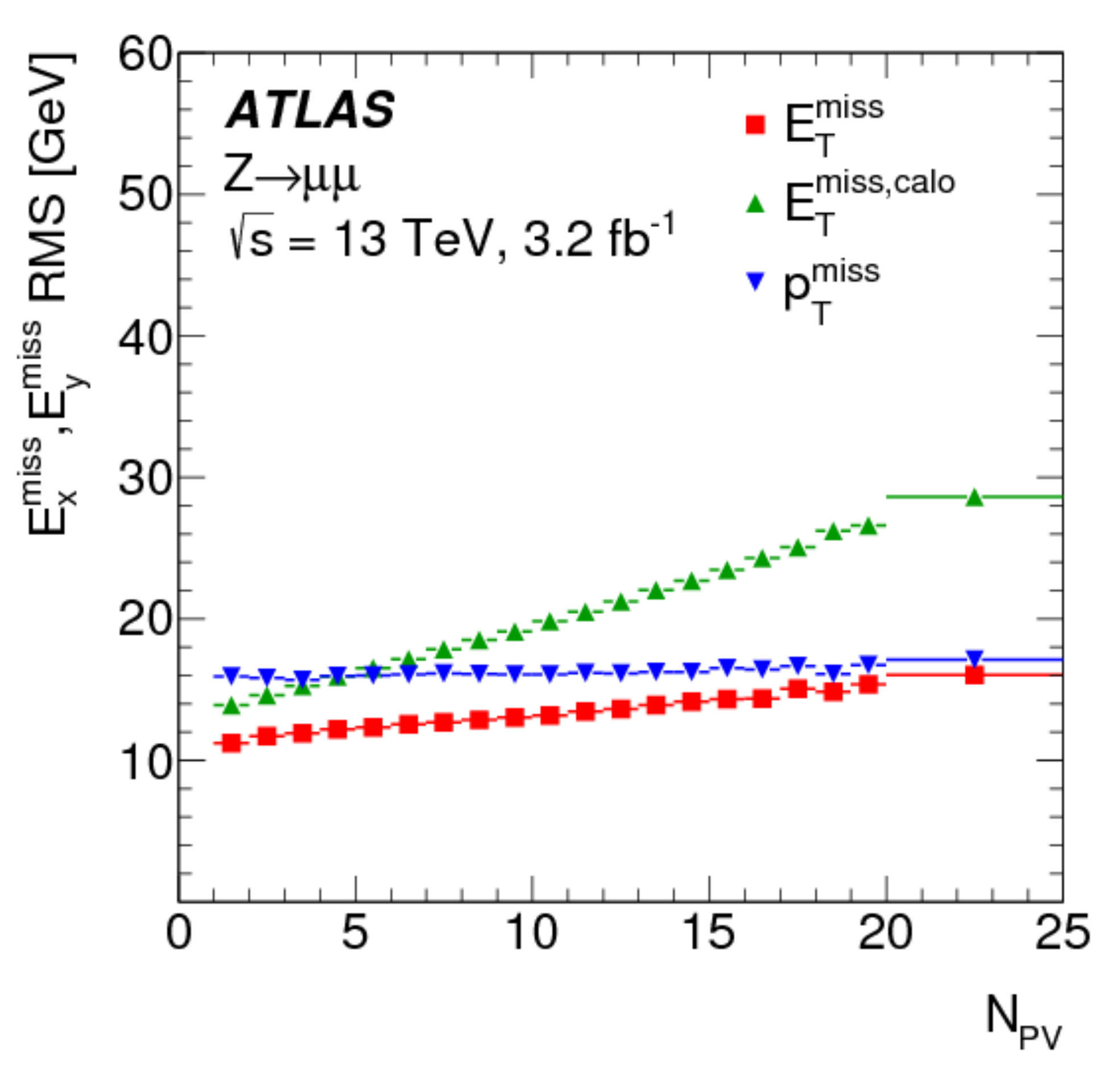}\label{fig:perf:zmm:reso_incl}}\qquad
\subfigure[]{\includegraphics[width=\fighalfwidth]{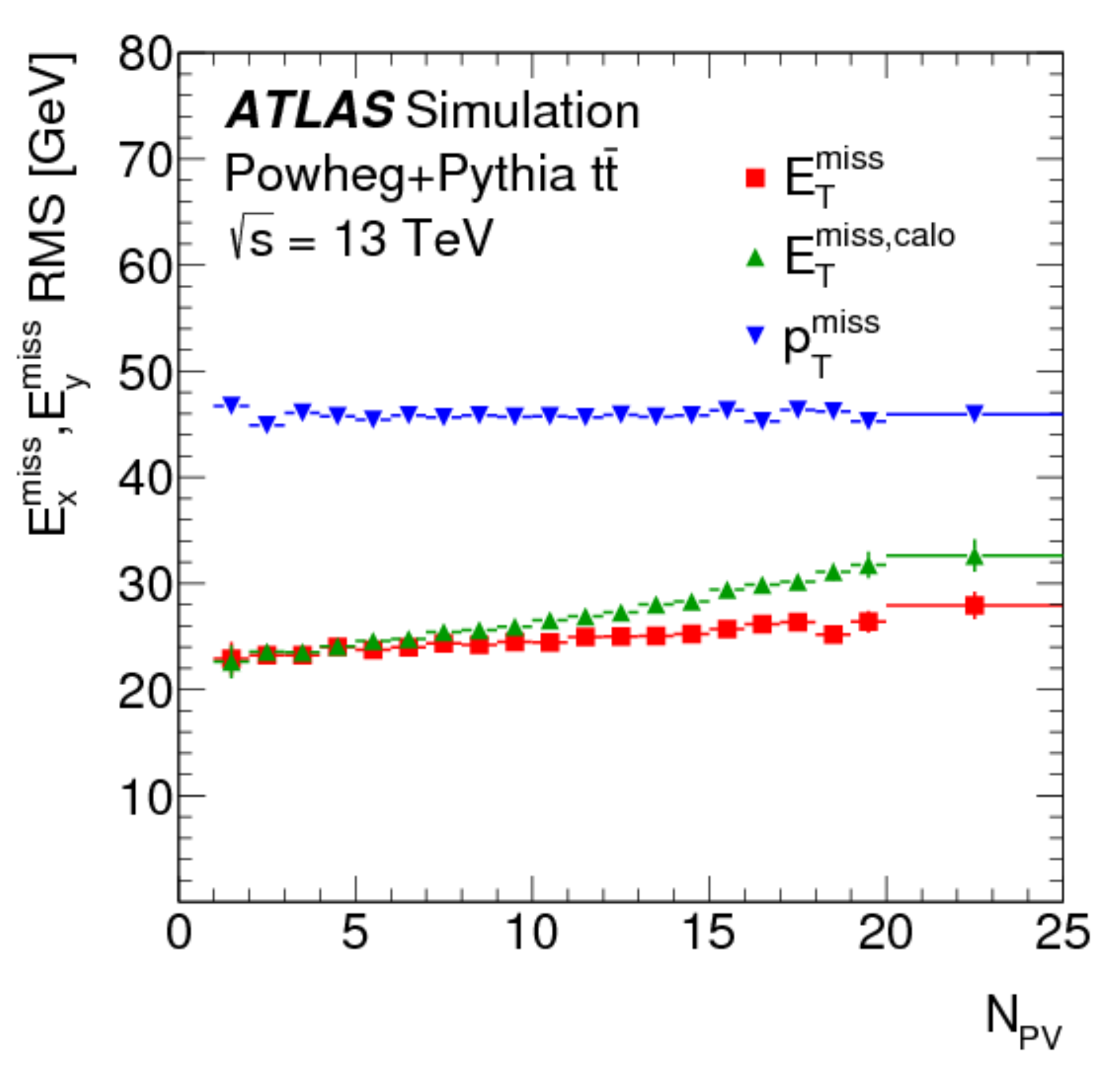}\label{fig:perf:tt:reco_incl}}\qquad
\caption{Comparison of the reference \met resolution with the resolutions of the track-only-based variant \mettrk described in \secRef{subsec:etmiss-track}, and the reconstruction variant \metcst employing a calorimeter-based soft term, as discussed in \secRef{subsec:etmiss-calo}. The resolutions are determined as described in \secRef{subsec:etmiss-resolution:method} and shown as a function of the \pu activity measured in terms of the number of reconstructed vertices \NPV for \subref{fig:perf:zmm:reso_excl} an exclusive \Zmm sample without jets with $\pT > \unit{20}{\GeV}$ and \subref{fig:perf:zmm:reso_incl} an inclusive \Zmm sample, both selected from data. In \subref{fig:perf:tt:reco_incl}, the resolution of the \met reconstruction-variants in a final state with significant jet activity and $\pTnu > 0$ is compared using \MC simulations of \ttbar production.}
\label{fig:performance:track_reso}
\end{figure}

\subsubsection{Comparisons of \texorpdfstring{\met}{Etmiss} resolution}
\label{subsec:etmiss-variants-perf-reso}

\FigRef{fig:performance:track_resos} compares the \metcst and \mettrk resolutions with the one obtained from the reference \met, for the inclusive \Zmm sample in data. 
Each is shown as a function of \sumet corresponding to the reference \met, giving an estimate of the total \hs activity.
The low-\sumet region is dominated by events with \nojets, where the contribution of \mettcsft in \metcst yields a poorer resolution than for \met, 
and where \met and \mettrk have identical performance. 
The high-\sumet region is dominated by events with higher jet multiplicity, where \mettrk resolution is degraded relative to the reference \met by the incomplete measurement of jets. 

\SubfigRef{fig:performance:track_reso}{fig:perf:zmm:reso_excl} compares the \metcst and \mettrk resolution as functions of the \pu activity measured by \NPV, with the one obtained from the reference \met for the exclusive \Zmm samples with \nojets in data.   
The \metcst resolution is dominated by \pu and shows significantly degraded performance relative to \mettrk and the reference \met.
The exclusive use of only tracks from the \hs vertex for both \mettrk and \met yields the same stability against \pu. 

In events with jet activity, the degraded \mettrk resolution is observable, especially outside the region of highest \pu activity, as seen in \subfigRef{fig:performance:track_reso}{fig:perf:zmm:reso_incl} for the \met resolution obtained with the inclusive \Zmm sample in data for $\NPV \lesssim 15$.  
This is even more obvious in final states with relatively high jet multiplicity and genuine missing transverse momentum, like for the \ttbar-production sample from \MC simulations. 
As shown in \subfigRef{fig:performance:track_reso}{fig:perf:tt:reco_incl} for this final state, both the reference \met and the calorimeter-based \metcst have a significantly better resolution than \mettrk, 
at the price of some sensitivity to \pu, which is absent for \mettrk. 
The \NPV dependence of the resolution is enhanced in \metcst, due to the increased contribution from soft calorimeter signals without \pu suppression at higher \NPV. 

\subsubsection{Comparisons of \texorpdfstring{\met}{Etmiss} scale}
\label{subsec:etmiss-variants-perf-scale}

\begin{figure}[t!]\centering
\subfigure[]{\includegraphics[width=\fighalfwidth]{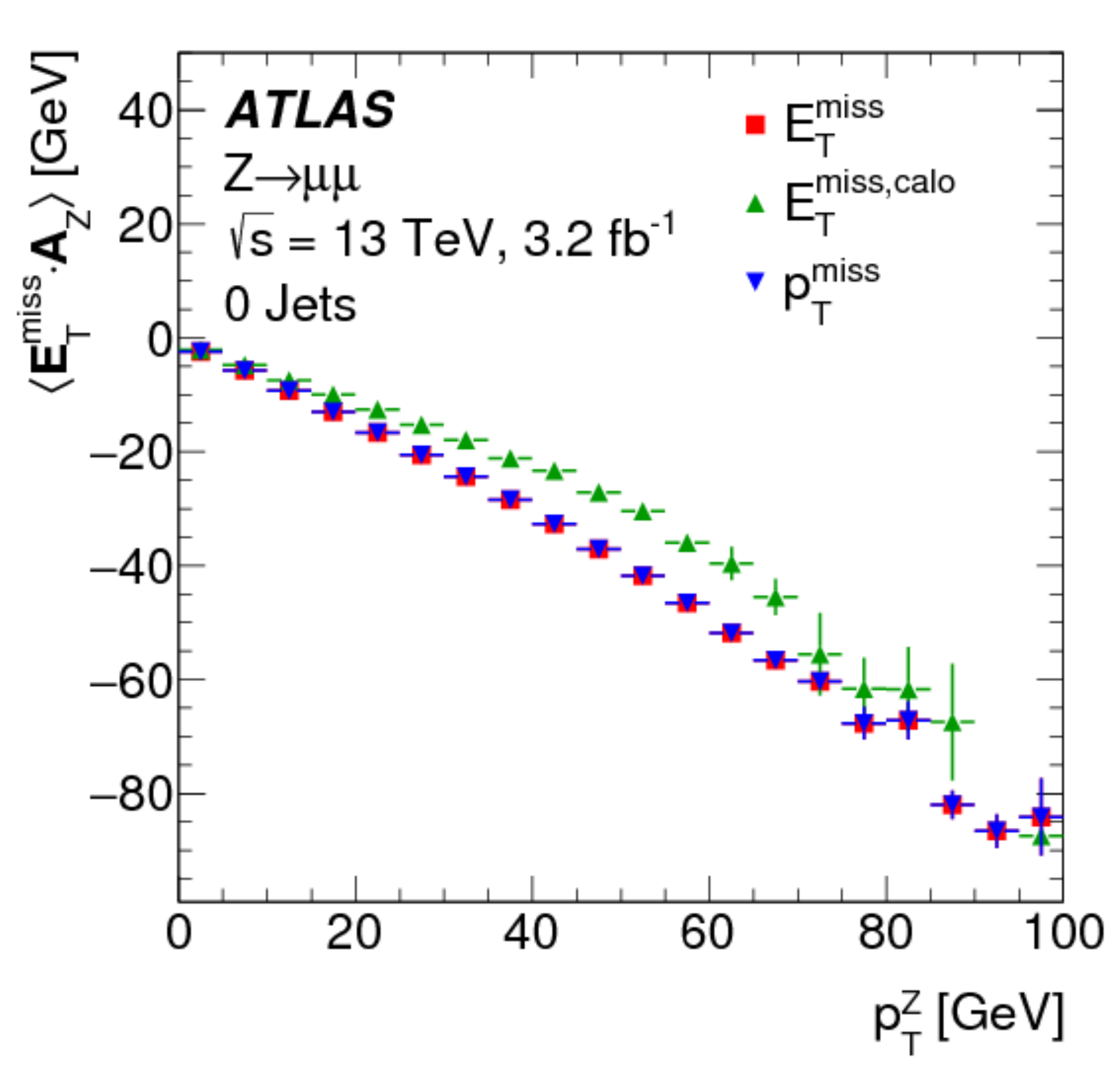}\label{fig:perf:zmm:scale_excl}}\quad
\subfigure[]{\includegraphics[width=\fighalfwidth]{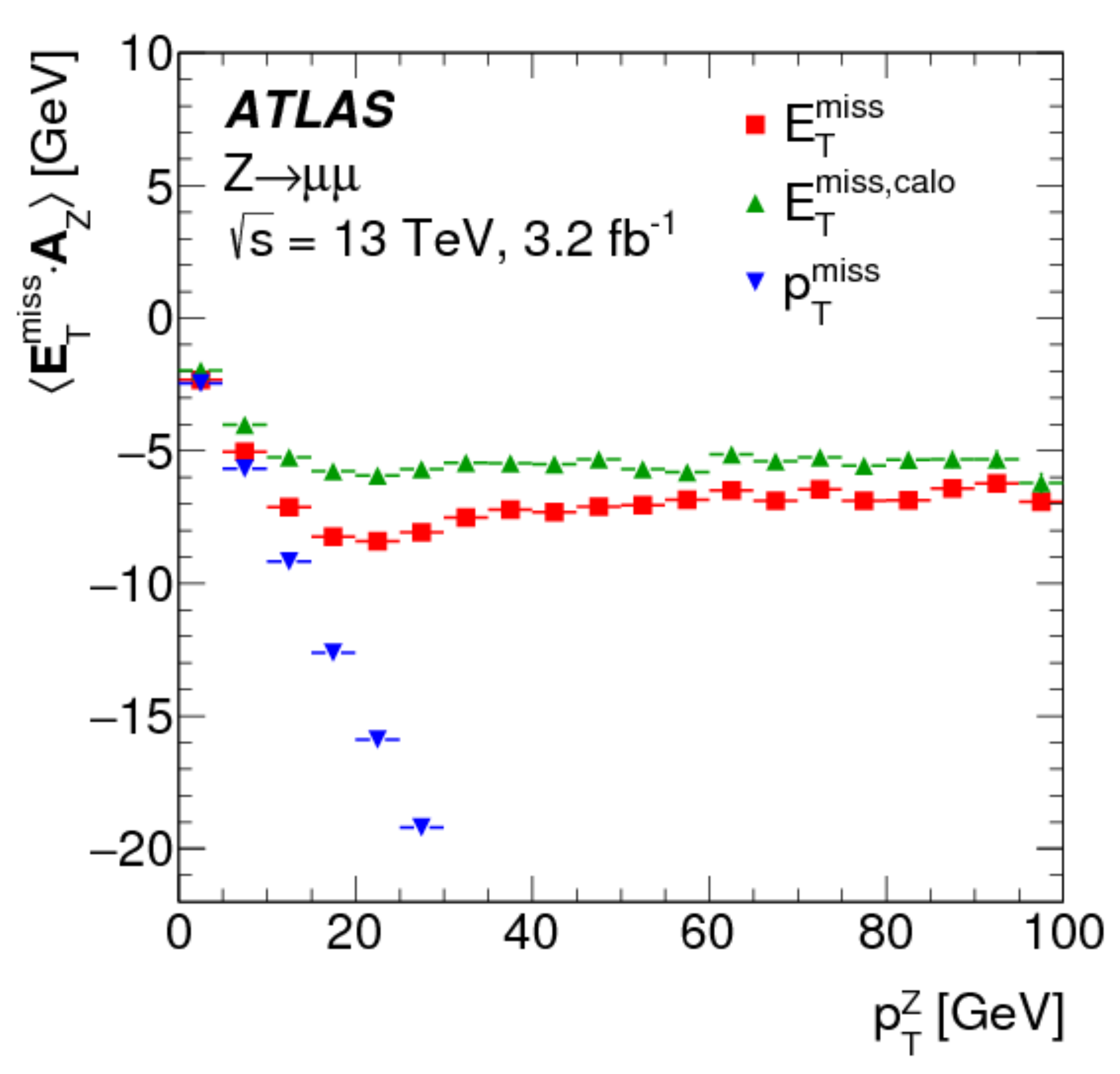}\label{fig:perf:zmm:scale_incl}}\qquad
\subfigure[]{\includegraphics[width=\fighalfwidth]{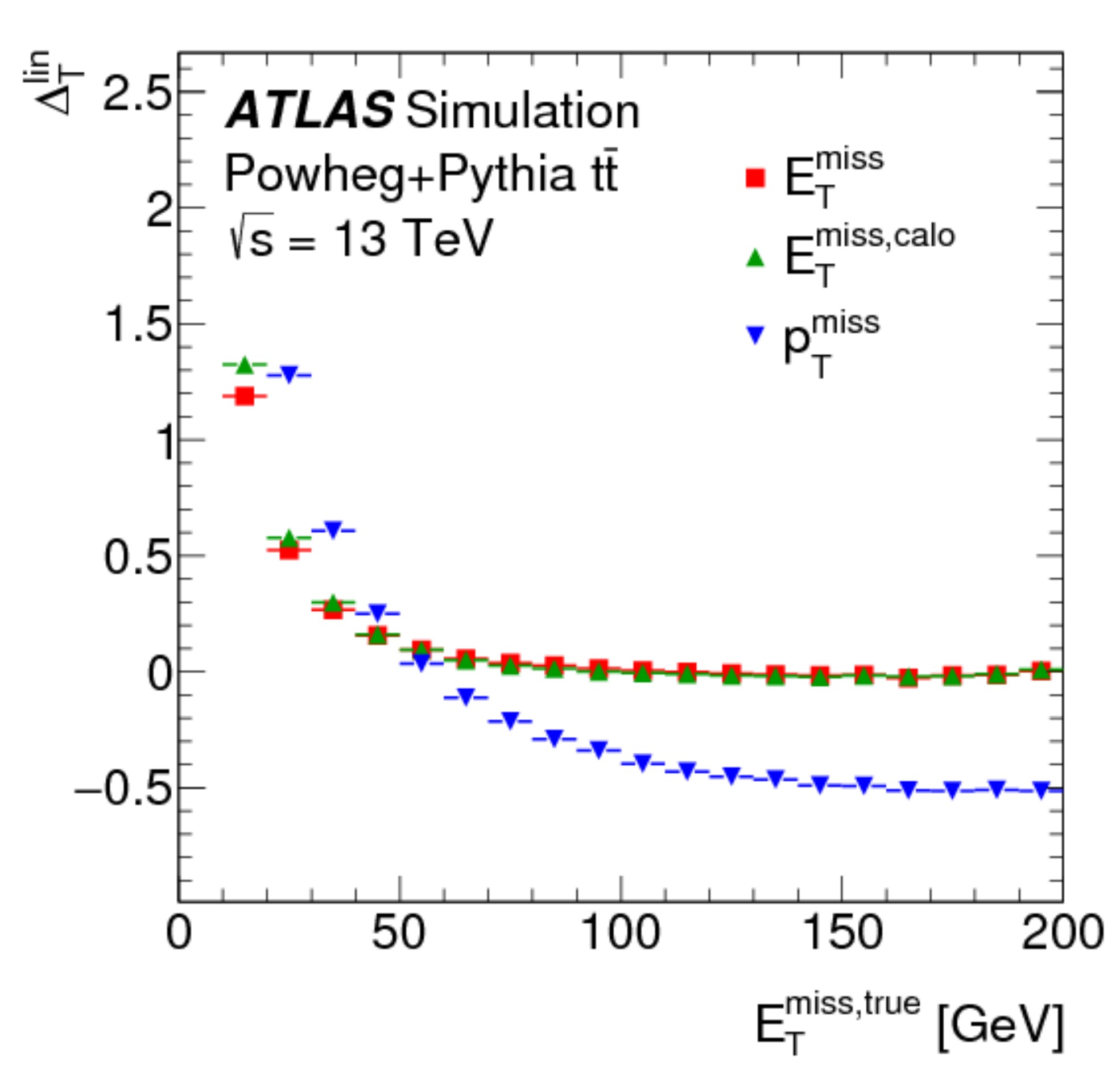}\label{fig:perf:tt:scale_incl}}\qquad
\caption{Comparison of the reference \met, the calorimeter-based \metcst and track-only-based \mettrk response in an \subref{fig:perf:zmm:scale_excl} exclusive and an \subref{fig:perf:zmm:scale_incl} inclusive \Zmm sample from data. The projections of the respective \metvec, \metvcst, and \metvtrk onto the direction of \ptZvec, calculated according to \eqMultiRef{eq:az}{and}{eq:projpar},  are shown as a function of \ptZ. In \subref{fig:perf:tt:scale_incl}, the linearity of the reference \met, \metcst, and \mettrk scales, calculated according to \eqRef{eq:devlin}, is shown as a function of the true \mettrue for  the \ttbar-production \MC simulation sample.}
\label{fig:performance:diagnostic-linearity}
\end{figure}

Following the description in \secRef{subsec:etmiss-scale:method}, the \met response is evaluated for the reference \met, \metcst, and \mettrk using the respective projections of \metvec, \metvcst, and \metvtrk onto the direction of \ptZvec, according to \eqMultiRef{eq:az}{and}{eq:projpar}. 
\SubfigRef{fig:performance:diagnostic-linearity}{fig:perf:zmm:scale_excl} shows the average projection as a function of \ptZ for the exclusive \Zmm sample with \nojets in data. 
Both \met and \mettrk show the same increasingly incomplete reconstruction of the hadronic recoil in this sample for rising \ptZ. 
This reconstruction is slightly improved for \metcst, but still insufficient at higher \ptZ. 

In the inclusive \Zmm sample, shown in \subfigRef{fig:performance:diagnostic-linearity}{fig:perf:zmm:scale_incl}, 
the indication at lower \ptZ is that \metcst has a higher response and thus a better representation of the hadronic recoil, 
due to the more complete \metsft reconstruction and the lack of a \JVT-tagging requirement.
This effect is partly due to the observation bias in the response introduced by the relatively poor \metcst resolution, as discussed in \secRef{subsec:etmiss-basics}.
Both \met and \metcst show comparable response for $\ptZ \gtrsim \unit{60}{\GeV}$, owing to the \JVT cut-off at $\unit{60}{\GeV}$.
The slightly larger \metcst response of about \unit{1}{\GeV} reflects the contribution from neutral signals to the soft term.
The degraded response associated with \mettrk related to the exclusion of hard objects is clearly visible in this figure. 

\SubfigRef{fig:performance:diagnostic-linearity}{fig:perf:tt:scale_incl} shows the linearity of the various \met reconstruction approaches as a function of \mettrue for the \ttbar-production sample from \MC simulations. 
Beyond $\mettrue \approx \unit{60}{\GeV}$ both the reference \met and \metcst show the same good linearity, while the lack of a jet contribution to \mettrk shows a loss of response up to about 50\% at higher \mettrue. 
The overestimation of \mettrue by all three reconstruction variants at lower \mettrue reflects the observation bias in the response introduced  by the resolution.
The poorer resolution associated with \mettrk observed in \subfigRef{fig:performance:track_reso}{fig:perf:tt:reco_incl} for this sample leads to a faster rise of the response with decreasing \mettrue than for the reference \met and \metcst, which show a very similar dependence on \mettrue. 

\subsubsection{Summary of performance}
\label{subsec:etmiss-variants-perf-sum}

Both \metcst and \mettrk offer alternative measures for \met. 
The calorimeter-based \metcst uses \topos calibrated at the \LCW scale for the soft term, which are neither part of the signal nor otherwise overlapping with the signals of other hard objects contributing to \met.
This introduces a \pu dependence into \metcst, due to the lack of \pu suppression of calorimeter signals outside of reconstructed hard objects.
It features a slightly modified jet contribution without the \JVT-based selection used in case of the reference \met reconstruction, to allow the cancellation of jet-like \pT-flow from \pu in \mettcsft by \pu jets in its hard term.
The \metcst response in the inclusive \Zmm sample is better than the reference \met response, in particular in the region of  small hadronic recoil ($\ptZ \lesssim \unit{20}{\GeV}$).
It is comparable to the reference in \ttbar final states.   
The observed \metresxy, in particular in \Zmm without jets, is significantly more affected by \pu than is the reference \met or the track-only-based \mettrk.
In final states with a considerable number of jets,  like \ttbar, \metcst performs nearly as well as the reference \met, with a slight degradation of the \met resolution at highest \pu activities.
This variant is useful for physics analyses least sensitive to the soft-term contribution to \met resolution but requiring a linear \met response.   

The track-only-based \mettrk displays a degraded response for the inclusive \Zmm sample, which is expected from the exclusive use of \hs-vertex tracks. 
As expected, resolution is not affected by \pu in the considered final states, but is poorer than, or at most comparable to, the reference \met algorithm.
Nevertheless, \mettrk provides a stable observable for event and phase-space selections in analyses sensitive to \met resolution. 

\FloatBarrier

\section{Conclusion}
\label{sec:conclusion}
The performance and features of the missing transverse momentum reconstruction in \pp collision data at the \LHC, acquired in 2015 with the \ATLAS detector at \sqrts{13} and corresponding to about \ilum, are evaluated for selected event samples with (\Wen, \Wmunu, \ttbar) and without (\Zmm) genuine \met.
The comparison of the data from the detector with the corresponding \MC simulations generally yields good agreement in the covered phase space.
The systematic uncertainty contribution from the soft event to the reconstructed \met is determined with \Zmm final states without jets.
It is calculated from the \datatomc-simulations comparison of the parallel and perpendicular projections of the missing transverse momentum vector \metvec onto the vector sum of the transverse momenta of the hard objects \pThardVec.
The parallel projections yield the uncertainty of the \met scale, evaluated as a function of the total transverse momentum of the hard objects (\pthard).
The widths of the distributions of the parallel and perpendicular projections yield the respective systematic uncertainties of the \met resolution.
Simulation tends to underestimate the perpendicular resolution and overestimate the scale and parallel resolution, in each case differing from data by at most 10\%.

The performance evaluation of \met response and resolution for the inclusive \Zmm sample shows that data and \MC simulations agree within the systematic uncertainties.
The \met response shows an underestimation of the soft contributions to \met. 
A degradation of the \met resolution is observed for increasing \sumet and \NPV, due to \pu and detector resolution effects.
Additional performance measures considered in these studies include the estimate of tails in the \met distribution.
As expected from the universality of the hadronic recoil, the integral tail fraction of the \met distribution is identical for inclusive \Zboson and \Wboson boson production, independent of the leptonic decay mode.
The \ttbar final states feature a higher jet multiplicity and show larger tails reflecting a higher sensitivity to residual \pu surviving in the jet contribution to \met, 
in terms of the inclusion of \pu jets as well as the increased fluctuations of the jet response introduced by \pu.

From the performance studies presented in this paper, the object-based \met reconstruction in \ATLAS, which was developed for \LHC \runOne and used in a large number of physics analyses, 
can be used with the discussed refinements and adjustments for \runTwo as well.

\section*{Acknowledgements}


We thank CERN for the very successful operation of the LHC, as well as the
support staff from our institutions without whom ATLAS could not be
operated efficiently.

We acknowledge the support of ANPCyT, Argentina; YerPhI, Armenia; ARC, Australia; BMWFW and FWF, Austria; ANAS, Azerbaijan; SSTC, Belarus; CNPq and FAPESP, Brazil; NSERC, NRC and CFI, Canada; CERN; CONICYT, Chile; CAS, MOST and NSFC, China; COLCIENCIAS, Colombia; MSMT CR, MPO CR and VSC CR, Czech Republic; DNRF and DNSRC, Denmark; IN2P3-CNRS, CEA-DRF/IRFU, France; SRNSFG, Georgia; BMBF, HGF, and MPG, Germany; GSRT, Greece; RGC, Hong Kong SAR, China; ISF, I-CORE and Benoziyo Center, Israel; INFN, Italy; MEXT and JSPS, Japan; CNRST, Morocco; NWO, Netherlands; RCN, Norway; MNiSW and NCN, Poland; FCT, Portugal; MNE/IFA, Romania; MES of Russia and NRC KI, Russian Federation; JINR; MESTD, Serbia; MSSR, Slovakia; ARRS and MIZ\v{S}, Slovenia; DST/NRF, South Africa; MINECO, Spain; SRC and Wallenberg Foundation, Sweden; SERI, SNSF and Cantons of Bern and Geneva, Switzerland; MOST, Taiwan; TAEK, Turkey; STFC, United Kingdom; DOE and NSF, United States of America. In addition, individual groups and members have received support from BCKDF, the Canada Council, CANARIE, CRC, Compute Canada, FQRNT, and the Ontario Innovation Trust, Canada; EPLANET, ERC, ERDF, FP7, Horizon 2020 and Marie Sk{\l}odowska-Curie Actions, European Union; Investissements d'Avenir Labex and Idex, ANR, R{\'e}gion Auvergne and Fondation Partager le Savoir, France; DFG and AvH Foundation, Germany; Herakleitos, Thales and Aristeia programmes co-financed by EU-ESF and the Greek NSRF; BSF, GIF and Minerva, Israel; BRF, Norway; CERCA Programme Generalitat de Catalunya, Generalitat Valenciana, Spain; the Royal Society and Leverhulme Trust, United Kingdom.

The crucial computing support from all WLCG partners is acknowledged gratefully, in particular from CERN, the ATLAS Tier-1 facilities at TRIUMF (Canada), NDGF (Denmark, Norway, Sweden), CC-IN2P3 (France), KIT/GridKA (Germany), INFN-CNAF (Italy), NL-T1 (Netherlands), PIC (Spain), ASGC (Taiwan), RAL (UK) and BNL (USA), the Tier-2 facilities worldwide and large non-WLCG resource providers. Major contributors of computing resources are listed in Ref.~\cite{ATL-GEN-PUB-2016-002}.

\clearpage
\appendix
\section*{Appendix}
\addcontentsline{toc}{section}{Appendix}
\setcounter{section}{1}
\renewcommand{\thesubsection}{\Alph{section}}
\subsection{Glossary of terms}
\label{app:nomenclature}
In this paper several acronyms and qualifiers are used to describe the reconstruction of \met and related observables. 
This brief glossary of terms is intended to help with the nomenclature. 
All terms should be interpreted in the context of \met reconstruction and may have other interpretations in other contexts.
\begin{description}
\item{\textbf{\met reconstruction}:} Using this nomenclature usually encompasses the reconstruction of a set of observables comprising the missing transverse momentum vector \metvec, its components \metxy, and its absolute value \met. 
In addition, the scalar sum \sumet of the \pT of all kinematic objects contributing to \metvec is calculated.  
The calculation of these variables is described in detail in \eqMultiRef{eq:metbasics:comp}{to}{eq:metbasics:phi} in \secRef{subsec:etmiss-basics}.
\item{\textbf{\HS} and \textbf{\pvs}:} The hardest \pp interaction in a given event is referred to as the \emph{hard scatter}.
It is normally associated with a reconstructed \emph{\hs vertex}, which is considered the hardest vertex among all reconstructed \emph{\pvs} in this event. 
The \hs vertex is defined as the one with the largest sum of \pTj{2} of tracks associated with it. 
The other primary vertices are assumed to be produced by \ipu interactions. 
The variable \NPV denotes the number of reconstructed primary collision vertices in the event. 
\item{\textbf{Hard event} and \textbf{hard term}:} The reconstruction of \emph{hard objects} includes individual particles such as electrons, photons, muons and \tauleps,  and jets. 
In all cases the final objects are characterised by a kinematic threshold and reconstruction quality requirements. 
Both the reconstructed charged-particle tracks from the \ID and \topos from the calorimeter are used as the input signals for these objects. 
In the context of \met{} reconstruction, the use of the same detector signals by different hard objects is excluded, see details in \secRef{sec:etmiss-reco}.
The finally accepted hard event objects give rise to the \emph{hard term} in \met{} reconstruction.
\item{\textbf{Soft event} and \textbf{soft term}:} All detector signals recorded for one triggered event and not used by the hard objects discussed above can be considered as \emph{soft signals} contributing to \met. 
They include signals or signal traces from scattered soft particles arising from the underlying event accompanying the \hs interaction, or from statistically completely independent \pu interactions producing 
diffuse particle emissions in the same bunch crossing. 
In addition, signals from particles and jets which do not satisfy the hard-object quality criteria, or are below the kinematic threshold, can be included in the soft event.  
The reference \met reconstruction configuration for the results presented in \secMultiRef{sec:perf}{and}{sec:uncertainties}  uses reconstructed \ID tracks from the soft event to form the \emph{soft term} in \met reconstruction, 
with the track selection details outlined under priority (\ref{ref:track}) in \tabRef{tab:contrib:default} in \secRef{sec:etmiss-reco}.
Alternative configurations employ \topos from the calorimeter, see \tabRef{tab:contrib:alt} in \appRef{app:composition}. 
\end{description}

\addtocounter{section}{1}
\subsection{Alternative \texorpdfstring{\met}{Etmiss} composition}
\label{app:composition}
\setcounter{table}{0}
\renewcommand{\thetable}{\thesection-\arabic{table}}
\setcounter{figure}{0}
\renewcommand{\thefigure}{\thesection-\arabic{figure}}

\TabRef{tab:contrib:alt} summarises the \met reconstruction configurations employing only \ID tracks, or using \topos for the soft term. 

\renewcommand{\arraystretch}{1.35}

\begin{table}[ht!]\centering\small
	\caption{Representations of \met and \sumet calculated from (\ref{ref:idtracks}) reconstructed charged-particle tracks from the \ID or (\ref{ref:topos}) using a soft term from \topos in the calorimeter only.
\label{tab:contrib:alt}}

\renewcommand{\arraystretch}{1.45}

\setlength{\fboxsep}{1pt}
\setlength{\fboxrule}{0.5pt}

\renewcommand{\multirowsetup}{} 
\setcounter{myrefctr}{0}

\begin{tabular}{p{\rwidth}>{$}p{\nwidth}<{$}>{$}p{\bwidth}<{$}>{\centering}p{\dwidth}p{\twidth}}
	\hline\hline
       	\multirow{2}{*}{$\#$}    & \MCC{4}{Objects contributing to \met and \sumet}              \tabularnewline \cline{2-5} 
	           	                   & \LCC{1}{Type} & \LCC{1}{Selections} & Variables & Comments \tabularnewline \hline
	\LCC{1}{\myrefstep{ref:idtracks}}												&   
		\ID\ \track 														& 
		\minitab[l]{\mstackfour{\abseta < 2.5}{\pT > \unit{400}{\MeV}}{\absdzero < \unit{1.5}{\mm}}{\abszzsth < \unit{1.5}{\mm}}}	& 
		\minitab[c]{\mstacktwo{\pmiss}{\sumettrk}}                              							&
		charged-particle-based estimators for \met, \sumet using all reconstructed tracks from the \hs vertex passing requirements for high-quality reconstruction in addition to kinematic selections
		\tabularnewline\hline
	\LCC{1}{\myrefstep{ref:topos}}                  										& 
		\text{\topo}														& 
		\ecluslcw > 0 \text{\ (soft\ term)}											& 
		\minitab[c]{\mstacktwo{\mettcsft}{\sumettcsft}}                                                                         &
		Variant reconstructing \metcst (\sumetcalo) using a soft term \mettcsft (\sumettcsft) reconstructed from \topos not used by, or not overlapping with, the hard objects used for the hard term composed of items (\ref{ref:epm})--(\ref{ref:jet}) in \tabRef{tab:contrib:default}, with the jet selection described in \secRef{subsec:etmiss-calo} applied
		\tabularnewline\hline\hline
 \end{tabular}
\end{table}

\addtocounter{section}{1}
\subsection{Jet selection}
\label{app:jet}
\setcounter{table}{0}
\renewcommand{\thetable}{\thesection-\arabic{table}}
\setcounter{figure}{0}
\renewcommand{\thefigure}{\thesection-\arabic{figure}}

\begin{figure}[t!]\centering
\subfigure[]{\includegraphics[width=\fighalfwidth]{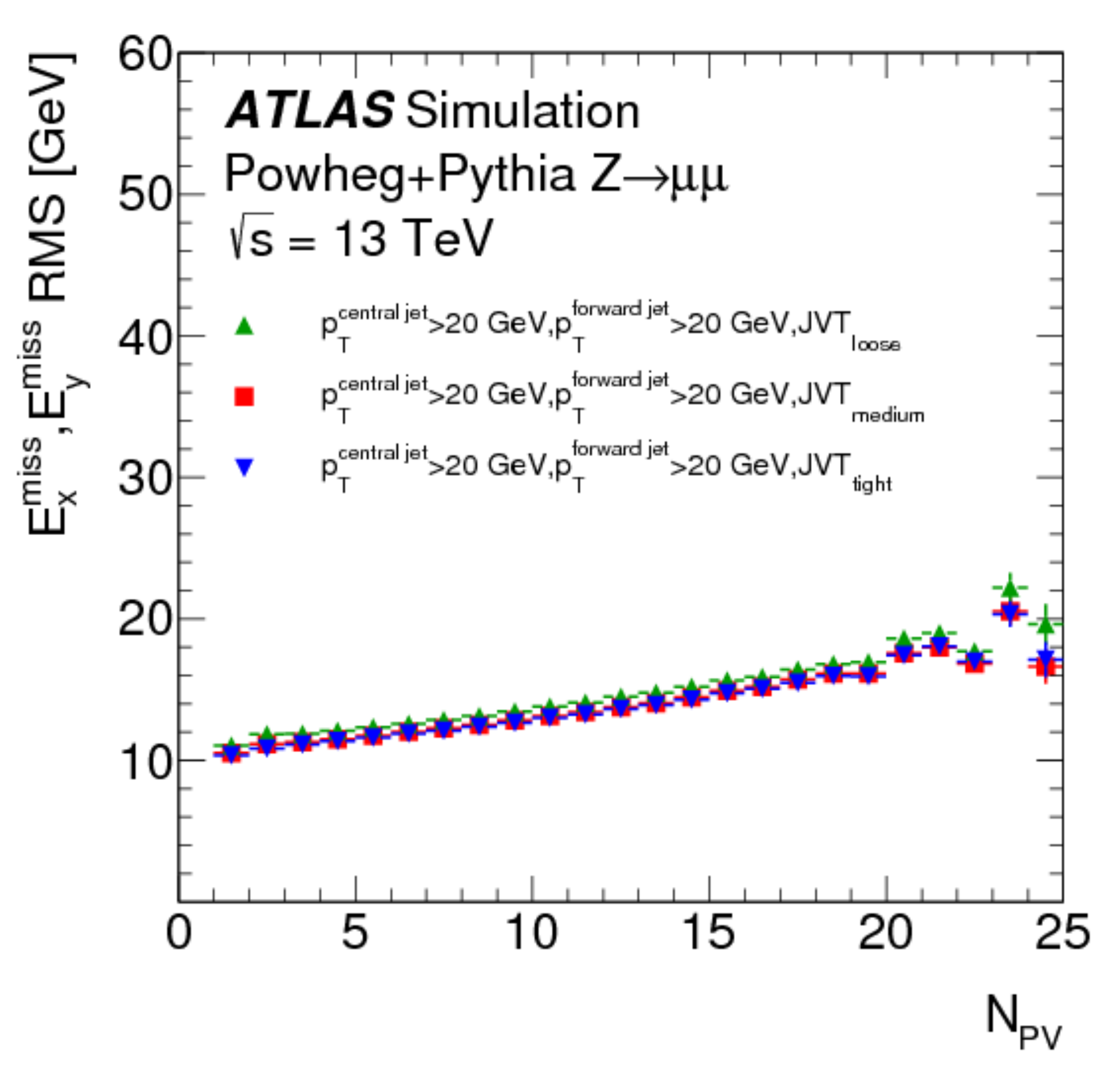}\label{fig:perf:jets:resojvt}}
\subfigure[]{\includegraphics[width=\fighalfwidth]{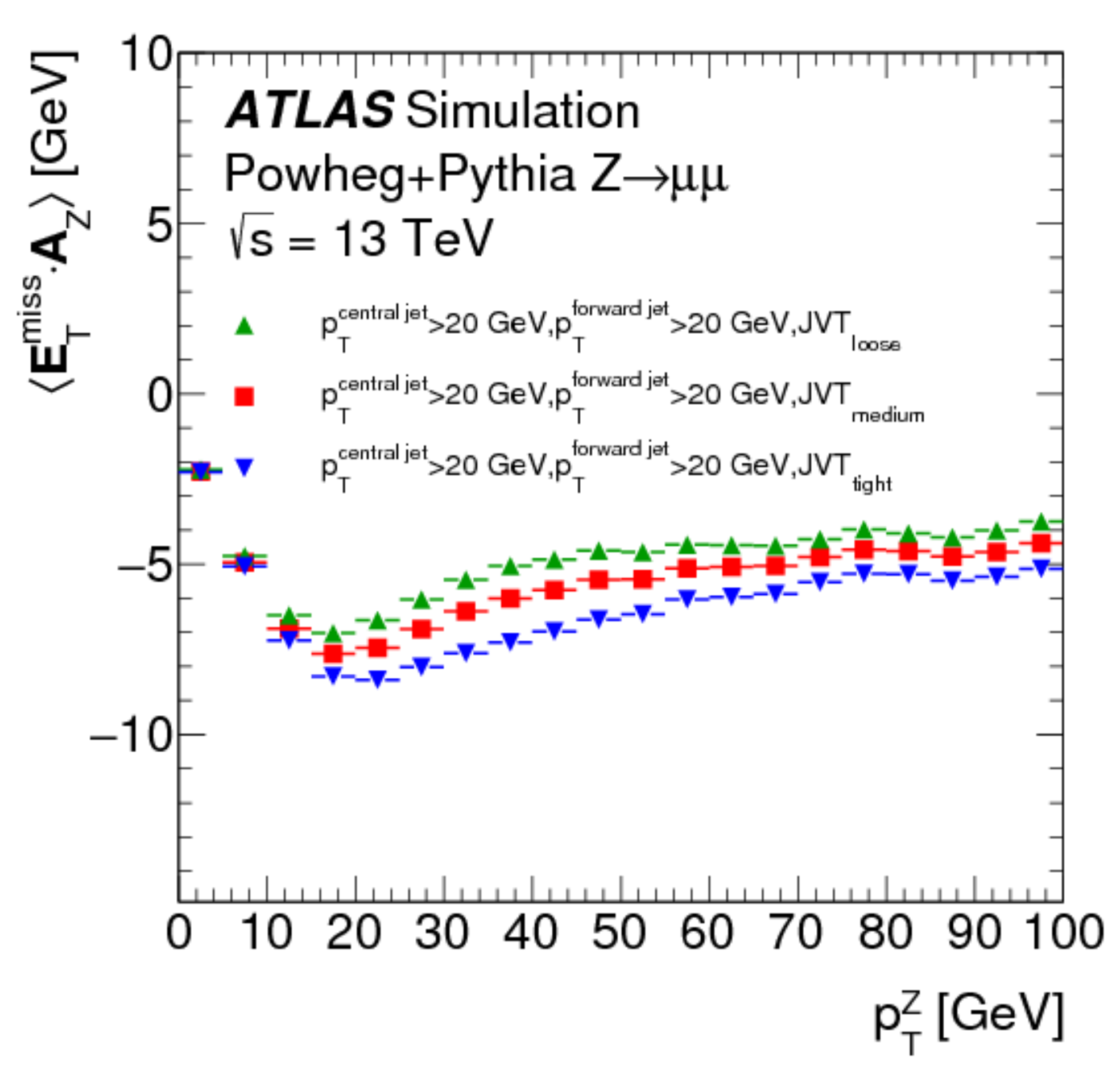}\label{fig:perf:jets:scalejvt}}
\caption{\met resolution and response for the different choices of \JVT discussed in the text, as measured for \Zmm events in \MC simulation.  The \subref{fig:perf:jets:resopt} \met resolution is shown as function of the \pu activity measured by the number of primary vertices \NPV, and the \subref{fig:perf:jets:scalept} \met response is shown as function of the transverse momentum \ptZ of the \Zboson boson.  A global selection of $\pT > \unit{20}{\GeV}$ is applied to the transverse momentum of jets within $\abseta < 2.4$ (\pTj{\text{central\ jet}}) and for forward jets with $\abseta \geq 2.4$ (\pTj{\text{forward\ jet}}).}
\label{fig:performance:jetComparejvt}
\end{figure}

As discussed in \secRef{subsec:jet-sel}, jets that are not rejected by the signal ambiguity resolution and have $\pT > \unit{60}{\GeV}$ contribute to \met reconstruction. 
Jets with less transverse momentum that fall within $\abseta < 2.4$ are subjected to further selection based on \JVT calculated by a track-based jet vertex tagger \cite{Aad:2015ina}.
Three values for \JVT, each representing a different efficiency for the reconstruction of non-\pu jets, were considered in the course of the optimisation of the \JVT-based selection,
\begin{center}
\begin{minipage}{0.95\textwidth}
$\JVT_{\text{tight}}      > 0.11$ \dotfill\ tight selection with high \pu rejection power at lower signal efficiency;\\
$\JVT_{\text{medium}} > 0.59$ \dotfill\ medium selection with good signal efficiency and \pu rejection power;\\
$\JVT_{\text{loose}}     > 0.91$ \dotfill\ loose selection with lower \pu rejection power and higher signal efficiency.
\end{minipage}
\end{center}
The effects of these selections on the \met resolution \metresxy and response are shown in \figRef{fig:performance:jetComparejvt}, for \Zmm events in \MC simulation. 
\SubfigRef{fig:performance:jetComparejvt}{fig:perf:jets:resojvt} shows that the \pu dependence of \metresxy is not significantly affected by the choice for \JVT.
However, the \met response measured by the projection given in \eqRef{eq:projpar} in \secRef{subsec:etmiss-scale:method} and evaluated as a function of \ptZ in the same sample, shows significant sensitivity to the choice of \JVT, as seen in \subfigRef{fig:performance:jetComparejvt}{fig:perf:jets:scalejvt}.  

\begin{figure}[ht!]\centering
\subfigure[]{\includegraphics[width=\fighalfwidth]{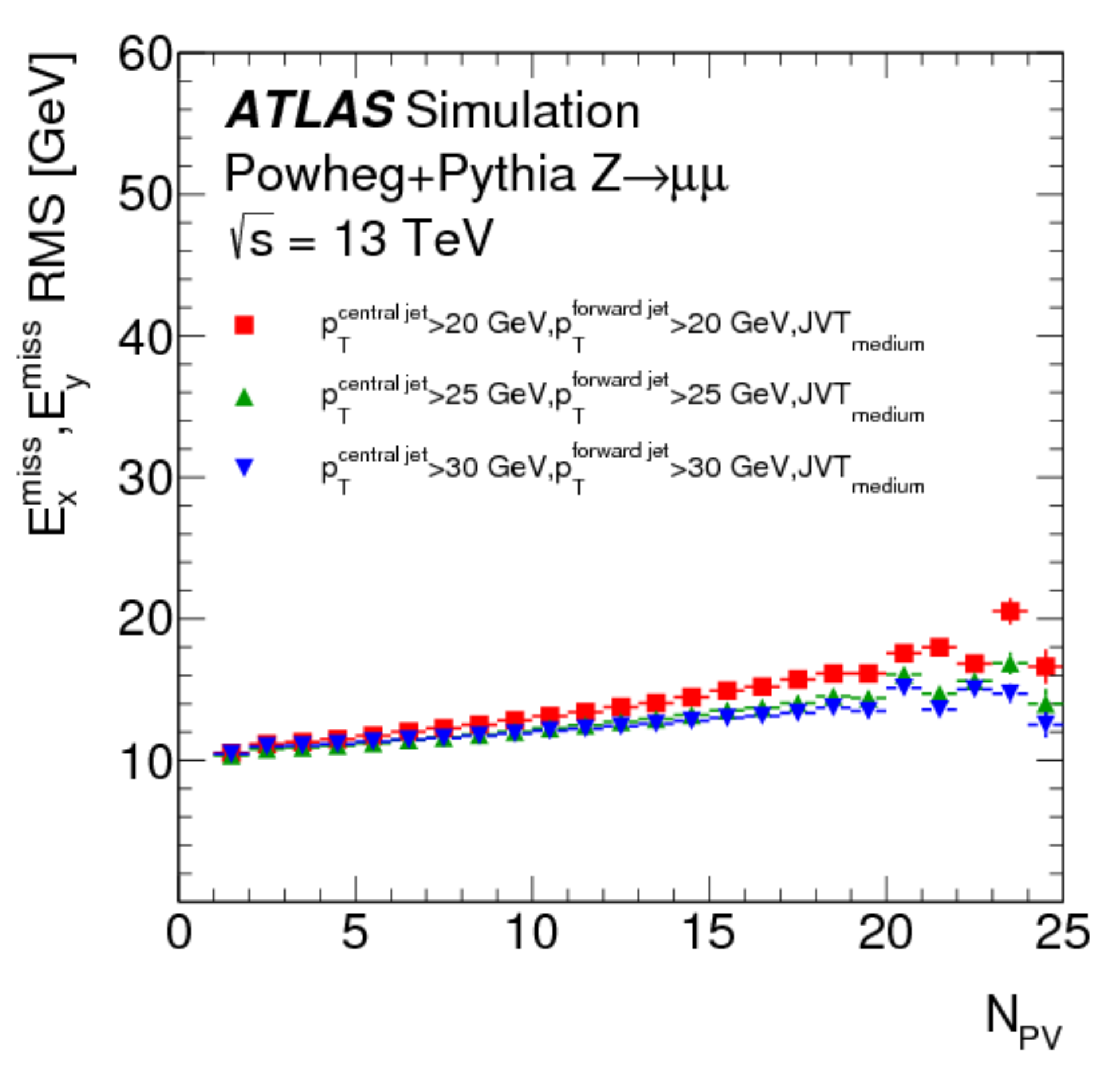}\label{fig:perf:jets:resopt}}
\subfigure[]{\includegraphics[width=\fighalfwidth]{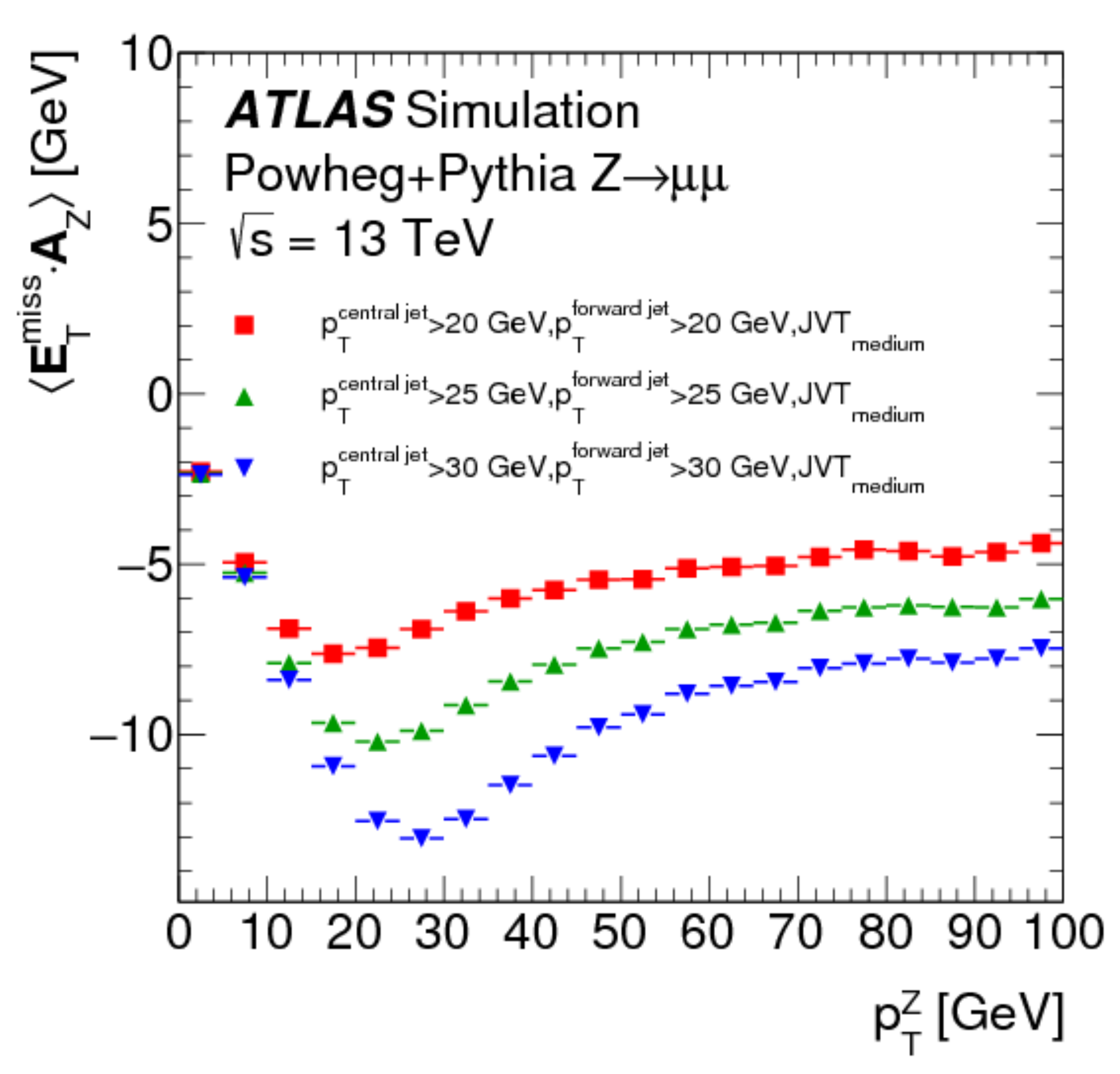}\label{fig:perf:jets:scalept}}
\caption{\met resolution and scale for different global jet-\pT thresholds, for \Zmm events in \MC simulation. The \subref{fig:perf:jets:resopt} \met resolution is shown as function of the \pu activity measured by the number of primary vertices \NPV, and the \subref{fig:perf:jets:scalept} \met response is shown as function of the transverse momentum \ptZ of the \Zboson boson. The same respective thresholds are applied to the transverse momentum of jets within $\abseta < 2.4$ (\pTj{\text{central\ jet}}) and for forward jets with $\abseta \geq 2.4$ (\pTj{\text{forward\ jet}}), with $\pTj{\text{central\ jet}}, \pTj{\text{forward\ jet}} > \unit{\{ 20, 25, 30 \}}{\GeV}$.} 
\label{fig:performance:jetComparept}
\end{figure}

In addition to the signal ambiguity resolution and the choice for \JVT, the contribution from jets in \met reconstruction is controlled by a kinematic threshold requiring the transverse momentum of the jet to be $\pT > \unit{20}{\GeV}$. 
The effects of variations of this threshold on \metresxy and the \met response are shown in \figRef{fig:performance:jetComparept}.
Increasing the threshold to \unit{30}{\GeV} for all jets satisfying the $\JVT_{\text{medium}}$ condition reduces the \pu dependence of the resolution shown in \subfigRef{fig:performance:jetComparejvt}{fig:perf:jets:resopt}, but leads to significant loss of \met response, as seen in \subfigRef{fig:performance:jetComparejvt}{fig:perf:jets:scalejvt}. 
Depending on the sensitivities observed in a given physics analysis, the \pT-threshold choice for the jet contribution to \met reconstruction needs to be adjusted to meet the required performance.  

\begin{figure}[ht!]\centering
\subfigure[]{\includegraphics[width=\fighalfwidth]{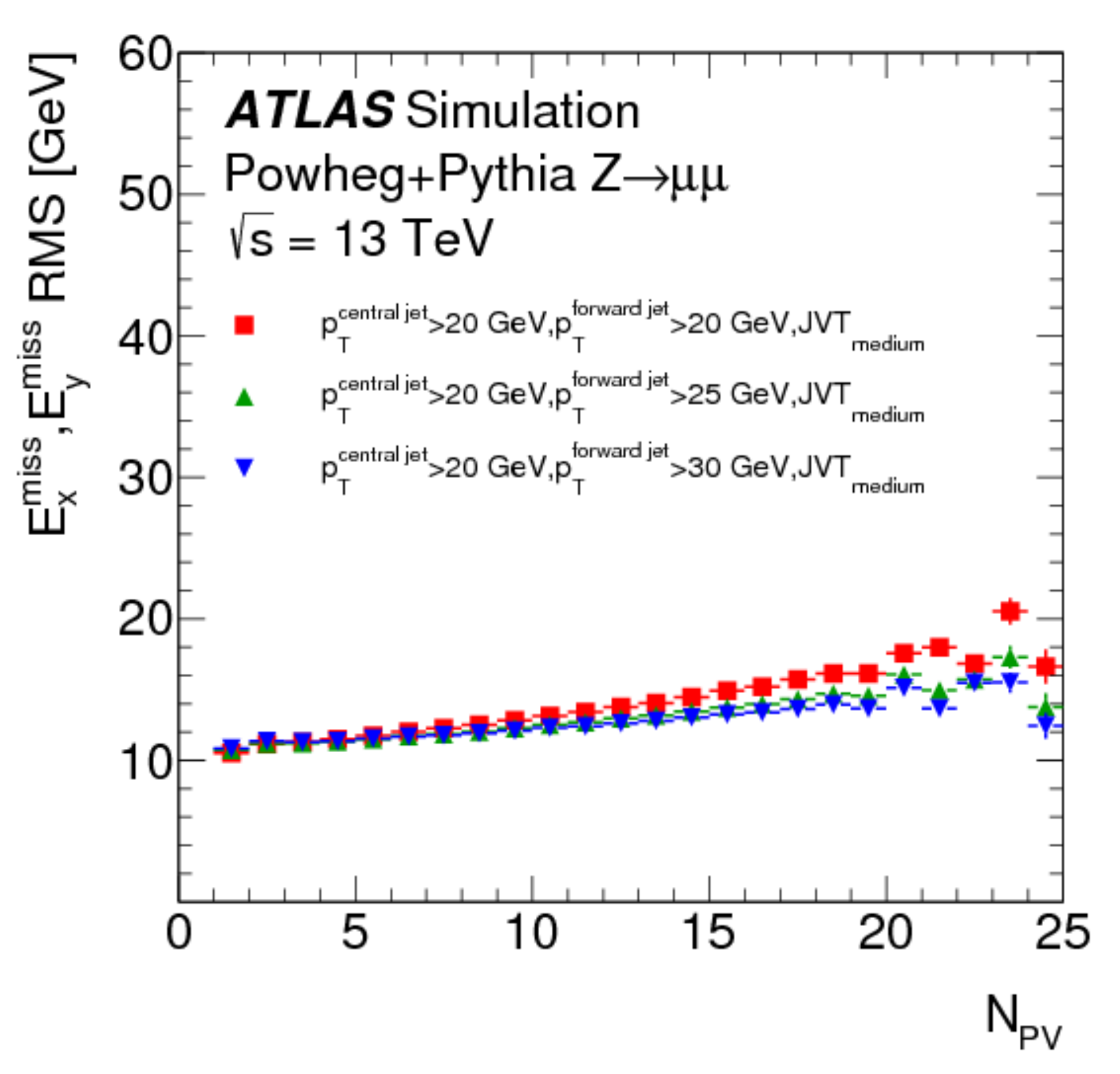}\label{fig:perf:jets:resofpt}}
\subfigure[]{\includegraphics[width=\fighalfwidth]{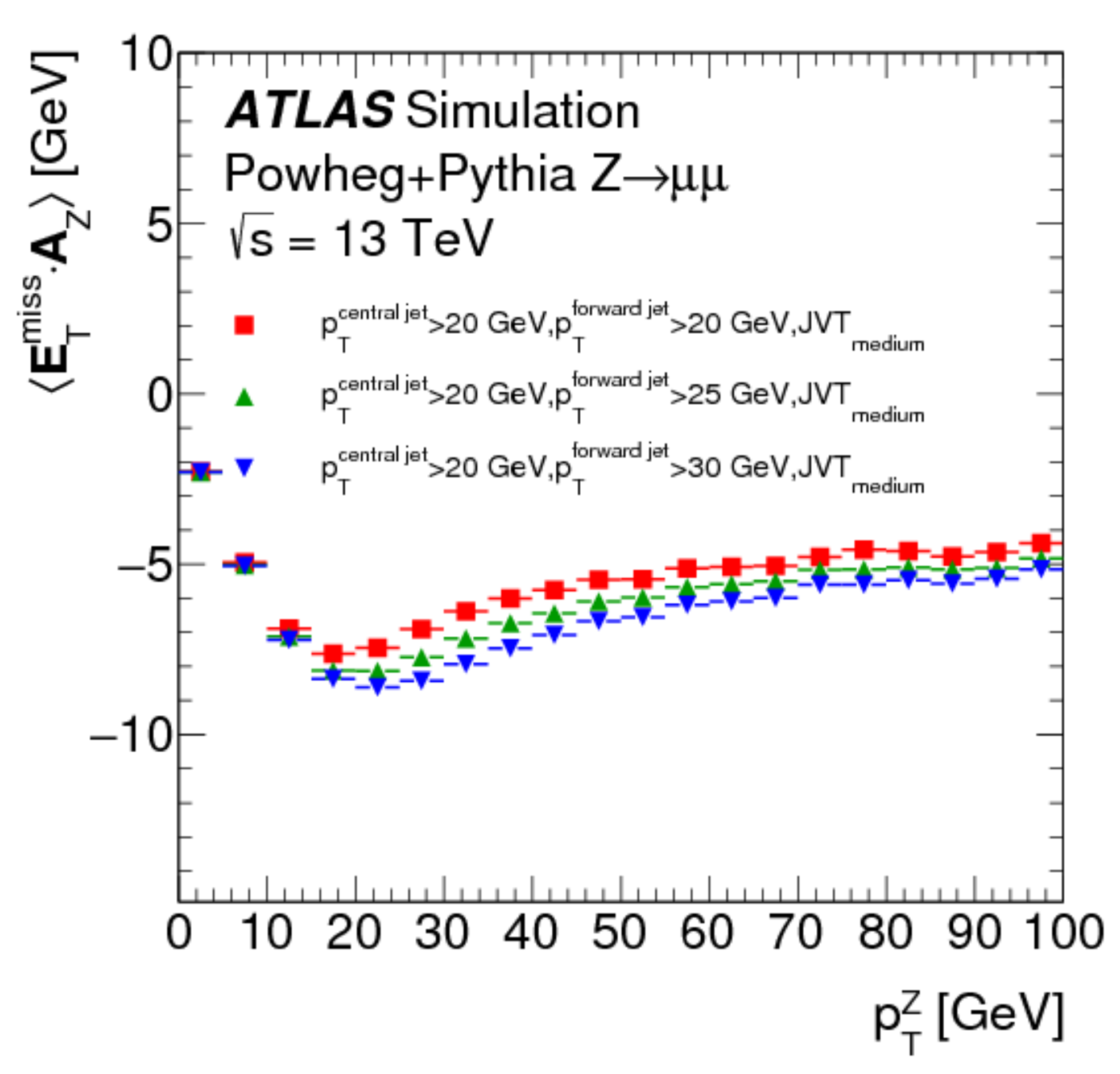}\label{fig:perf:jets:scalefpt}}
\caption{\met resolution and scale for different regional jet-\pT thresholds, for \Zmm events in \MC simulation. The \subref{fig:perf:jets:resopt} \met resolution is shown as function of the \pu activity measured by the number of primary vertices \NPV, and the \subref{fig:perf:jets:scalept} \met response is shown as function of the transverse momentum \ptZ of the \Zboson boson. The same threshold is applied to the transverse momentum of jets within $\abseta < 2.4$ ($\pTj{\text{central\ jet}} > \unit{20}{\GeV}$), while for forward jets with $\abseta \geq 2.4$ threshold variations ($\pTj{\text{forward\ jet}} > \unit{\{20, 25,30\}}{\GeV}$) are studied.}
\label{fig:performance:jetComparefpt}
\end{figure}

Extending the \pT threshold studies with the option of regional thresholds yields the performance results presented in \figRef{fig:performance:jetComparefpt}. 
In this case jets within $\abseta < 2.4$ are subjected to the $\pTj{\text{central\ jet}} > \unit{20}{\GeV}$ selection, while jets outside of this $\eta$ range are filtered using $\pTj{\text{forward\ jet}} > \unit{\{20, 25, 30\}}{\GeV}$. 
This leads to the improvements in the \pu dependence of \metresxy shown in \subfigRef{fig:performance:jetComparefpt}{fig:perf:jets:resofpt}, which are very similar to the ones observed in \subfigRef{fig:performance:jetComparept}{fig:perf:jets:resopt} for a global jet-\pT threshold variation. 
The comparison indicates that the main \pu contribution to \metresxy is introduced by forward jets, for which no \JVT-based \pu-mitigation is available.

Increasing the \pT threshold only for forward jets reduces the average loss of response observed in case of the global \pT threshold increase. 
This can be seen by comparing the results shown in \subfigRef{fig:performance:jetComparefpt}{fig:perf:jets:scalefpt} for regional \pT thresholds with the ones shown in \subfigRef{fig:performance:jetComparept}{fig:perf:jets:scalept} for global thresholds.      
Like for the \JVT threshold selection, the choice of the appropriate global or regional \pT-threshold depends on the \met reconstruction performance required in the context of a given analysis.

\FloatBarrier
\printbibliography
\clearpage
 
\begin{flushleft}
{\Large The ATLAS Collaboration}

\bigskip

M.~Aaboud$^\textrm{\scriptsize 34d}$,    
G.~Aad$^\textrm{\scriptsize 99}$,    
B.~Abbott$^\textrm{\scriptsize 125}$,    
O.~Abdinov$^\textrm{\scriptsize 13,*}$,    
B.~Abeloos$^\textrm{\scriptsize 129}$,    
S.H.~Abidi$^\textrm{\scriptsize 165}$,    
O.S.~AbouZeid$^\textrm{\scriptsize 143}$,    
N.L.~Abraham$^\textrm{\scriptsize 153}$,    
H.~Abramowicz$^\textrm{\scriptsize 159}$,    
H.~Abreu$^\textrm{\scriptsize 158}$
% ...
\bigskip
\\

% ...
$^{13}$Institute of Physics, Azerbaijan Academy of Sciences, Baku; Azerbaijan.\\
% ...
$^{34}$$^{(a)}$Facult\'e des Sciences Ain Chock, R\'eseau Universitaire de Physique des Hautes Energies - Universit\'e Hassan II, Casablanca;$^{(b)}$Centre National de l'Energie des Sciences Techniques Nucleaires (CNESTEN), Rabat;$^{(c)}$Facult\'e des Sciences Semlalia, Universit\'e Cadi Ayyad, LPHEA-Marrakech;$^{(d)}$Facult\'e des Sciences, Universit\'e Mohamed Premier and LPTPM, Oujda;$^{(e)}$Facult\'e des sciences, Universit\'e Mohammed V, Rabat; Morocco.\\
% ...
$^{99}$CPPM, Aix-Marseille Universit\'e, CNRS/IN2P3, Marseille; France.\\
% ...
$^{125}$Homer L. Dodge Department of Physics and Astronomy, University of Oklahoma, Norman OK; United States of America.\\
% ...
$^{129}$LAL, Universit\'e Paris-Sud, CNRS/IN2P3, Universit\'e Paris-Saclay, Orsay; France.\\
% ...
$^{143}$Santa Cruz Institute for Particle Physics, University of California Santa Cruz, Santa Cruz CA; United States of America.\\
% ...
$^{153}$Department of Physics and Astronomy, University of Sussex, Brighton; United Kingdom.\\
% ...
$^{158}$Department of Physics, Technion, Israel Institute of Technology, Haifa; Israel.\\
$^{159}$Raymond and Beverly Sackler School of Physics and Astronomy, Tel Aviv University, Tel Aviv; Israel.\\
% ...
$^{165}$Department of Physics, University of Toronto, Toronto ON; Canada.\\
% ...
$^{a}$ Also at  Department of Physics, University of Malaya, Kuala Lumpur; Malaysia.\\
$^{b}$ Also at Borough of Manhattan Community College, City University of New York, NY; United States of America.\\
$^{c}$ Also at Centre for High Performance Computing, CSIR Campus, Rosebank, Cape Town; South Africa.\\
$^{d}$ Also at CERN, Geneva; Switzerland.\\
$^{e}$ Also at CPPM, Aix-Marseille Universit\'e, CNRS/IN2P3, Marseille; France.\\
% ...
$^{*}$ Deceased

\end{flushleft}


\end{document}